\documentclass[article]{aa}  

\usepackage{graphicx}
\usepackage{subfig}
\usepackage{txfonts}
\usepackage{color}

\newcommand{\msun}{\mbox{$M_\odot$}}

\begin{document}

   \title{Chemical evolution 
   during the formation of a protoplanetary disk}

   \author{A. Coutens \inst{1} 
   \and B. Commer\c con \inst{2}
   \and V. Wakelam \inst{1}
          }

   \institute{Laboratoire d'Astrophysique de Bordeaux, Univ. Bordeaux, CNRS, B18N, all\'ee Geoffroy Saint-Hilaire, 33615 Pessac, France \\
              \email{audrey.coutens@u-bordeaux.fr}
         \and Centre de Recherche Astrophysique de Lyon UMR5574, ENS de Lyon, Univ. Lyon1, CNRS, Universit\'e de Lyon, 69007 Lyon, France  
             }

   \date{Received xxx; accepted xxx}
 
  \abstract
  {The chemical composition of protoplanetary disks is expected to impact the composition of the forming planets. Characterizing the diversity of chemical composition in disks and the physicochemical factors that lead to this diversity is consequently of high interest.}
  {The aim of this study is to investigate the chemical evolution from the prestellar phase to the formation of the disk, and to determine the impact that the chemical composition of the cold and dense core has on the final composition of the disk.}
  {We performed 3D nonideal magneto-hydrodynamic (MHD) simulations of a dense core collapse using the adaptive-mesh-refinement RAMSES code. For each particle ending in the young rotationally supported disk, we ran chemical simulations with the three-phase gas-grain chemistry code Nautilus. Two different sets of initial abundances, which are characteristic of cold cores, were considered. The final distributions of the abundances of common species were compared to each other, as well as with the initial abundances of the cold core.}
  {We find that the spatial distributions of molecules reflect their sensitivity to the temperature distribution. The main carriers of the chemical elements in the disk are usually the same as the ones in the cold core, except for the S-bearing species, where HS is replaced by H$_2$S$_3$, and the P-bearing species, where atomic P leads to the formation of PO, PN, HCP, and CP. However, the abundances of less abundant species change over time. This is especially the case for ``large'' complex organic molecules (COMs) such as CH$_3$CHO, CH$_3$NH$_2$, CH$_3$OCH$_3$, and HCOOCH$_3$ which see their abundances significantly increase during the collapse. These COMs often present similar abundances in the disk despite significantly different abundances in the cold core. In contrast, the abundances of many radicals decrease with time. A significant number of species still show the same abundances in the cold core and the disk, which indicates efficient formation of these molecules in the cold core. This includes H$_2$O, H$_2$CO, HNCO, and ``small'' COMs such as CH$_3$OH, CH$_3$CN, and NH$_2$CHO. We computed the MHD resistivities within the disk for the full gas--grain chemical evolution and find  results in qualitative agreement with the literature assuming simpler chemical networks.}
  {In conclusion, the chemical content of prestellar cores is expected to affect the chemical composition of disks. The impact is more or less important depending on the type of species. Users of stand-alone chemical models of disks should pay special attention to the initial abundances they choose. } 
 

   \maketitle
%

\section{Introduction}

Planets are known to form in the disk of gas and dust that arises during the star formation process following the accretion of material onto the protostar. The chemical composition of the disk should consequently determine the composition of the planetesimals that will lead to the future planets \citep[e.g.,][]{Thiabaud2015,Kamp2020}. Characterizing the disk composition in young stellar objects is therefore of significant interest for our understanding of planet formation. 

To make an inventory of the molecules present in disks, observational studies have been carried out towards several protoplanetary disks \citep[e.g.,][]{Dutrey2014,Loomis2020}. Although some complex organic molecules (COMs) such as methanol (CH$_3$OH) and methyl cyanide (CH$_3$CN) have been detected in the gas phase with the Atacama Large Millimeter/submillimeter Array (ALMA), their detected emission does not stem from the midplane, but from higher regions of the disk \citep{Oberg2015,Walsh2016}. Many species remain undetected compared to what is found in much younger protostars, as these species are expected to be frozen out on grains due to the low temperatures in the midplane of disks. The composition of ices was investigated with the {\it Spitzer Space Telescope} and the {\it Infrared Space Observatory}, but the detections were limited to the most abundant species (H$_2$O, CO, CO$_2$, CH$_4$, CH$_3$OH, NH$_3$, e.g., \citealt{Boogert2008,Oberg2008}) and associated to protostellar envelopes and cold cores instead of disks. It is consequently necessary to wait for the {\it James Webb Space Telescope} to hopefully detect a large variety of species in the ices of disks. 

Another efficient way to investigate the chemical composition of disks consists in running chemical simulations for realistic evolutions of the physical conditions from the molecular cloud phase up to formation of the disk. Two approaches are usually considered for the physical evolution: two dimensional (2D) semi-analytic models (e.g., \citealt{Visser2009,Visser2011,Drozdovskaya2014,Drozdovskaya2016}) and multi-dimensional (2D or 3D) hydrodynamic or magneto-hydrodynamic (MHD) simulations (e.g., \citealt{vanWeeren2009,Furuya2012,Hincelin2013,Hincelin2016}). The advantage of chemical models is that they also allow us to explore  the chemical diversity in more detail and in particular to characterize the factors that impact chemical composition. Physical changes are known to affect the chemistry. It remains to be known whether the chemical composition of disks is mainly determined by their physical evolution or if the composition of cold cores can play a major role. 
Cold cores have been shown to present a diversity of chemical compositions \citep{Lefloch2018, Ruaud2018}, which begs the question of whether or not this early diverse composition can be inherited by the disk. 

In this study, we used 3D MHD simulations of a dense core collapse to investigate the chemical evolution that occurs from the cold core phase to the formation of a young disk. We tested two different sets of initial abundances and compared the abundances of the particles ending in the disk to characterize their impact on the composition of the disk. The paper is organized as follows. The physical and chemical simulations are described in Sections 2 and 3, respectively. The results are presented in Section 4, and then discussed in Section 5. Finally, we conclude in Section 6.


\section{Physical simulations}

\subsection{Physical model} 

We performed 3D dense-core-collapse calculations using the adaptive-mesh-refinement (AMR) \ttfamily{RAMSES}~\rm code \citep{Teyssier2002}, which integrates the equation of resistive MHD \citep{Teyssier2006,Fromang2006,Masson2012}. \ttfamily{RAMSES}~\rm uses a finite-volume second-order Godunov scheme to integrate the conservative MHD equations and a constrained transport algorithm for the induction equation \citep{Evans1988}. In this study, our models only account for ambipolar diffusion, which has been found to be the most efficient resistive effect regulating the disk formation \citep{Hennebelle2016,Masson2016}. 

We use \cite{Boss1979} initial conditions, which consist of a uniform temperature and density sphere of mass $M_0=1$~\msun. The initial temperature, density, and radius are $T_0=10$~K, $\rho_0=2.3\times10^{-18}$~g~cm$^{-3}$ ($\sim$ 6\,$\times$\,10$^5$ cm$^{-3}$), and $R_0=$~3 940~au. The initial ratio of thermal energy to gravitational energy is $\alpha=0.4$. We do not apply density perturbation or solid body rotation, but we introduce initial turbulent velocity perturbations following a Kolmogorov power spectrum. More details on the initial setup of the turbulent velocity field can be found in \citet{Hennebelle2011}, \citet{Commercon2011}, and \citet{Joos2013}. The amplitude of the velocity perturbations are scaled such that the initial turbulent Mach number is trans-sonic: $\mathcal{M}_\mathrm{turb,0}=v_\mathrm{rms,0}/c_\mathrm{s,0}=0.9$, with $v_\mathrm{rms,0}$ being the initial rms turbulent velocity and $c_\mathrm{s,0}$ the initial isothermal sound speed at 10~K. To mimic the thermal behavior of the collapsing dense core during the initial collapse and first Larson core formation \citep{Larson1969,Commercon2011b,Vaytet2013}, we use a barotropic equation of state (EOS)
\begin{equation}
    \frac{P}{\rho} = c^2_\mathrm{s,0} \left( 1+ \left(\frac{\rho}{\rho_\mathrm{c}} \right)^{\gamma-1}\right),
\end{equation}
with $P$ the thermal pressure, $\rho$ the gas density, and $\gamma=5/3$ the ratio of specific heats. $\rho_\mathrm{c}=10^{-13}$ g cm$^{-3}$ represents the critical density at which the gas turns adiabatic. 
The gas temperature is given by the barotropic EOS (eq. 1), which relates the density to the temperature. Last, the initial magnetic field ($B_0$ $\sim$ 75 $\mu$G) is set to be uniform in the $z$-direction, with a mass-to-flux over critical mass-to-flux ratio of  $\mu=5$, where $\mu=(M_0/\phi)/(M/\phi)_\mathrm{crit}$, $\phi=\pi B_0 R_0^2$ is the magnetic flux, and $(M/\phi)_\mathrm{crit}=0.53/3\pi\times(5/G)^{1/2}$ \citep{Mouschovias1976}. To estimate the ambipolar diffusion resistivity, we use the equilibrium abundances tables from \cite{Marchand2016}.

The initial grid has a resolution of $2^6$ cells per direction, and the simulations box length is $8R_0$. Outside the initial dense core, the gas is in thermal equilibrium (i.e., T\,=\,10 K) but has a density of $\rho_0$/100. The grid is refined according to the Jeans length criterion, with at least 12 cells per Jeans length. The maximum resolution is $2^{16}$, which is $\simeq 0.5$~au. This type of initial conditions and numerical setup has been extensively used in previous studies \citep[e.g.,][]{Commercon2008,Commercon2012,Hincelin2016,Masson2016,Vaytet2018}.

We introduced $10^6$ tracer particles with an initial uniform distribution within the initial sphere, which allowed us to reconstruct their trajectories and extract the evolution of their density, temperature, and visual extinction with time. This information was subsequently used as input parameters for the chemical modeling (see Section \ref{sect_chem_model}).   The dust temperature is assumed to be equal to the gas temperature. The time steps are determined by the stability conditions to integrate the Euler equations for the gas dynamics. In order to compute the visual extinction, we used the column density estimator presented in \cite{Valdivia2014}. A similar strategy for the tracer particles was used in \cite{Hincelin2013,Hincelin2016}. 

Our physical model is well adapted to the early phases of star formation, but has a few limitations, which are detailed below.  First,  the barotropic EOS is a crude approximation of the radiative transfer modeling of the star--disk system, but remains efficient as long as no central object with an internal luminosity (the protostar) has formed. Second, we do not employ sink particles either to describe the central collapsing part, as is done in \citet{Hennebelle2020} for instance, because we restrict our study to the first Larson core stage. Sink particles and a more accurate radiation transport scheme (e.g., \citealt{Mignon2020}) will be used in future works investigating a longer time evolution.

\subsection{Characteristics of the disk at the final time of the simulation} 
\label{sect_disk_final}

\begin{figure*}[!ht]
\begin{center}
 \includegraphics[width=1\hsize]{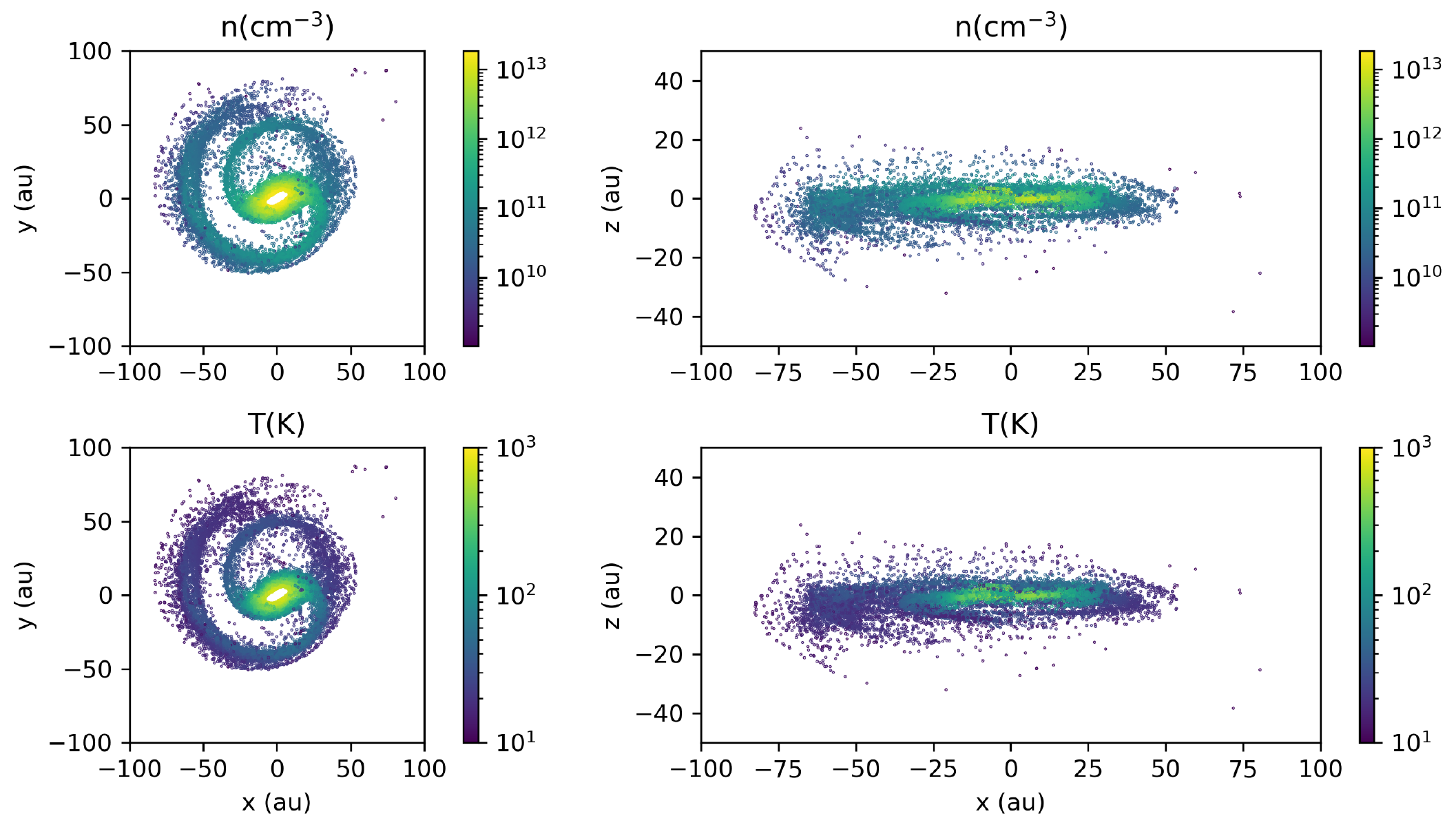} 
 \caption{Spatial distribution of the density (top) and temperature (bottom) of the 14 948 particles in the disk at the final time of the simulations (5.83\,$\times$\,10$^4$~yr). The left panels show the radial distribution of all the particles (stacked z axis) in the x-y plane, while the right panels show the vertical distribution in the x-z plane (stacked y axis).}
\label{fig_phys_cond}
\end{center}
\end{figure*}

We define $t$ = 0 as the starting time of the MHD simulations. The collapsing dense core reaches the first hydrostatic core (FHSC) stage at $t_\mathrm{FHSC}$ = 5.01\,$\times$\,10$^4$~yr. The calculations are continued for an additional 8.2\,$\times$\,10$^3$ yr. At the final time of the simulation (i.e., 5.83\,$\times$\,10$^4$ yr), the particles reach different components: the collapsing envelope, the outflow, the magnetic pseudo-disk \citep{Galli1993}, the rotationally supported disk, and the central core (i.e., the FHSC). In this paper, we focus on the rotationally supported disk only. The rotationally supported disk is the precursor of the protoplanetary disk. 
\citet{Hincelin2016} explained in their Section 3.1.1 the different criteria to determine the component the particles belong to. In this work, we use the coordinate system shifted to the FHSC center of mass with a vertical axis aligned with the angular momentum of the inner 100 au. For the rotationally supported disk, four criteria have to be fulfilled. The azimuthal velocity $v_\phi$ has to be at least twice larger than the radial velocity $v_r$, so that we do not consider particles that collapse too quickly. The azimuthal velocity should also be twice larger than the vertical velocity $v_z$. Particles with a smaller azimuthal velocity would rather come from the outflow cavity wall. The rotational support is at least twice the thermal support and the density is above 10$^9$ cm$^{-3}$. In the end, 14 948 out of the 10$^6$ tracer particles introduced in the simulation end in the disk. 
Figure \ref{fig_phys_cond} shows the spatial distribution of the temperature and the density of the particles in the disk at the final time of the simulation. The disk presents spiral arms and has a diameter of about 100 au. 
From the outer region to the inner region of the disk, the temperature goes from 11 K to 802 K, while the density ranges between 1\,$\times$\,10$^9$ cm$^{-3}$ and 2\,$\times$\,10$^{13}$ cm$^{-3}$. The upper panels of Figure \ref{fig_histo_structure} show the distribution of final density and temperature of the disk particles with histograms. The particles are relatively well spread radially, while they are vertically concentrated on the midplane (see Figure \ref{fig_histo_radius_vertical}). We count 81\% of particles with $|z|$ < 5 au, 15\% with 5 $\leq$ $|z|$ < 10 au, and 4\% with $|z|$ $\geq$ 10 au. For the same radius, no strong variation in density or temperature is seen vertically. The particles with the highest $|z|$ present low densities and temperatures, but mainly because their radii are also high.
The vertical structure of the disk is affected by the low resolution of the AMR grid within the disk, which may smooth the density and temperature distribution to the grid scale. There are typically 5-20 cells in the vertical extent of the disk depending on the radius. Describing the vertical structure of the disk would require a much higher resolution (and physical modeling) and is beyond the scope of the present study.

\subsection{Evolution of the physical conditions for the particles ending in the disk} 
\label{Sect_evol}

\begin{figure}[!ht]
\begin{center}
 \includegraphics[width=1.\hsize]{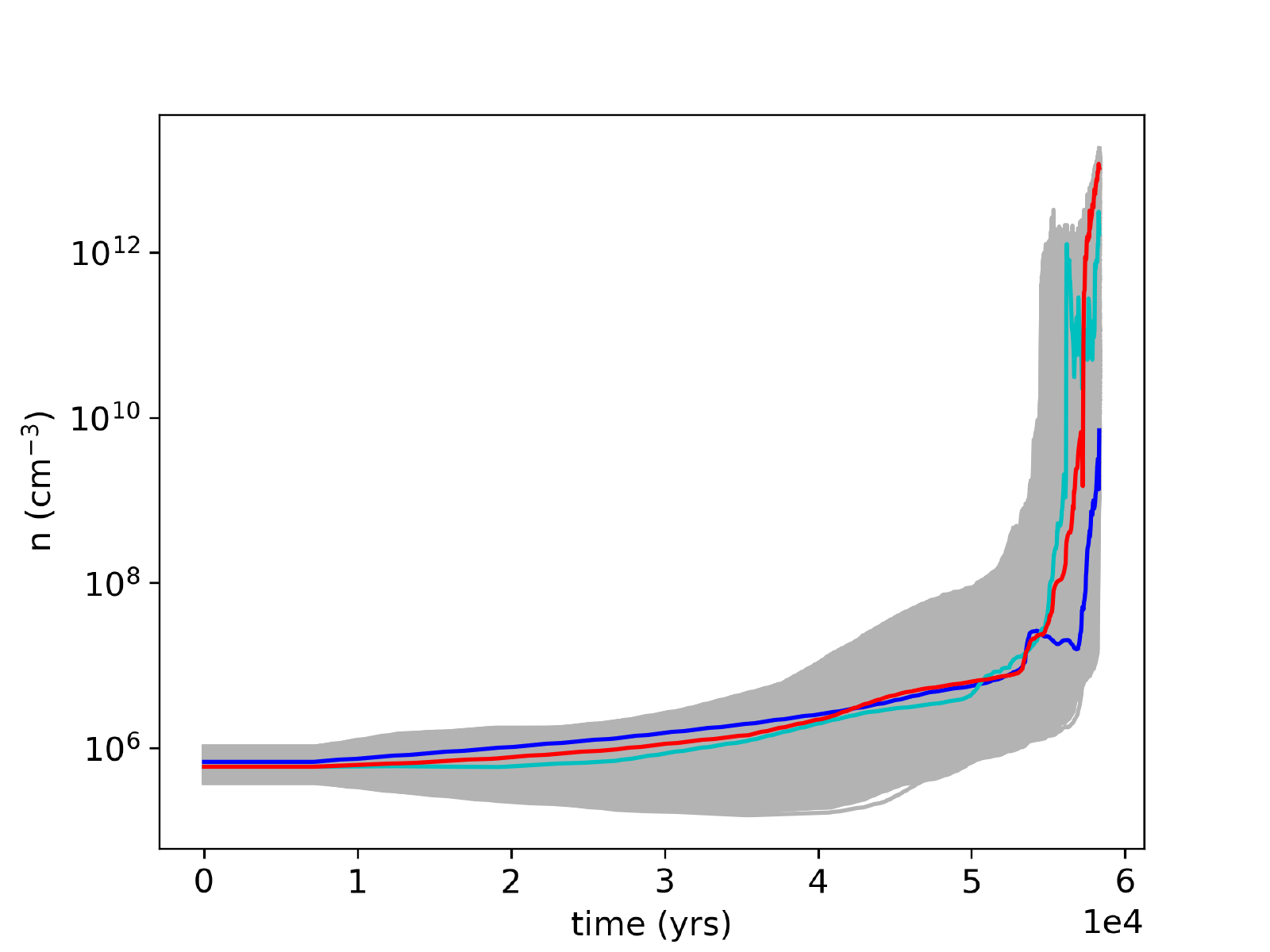}
 \includegraphics[width=1.\hsize]{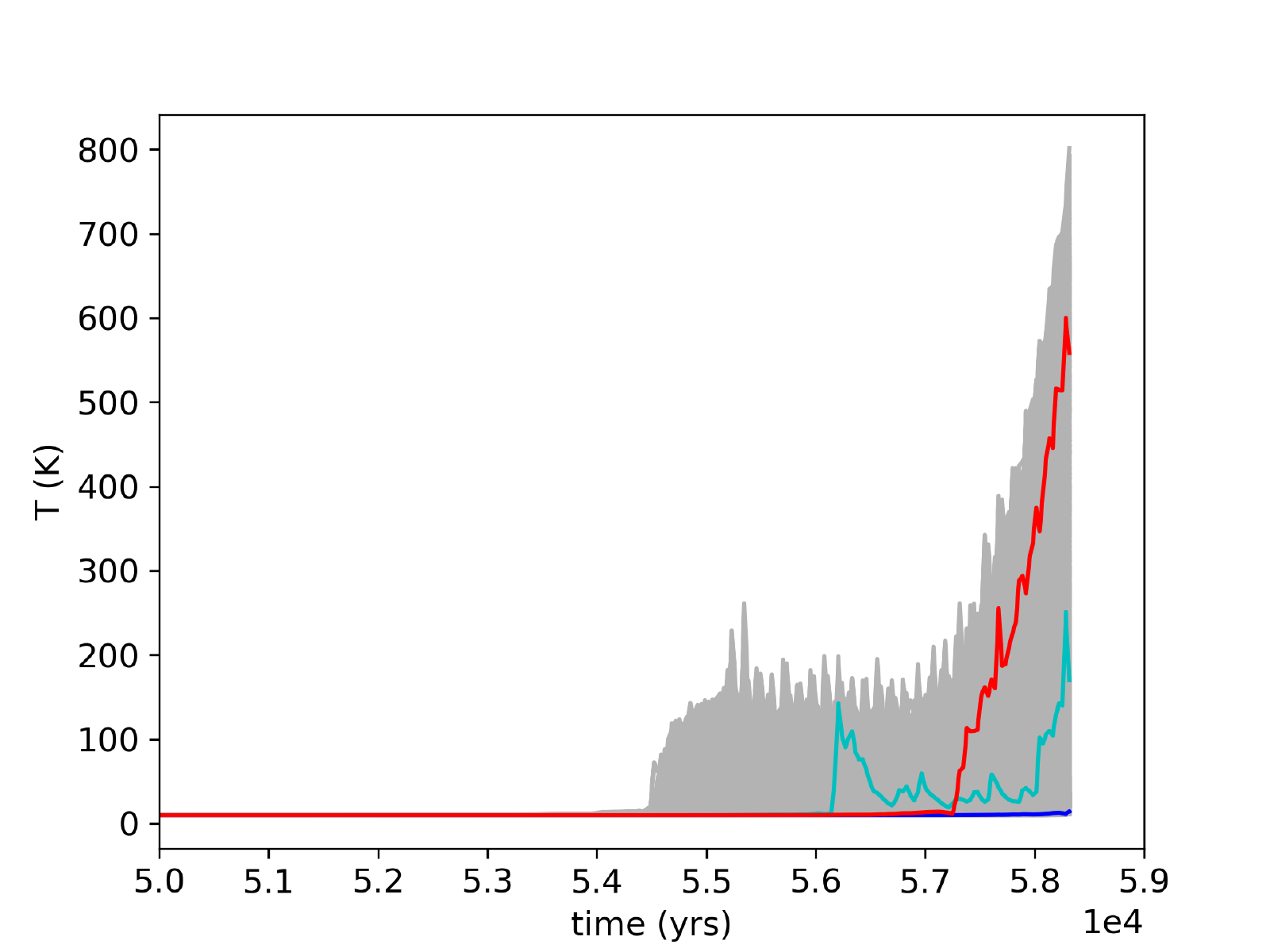}
 \caption{Density (top) and temperature (bottom) evolutions of all the particles ending in the disk in gray. The evolution of three random particles is shown in blue, cyan, and red. A different timescale is used in the two panels.}
\label{fig_phys_cond_part}
\end{center}
\end{figure}

\begin{figure}[!ht]
\begin{center}
 \includegraphics[width=1.\hsize]{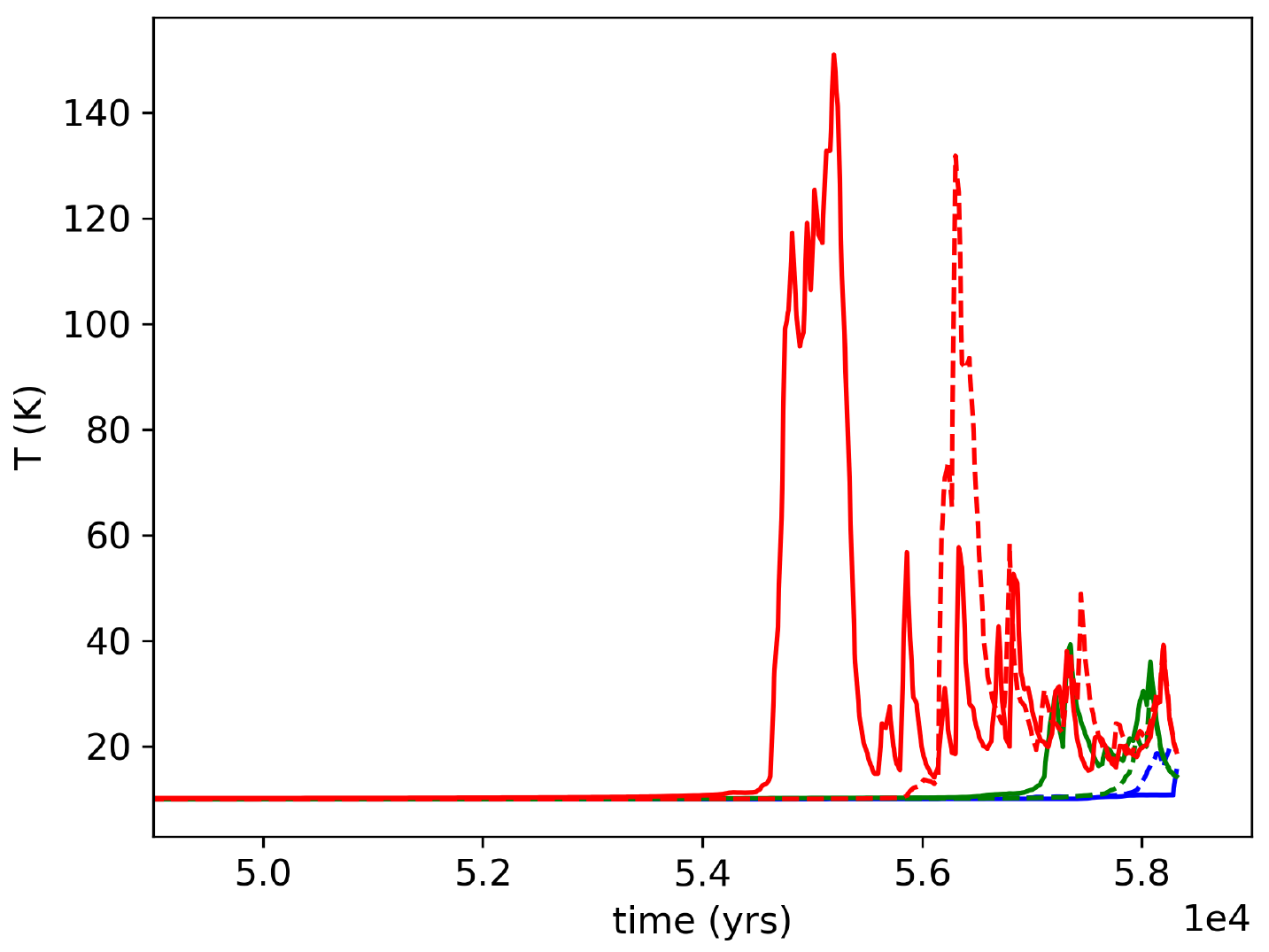}
 \caption{Temperature evolution of six particles ending in the outer region of the disk (beyond a radius of 30 au) with a temperature lower than 20 K. Two particles reach a maximum temperature above 100 K (red), two below 20 K (blue), and two between 20 and 100 K (green). }
\label{fig_phys_cond_part_20K}
\end{center}
\end{figure}

Figure \ref{fig_phys_cond_part} shows the evolution of density and temperature as a function of time of all the particles ending in the disk. 
The temperature remains low (10 K) for all particles until $\sim$5.4\,$\times$\,10$^4$ yr. The particles follow different trajectories and encounter different physical conditions. For example, among the 1 632 particles with a final temperature $<$ 20 K, 329 (20\%) remain at low temperature along the entire trajectory (see blue line in Figure \ref{fig_phys_cond_part}), while 145 (9\%) reach a temperature above 100 K 
at some point. The remaining 1 158 particles (71\%) experience a maximum temperature of between 20 and 100 K (see Figure \ref{fig_phys_cond_part_20K}). The distribution of the maximum density and temperature reached by the disk particles during their trajectories is presented with histograms in Figure \ref{fig_histo_structure} (lower panels).

\section{Chemical simulations}
\label{sect_chem_model}

\subsection{Chemical model}

The evolution of the physical conditions of the 14 948 particles was used as input parameters in the Nautilus three-phase gas-grain chemical model to calculate the chemical evolution of the particles ending in the disk. This model is fully described in \citet{Ruaud2016}. It includes gas-phase, grain-surface, and grain-mantle chemistry, along with adsorption of gas-phase species onto grain surfaces, thermal and nonthermal desorption of species from the grain surface into  the  gas  phase, and surface--mantle  and  mantle--surface  exchange  of species.  The diffusion energies of the species are equal to 0.4 times the binding energies for the surface and 0.8 times the binding energies for the mantle. The cosmic-ray ionization rate is fixed at 1.3\,$\times$\,10$^{-17}$ s$^{-1}$.
Direct photo-processes are negligible because the visual extinction is always higher than 30 mag. Chemical desorption \citep[see][]{Garrod2007} is included with an efficiency of 1\% for all species.  

The gas-phase network is derived from kida.uva.2014 and the grain network is described in \citet{Wakelam2016}. The gas and grain networks have been updated according to the following works: \citet{Wakelam2015}, \citet{Loison2016}, \citet{Hickson2016}, \citet{Wakelam2017}, \citet{Vidal2017}, and \citet{Loison2017}. In total, the network consists of 589 gas-phase species and 540 solid-phase species (either on the surface or in the ice mantle) with 13 384 reactions. 
It should be noted that this version of the chemical network is not complete regarding complex organic chemistry. Some COMs detected in several young solar-type protostars, such as for example glycolaldehyde and ethylene glycol (e.g., \citealt{Coutens2015}), are  not included in this version. Molecules recently detected for the first time in such objects (e.g., CH$_3$Cl, HONO, \citealt{Fayolle2017,Coutens2019}) are not included either. A chemical network is currently being developed to include more complex species (up to three-carbon species) and grain-surface reactions \citep[e.g.,][]{Manigand2020b}. The results presented here for the most complex species should consequently be seen more qualitatively than quantitatively. 

 The abundances are expressed with respect to the total density of hydrogen nuclei ($n_{\rm H}$ = $n$(H) + 2 $n$(H$_2$)). The abundances of molecules in ices presented in this paper always correspond to the sum of abundances obtained for the grain mantle and the grain surface. When only referring to the solid-phase species, ``s-'' precedes the species.

\subsection{Initial abundances}
\label{Sect_ab_init}

\begin{figure}[!t]
\begin{center}
 \includegraphics[width=1.\hsize]{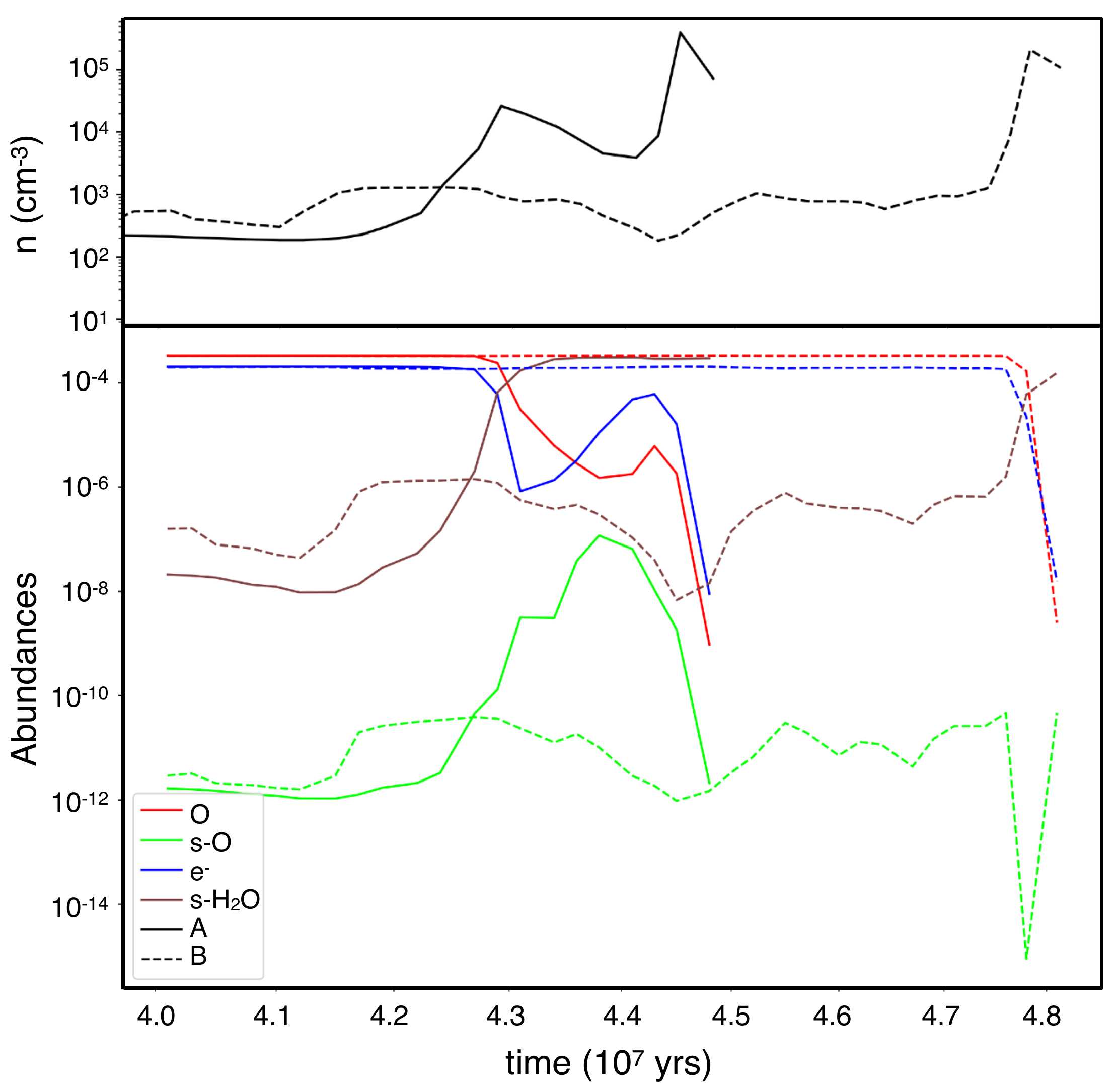}
 \caption{Upper panel:  Evolution of the density of the two SPH simulations of the formation of dense cores between 4\,$\times$\,10$^7$ and 5\,$\times$\,10$^7$ yr (solid line for case A, dashed line for case B). Lower panel: Evolution of the abundance of O (red), s-O (green), e$^-$ (blue), and s-H$_2$O (brown) for cloud A (solid lines) and cloud B (dashed lines). }
\label{fig_ab_evol_precollapse}
\end{center}
\end{figure}

To determine the impact of the initial conditions on the final chemical composition of the protoplanetary disk, we consider two different sets of initial abundances that we call A and B. These sets were obtained by \citet{Ruaud2018} with the chemical model described above for smoothed-particle hydrodynamics (SPH) simulations of dense core formation \citep{Bonnell2013}. The abundances were extracted at the time where the density has sufficiently increased to form a dense core. The two clouds that produced the two cores considered in this study underwent different physical histories. The physical conditions at the time we extracted the initial abundances for cores (maximum of density) are similar however. For A (B), the density is 4.0\,$\times$\,10$^5$ (2.1\,$\times$\,10$^5$) cm$^{-3}$, the temperature is 10 (11) K, and the visual extinction $A_V$ is 22 (18) mag. In case A, the cloud has experienced a relatively quiescent phase, while the history of B is more complex with significantly warmer and colder phases (see Figure \ref{fig_ab_init}). Cloud A has also experienced a moderately dense and cold phase (at 4.3\,$\times$10$^7$ yr) before reaching its maximum ``cold core'' density that significantly changed its composition. During this phase, a large fraction of the electron donors and the atomic oxygen have depleted onto the cold dust grains, which led to the formation of solid water (see Figure \ref{fig_ab_evol_precollapse}). Even when the density decreases (between the two density peaks), water remains on grains. Less atomic oxygen is available in the gas phase. The main consequence is that water and gas-phase carbon chains are more abundant in cloud A, while oxygen-rich organics and O$_2$ are more abundant in cloud B (see Figures \ref{fig_stat_1}-\ref{fig_stat_4}). 
As explained in \citet{Ruaud2018}, the SPH simulations follow the gas dynamics of the interstellar medium at galactic scales. Consequently, the timescales shown in Figures \ref{fig_ab_evol_precollapse} and \ref{fig_ab_init} do not represent the age of dense and cold cores. If we estimate the ages of the cold cores by calculating the time spent between the moment where the density reached 10$^3$ cm$^{-3}$ and the final density, the core formed in Cloud A is older than the one in Cloud B with respective ages of 2.1\,$\times$\,10$^6$ yr and 4\,$\times$\,10$^5$ yr.

The main carriers of the different elements are found on grains, and are summarized in Table \ref{tab_init_element}.. Indeed, at the initial time, the temperature is low and the density is relatively high, which leads to the depletion of many species. For oxygen, water is dominant (91\%) in case A, while in case B water only represents 48\% of the O elemental abundance and the rest is in the form of organic molecules (H$_2$CO, CH$_3$OH, HCOOH, and HCO). The main carbon carriers are O-bearing organic molecules in case B, while HCN accounts for 27\% in case A. The contributions of CH$_4$ are relatively similar in both cases, and CH$_3$OH is more abundant in case B. In Case A, a large range of species with very small contributions (less than 5\% of the C atomic abundance) explains the total amount of carbon. Regarding nitrogen, the main carrier is HCN in case A and NH$_3$ in case B. The sulfur reservoirs are dominated by HS and H$_2$S in both cases, but differ by their slightly less abundant contributors, CH$_3$SH and CH$_3$S in case A, and NS and OCS in case B. SiH$_4$ is the most abundant Si carrier followed by SiC$_8$H in case A, while SiC$_4$H and SiO dominate in case B. Atomic P is the major carrier of phosphorus. CP and HCP also contribute in case A, while PN is the second-most abundant carrier in case B. The carriers of Cl and F are the same in both cases: HCl and HF. 

Regarding the ionization, the most abundant ions are H$_3^+$, H$^+$, and S$^+$ in case A, while they are Fe$^+$, H$_3^+$, H$^+$, S$^+$, and Mg$^+$ in case B. 
In both cases, the abundances are about a few 10$^{-9}$.

\begin{table*}[!ht]
\begin{center}
\caption{Elemental reservoirs for the initial conditions (contribution of at least 5\% of the elemental abundance).}
\label{tab_init_element}
\begin{tabular}{c p{0.4\linewidth} p{0.4\linewidth}}
\hline \hline
 Element & Main reservoir, Case A & Main reservoir, Case B \\
 \hline
  O & s-H$_2$O (91\%) & s-H$_2$O (48\%), s-H$_2$CO (7\%), s-CH$_3$OH (7\%), s-HCOOH (2 $\times$ 6\%), s-HCO (5\%) \\ 
 \hline
 C & s-HCN (27\%), s-CH$_4$ (12\%), s-CH$_3$OH (6\%), s-C$_3$H$_8$ (3 $\times$ 4\%)
 & s-H$_2$CO (14\%), s-CH$_3$OH (13\%), s-HCOOH (11\%), s-CH$_4$ (10\%), s-HCO (10\%), s-CO (8\%), s-CO$_2$ (5\%), s-CH$_2$OH (5\%) \\
 \hline
 N &  s-HCN (79\%), s-NH$_3$ (6\%), s-N$_2$ (2 $\times$ 6\%) & s-NH$_3$ (53\%), s-NH$_2$CH$_2$OH (11\%), s-N$_2$ (2 $\times$ 5\%), s-HCN (5\%) \\
 \hline
 S &  s-HS (32\%), s-H$_2$S (31\%), s-CH$_3$SH (12\%), s-CH$_3$S (9\%) & s-HS (39\%), s-H$_2$S (29\%), s-NS (7\%), s-OCS (7\%) \\
 \hline
 Si &  s-SiH$_4$ (41\%), s-SiC$_8$H (17\%), s-SiC$_6$H (8\%), s-SiC$_4$H (7\%), s-HCSi (5\%) & s-SiC$_4$H (28\%), s-SiO (21\%), s-SiH$_4$ (16\%), s-SiC$_6$H (7\%) \\
 \hline
 P & s-P (83\%), s-CP (9\%), s-HCP (5\%) & s-P (86\%), s-PN (14\%) \\
  \hline
 Cl & s-HCl (99\%) & s-HCl (96\%) \\
  \hline
 F & s-HF (100\%) & s-HF (99\%) \\
 \hline
\end{tabular}
\end{center}
{\bf Note:} s- means that the species are on grains. The gas-phase abundances are negligible here.
\end{table*}%

\section{Results}
\label{sect_results}

The gas-grain chemical model Nautilus was run for the 14 948 disk particles and the two sets of initial abundances A and B, which are the chemical compositions at the maximum density of two clouds \citep{Ruaud2018}. The analysis of the chemical results focuses on three different aspects. First, we analyze the spatial distribution of molecules in the disk. We then study the chemical evolution during the formation of the disk. In particular, we investigate the impact of the initial abundances on the disk abundances. Finally, we explore the ionization and its role in the disk resistivity.

\subsection{Spatial distribution of molecules}
\label{sect_spatial_dist}

\begin{figure}[!t]
\begin{center}
 \includegraphics[width=1\hsize]{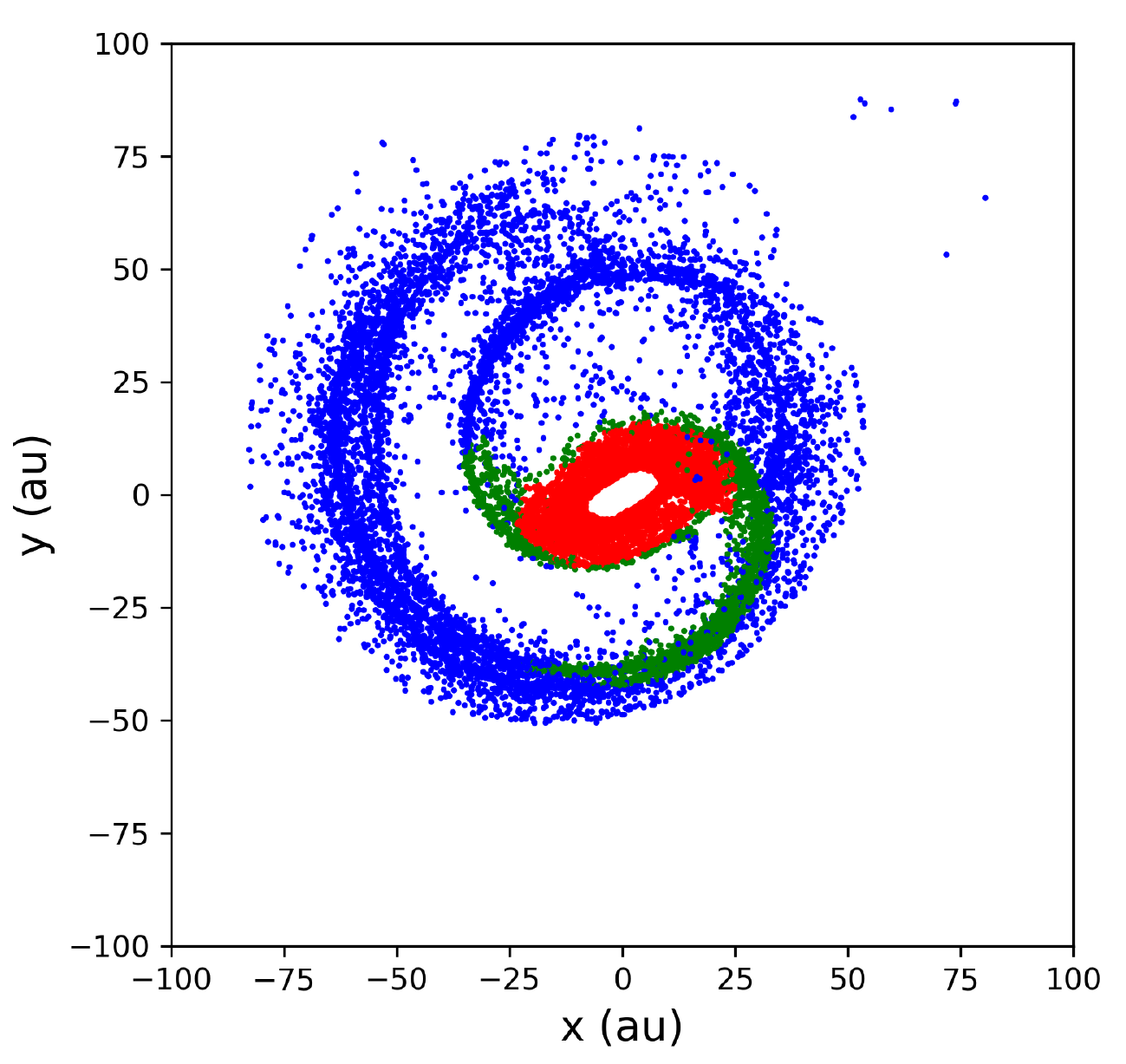} 
 \caption{Regions of the disk used for the classification of molecules according to their spatial distributions at the final time of the simulations (5.83\,$\times$\,10$^4$~yr): the inner regions ($n \gtrsim 10^{12}$ cm$^{-3}$, $T \gtrsim 100$ K, red), the spiral arms ($n \lesssim 10^{11}$ cm$^{-3}$, $T \lesssim 40$ K, blue), and the transition region  (10$^{11}$ cm$^{-3}$ $\lesssim n \lesssim$ 10$^{12}$ cm$^{-3}$, 40 K  $\lesssim T \lesssim$ 100 K, green).}
\label{fig_areas}
\end{center}
\end{figure}

Figures \ref{fig_spatial_distrib}--\ref{fig_spatial_distribf} show the spatial distribution of various molecules with non-negligible abundances in the x-y plane of the disk (stacked z axis) at the end of the simulations. Although the abundances of the same molecules can vary between cases A and B, their spatial distribution is usually very similar. Exceptions are found for HCO, HNO, OH, SO, SO$_2$, and CH$_3$. 
The spatial distribution does vary strongly from molecule to molecule. While the abundances of some species peak towards the center, others are more abundant in the colder regions that correspond to the spiral arms. Some molecules are also found to peak in the gas phase in an intermediary region. 
We classified in Table \ref{tab_distrib_molecules} the molecules depending on the region(s) where their gas-phase and total (gas + solid) abundances peak (within about one order of magnitude). We considered the three regions to be: the inner region of the disk ($n \gtrsim 10^{12}$ cm$^{-3}$, $T \gtrsim 100$ K), the spiral arms ($n \lesssim 10^{11}$ cm$^{-3}$, $T \lesssim 40$ K), and the transition region that takes the shape of a ring surrounding the inner region and includes the beginning of the spiral arms (10$^{11}$ cm$^{-3}$ $\lesssim n \lesssim$ 10$^{12}$ cm$^{-3}$, 40 K  $\lesssim T \lesssim$ 100 K, see Figure \ref{fig_areas}). 
It should be noted that molecules can present variations of their abundances inside these regions and consequently molecules in the same category do not necessarily present the same spatial distribution. For example, the spatial extent of the molecule can vary in the inner regions or the molecule peaks in different regions of the spiral arms. 
In regard to this classification, we observe different trends:
\begin{itemize}
\item[$\bullet$] Except for a few ions, most of the molecules that peak in the spiral arms are also abundant in the transition region.
\item[$\bullet$] The molecules with their highest gas-phase abundance in the spiral arms + transition region or the transition region only have their total abundance peak in the spiral arms + transition region. These molecules are only present in the cold regions either in the gas phase or in the ices. Their absence from the inner regions can be explained by efficient destruction reactions at relatively high temperatures ($\gtrsim$ 100\,K).
\item[$\bullet$] No molecules appear in the transition region only when we consider total abundance peaks. The transition region observed for some molecules in the gas phase can be explained by the release of these species trapped on grains below a certain temperature, but they are easily destroyed in the gas phase at warmer temperatures. Indeed, this category of molecules includes a lot of radicals.
\item[$\bullet$] Most of the molecules with their highest gas-phase abundance in the inner regions present their total abundance peak in all the regions. This means that these species are also abundant on grains in the region where the temperature is low. 
Thermal desorption explains their high gas-phase abundances in the inner regions. Many COMs and water belong to this category.
\end{itemize}

\begin{table*}[!ht]
\begin{center}
\caption{Classification of molecules according to their spatial distributions. \label{tab_distrib_molecules}}
\begin{tabular}{p{2.5cm}p{7cm}p{7cm}}
\hline \hline
 Component & Molecules with the highest gas-phase abundances &  Molecules with the highest total (gas + solid) abundances \\ 
 \hline
Spiral arms only & CH$_5$$^+$, H$_3$$^+$
& CH$_3$$^\ddagger$, CH$_5$$^+$, H$_3$$^+$, HCO$^\ddagger$, OH$^\ddagger$, NH, NH$_2$ \\
\hline
Spiral arms  +  transition region & c-C$_3$H, C$_2$H$_7$$^+$, C$_4$H, CCH, HCO$^+$, l-C$_3$H, N$_2$H$^+$, NH$_2$, O$_2$, PCH$_3$$^+$ 
& c-C$_3$H, c-C$_3$H$_2$, C$_2$H$_7$$^+$, C$_4$H, CCH, CCS, CH$_2$OH, CH$_3$$^\ddagger$, CH$_3$O, CH$_3$S, CN, HCO$^\ddagger$, HCO$^+$, HCS, HNO$^\ddagger$, l-C$_3$H, l-C$_3$H$_2$, N$_2$H$^+$, O$_2$, OH$^\ddagger$, PCH$_3$$^+$  \\
\hline
Transition region only & c-C$_3$H$_2$, CCS, CH$_2$OH, CH$_3$$^\dagger$, CH$_3$O, CH$_3$S, CN, HCO$^\ddagger$, HCS, l-C$_3$H$_2$, OH &             \\ 
\hline
Inner disk (with or without the transition region) & C$_3$H$_8$, CH$_2$NH, CH$_3$CCH, CH$_3$CHO, CH$_3$CN, CH$_3$COCH$_3$, CH$_3$NH$_2$, CH$_3$OCH$_3$, CH$_3$OH, CH$_3$SH, CO$_2$, CP, CS, H$_2$CCO, H$_2$CO, H$_2$CS, H$_2$O, H$_2$S, H$_2$S$_3$, H$_5$C$_2$O$_2$$^+$, HC$_3$N, HCl, HCN, HCO$^\ddagger$, HCOOCH$_3$, HCOOH, HCP,  HCSi, HF, HNC, HNCO, HNO, HOOH, HSO, N$_2$O, NH, NH$_2$CH$_2$OH, NH$_2$CHO, NH$_3$, NO, NS, OCS, SiC$_4$H, SiC$_6$H, SiC$_8$H, SiH$_4$, SiO, SO, SO$_2$      
&   H$_5$C$_2$O$_2$$^+$, HNO$^\ddagger$, N$_2$O, SO$^\ddagger$, SO$_2$$^\ddagger$  \\ 
\hline
All regions & CH$_3$$^\ddagger$, CH$_4$, CO, N$_2$, P, PN, PO   
& C$_3$H$_8$, CH$_2$NH, CH$_3$CCH, CH$_3$CHO, CH$_3$CN, CH$_3$COCH$_3$, CH$_3$NH$_2$, CH$_3$OCH$_3$, CH$_3$OH, CH$_3$SH, CH$_4$, CO, CO$_2$, CP, CS, H$_2$CCO, H$_2$CO, H$_2$CS, H$_2$O, H$_2$S, H$_2$S$_3$, HC$_3$N, HCl, HCN, HCOOCH$_3$, HCOOH, HCP, HCSi, HF, HNC, HNCO, HOOH, HSO, N$_2$, NH$_2$CH$_2$OH, NH$_2$CHO, NH$_3$, NO, NS, OCS, P, PN, PO, SiC$_4$H, SiC$_6$H, SiC$_8$H, SiH$_4$, SiO, SO$^\ddagger$, SO$_2$$^\ddagger$ \\
\hline
\end{tabular}
\end{center}
{\bf Note:} We consider that the variation of abundance between two components is significant when the abundances change by at least an order of magnitude. $^\ddagger$ These species appear in two categories, as their spatial distribution is sensitive to the chosen set of initial abundances.
\end{table*}%

\begin{figure*}[!t]
\begin{center}
\includegraphics[width=\hsize]{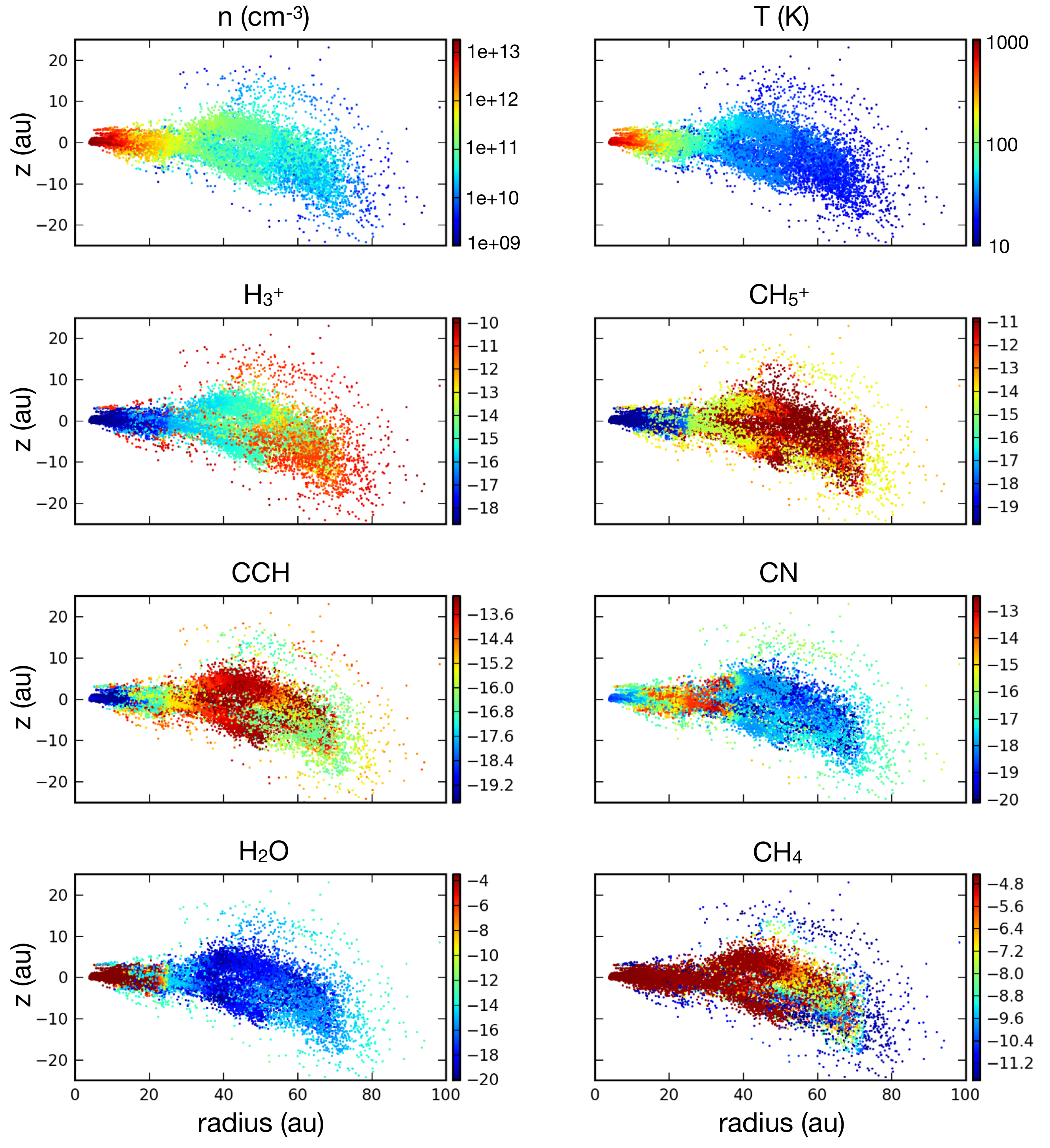}
\caption{Distribution of the density (upper left), temperature (upper right), and gas-phase abundances of six molecules (H$_3^+$, CH$_5^+$, CCH, CN, H$_2$O, and CH$_4$) as a function of  radius ($\sqrt{x^2+y^2}$) and the height $z$ at the final time of the simulations (5.83\,$\times$\,10$^4$~yr) for the set of initial abundances A. For the molecules, the color scales represent log$_{10}$(n(X)/n$_{\rm H}$). }
\label{fig_vertical_species}
\end{center}
\end{figure*}

Vertically, we do not see strong variations in abundances. Figure \ref{fig_vertical_species} shows the distribution as a function of the radius and height ($z$) of the gas-phase abundances of six species that belong to the different components discussed above (H$_3^+$ and CH$_5^+$ for the spiral arms only, CCH for the spiral arms  +  transition region, CN for the transition region only, H$_2$O for the inner disk, and CH$_4$ for all regions). As explained in Section \ref{sect_disk_final}, the vertical structure of the disk is affected by the low resolution of the AMR grid within the disk. The disk is also embedded, so no chemical stratification due to UV effects is expected contrary to evolved disks. Figure \ref{fig_vertical_species} clearly shows that the spatial distributions of the molecules are correlated with the temperature and density.

\subsection{Impact of the initial conditions on the final chemical composition of the disk}

\subsubsection{Final reservoirs}

\begin{table*}[!ht]
\begin{center}
\caption{Elemental reservoirs at the end of the simulations (contribution of at least 5\% of the elemental abundance).}
\label{tab_final_element}
\begin{tabular}{c p{0.4\linewidth} p{0.4\linewidth}}
\hline \hline
 Element & Main reservoir, Case A & Main reservoir, Case B \\
 \hline
  O & H$_2$O (91\%) & H$_2$O (49\%), H$_2$CO (9\%), CH$_3$OH (7\%), HCOOH (2 $\times$ 6\%) \\ 
 \hline
 C & HCN (28\%), CH$_4$ (15\%), CH$_3$OH (7\%), C$_3$H$_8$ (3 $\times$ 5\%)
 & H$_2$CO (16\%), CH$_3$OH (13\%),  CH$_4$ (12\%), HCOOH (12\%), CO (7\%), CO$_2$ (6\%) \\
 \hline
 N &  HCN (81\%), NH$_3$ (6\%), N$_2$ (2 $\times$ 6\%) & NH$_3$ (58\%), NH$_2$CH$_2$OH (10\%), N$_2$ (2 $\times$ 7\%), HCN (5\%) \\
 \hline
 S &  H$_2$S (35\%), CH$_3$SH (11\%), H$_2$S$_3$ (3 $\times$ 10\%), CH$_3$S (5\%) & H$_2$S (30\%), H$_2$S$_3$ (3 $\times$ 11\%), OCS (7\%), NS (7\%) \\
 \hline
 Si &  SiH$_4$ (44\%), SiC$_8$H (16\%), SiC$_6$H (8\%), SiC$_4$H (7\%), HCSi (5\%) & SiC$_4$H (27\%), SiO (21\%), SiH$_4$ (18\%), SiC$_6$H (7\%) \\
 \hline
 P & P (80\%), CP (11\%), HCP (7\%) & P (40\%), PO (26\%), PN (14\%), HCP (10\%), CP (9\%) \\
  \hline
 Cl & HCl (100\%) & HCl (99\%) \\
  \hline
 F & HF (100\%) & HF (100\%) \\
 \hline
\end{tabular}
\end{center}

{\bf Note:} The gas and ice abundances are summed.
\end{table*}%

The final reservoirs of the different elements are summarized in Table \ref{tab_final_element}. To do so, we averaged the abundances found for all the disk particles. 
The reservoirs of oxygen are very similar to the initial ones. Only in case B, does HCO decrease from 5\% to 1\%.
The same is seen for the carbon. In case B, only HCO and CH$_2$OH become less abundant. As we see later, these two radicals are consumed to form more COMs. The main carriers of nitrogen, silicon, chlorine, and fluorine do not change either. A change is seen for phosphorus, but only in case B. Atomic P is still the main carrier, but the contributions of PO, HCP, and CP significantly increase at the final time. Sulfur is the only element for which major changes are observed in both cases.  At the end of the simulations, HS is no longer the main carrier, while the contribution of H$_2$S$_3$ has strongly increased. However, the other carriers are the same as in the initial conditions. H$_2$S, CH$_3$SH, and CH$_3$S still contribute to more than 5\% in case A, while H$_2$S, NS, and OCS remain similarly abundant in case B.
Based on these results, we can conclude that the main carriers of the different elements in the disk are generally inherited from the molecular cloud, but this does not necessarily mean that the chemical content does not evolve during the formation of the disk, as we see in the following section.

\subsubsection{Evolution of the chemical composition}
\label{Sect_evol}

\begin{table*}[!ht]
\begin{center}
\caption{Classification of molecules according to their sensitivity to the initial abundances.}
\label{tab_classification}
\begin{tabular}{c p{0.12\linewidth} p{0.35\linewidth} p{0.35\linewidth}}
\hline \hline
$\#$ & Conditions & [X]$_{i}$ = [X]$_{f}$  & [X]$_{i}$ $\neq$ [X]$_{f}$ \\
\hline
1 & [X]$_{i,A}$~=~[X]$_{i,B}$  \ [X]$_{f,A}$~~=~~[X]$_{f,B}$ & c-C$_3$H, c-C$_3$H$_2$, C$_3$H$_8$, CCS, CH$_2$OH, CH$_3$CN, CH$_3$OH, CH$_3$S, CH$_3$SH, CH$_4$, CH$_5$$^+$, H$_2$CO, H$_2$CS,  H$_2$O, H$_2$S, HC$_3$N, HCS, HCl, HF, l-C$_3$H$_2$, N$_2$, P, SiC$_4$H, SiC$_6$H, SiC$_8$H, SiH$_4$
& C$_4$H, CCH, CH$_3$, CN, CS, H$_3$$^+$, H$_5$C$_2$O$_2$$^+$, l-C$_3$H, N$_2$H$^+$ \\
\hline
2 & [X]$_{i,A}$~=~[X]$_{i,B}$ \ [X]$_{f,A}$~~$\neq$~~[X]$_{f,B}$  & & CH$_3$O, HCO, HCO$^+$, OH, PCH$_3$$^+$ \\
\hline
3 & [X]$_{i,A}$~$\neq$~[X]$_{i,B}$ \ [X]$_{f,A}$~~=~~[X]$_{f,B}$ & & CH$_3$CCH, CH$_3$CHO, CH$_3$OCH$_3$, H$_2$CCO, HCOOCH$_3$, OCS \\
\hline
4 & [X]$_{i,A}$~$\neq$~[X]$_{i,B}$ \ [X]$_{f,A}$~~$\neq$~~[X]$_{f,B}$ & CH$_2$NH, CH$_3$COCH$_3$$^b$, CO$^b$, CO$_2$, H$_2$S$_3$, HCN, HCOOH, HCP$^a$, HCSi, (HNC), HNCO, HNO$^b$, HOOH, HSO, N$_2$O, NH$_2$CHO, NH$_2$CH$_2$OH, (NH$_3$), NS, PN, PO$^a$, SiO, SO, SO$_2$ 
& C$_2$H$_7$$^+$, CH$_3$COCH$_3$$^a$,  CH$_3$NH$_2$, CO$^a$, HCP$^b$, HNO$^a$, NH, NH$_2$, NO, O$_2$, PO$^b$  \\
\hline
\end{tabular}
\end{center}

{\bf Note:} [X]$_{i,x}$ and [X]$_{f,x}$ correspond, respectively, to the initial and final abundances for the set of initial abundances $x$ (A or B). We consider the total gas and ice abundances. On the last row, some molecules are found with both [X]$_{i}$ = [X]$_{f}$  and [X]$_{i}$ $\neq$ [X]$_{f}$. The category depends on the set of initial abundances. To differentiate them, we used $^a$ for set A and $^b$ for set B. NH$_3$ and HNC (within brackets) have been moved to a different category, as their initial abundances differ by almost an order of magnitude.
\end{table*}%

Here we investigate the evolution of the chemical content during the formation of the disk more thoroughly. The histograms presented in Figures \ref{fig_histo}-\ref{fig_histof} show the abundance distribution for all the disk particles at the end of the simulations, while the initial abundances and the mean, median, minimum, and maximum abundances at the final time are summarized in Table \ref{tab_stat_abundances} and shown in Figures \ref{fig_stat_1}-\ref{fig_stat_4}. The distributions obtained for the set of initial abundances A and B are compared with each other as well as with their respective initial abundances to determine the impact of the initial conditions on the chemical composition of the disk. The distributions of the final abundances are considered similar if they overlap by more than 75\% over bins of one order of magnitude. The initial abundances are considered similar if they do not differ by more than one order of magnitude. The final abundances are considered similar to the initial abundance if the median of the final abundances and the initial abundance differ by less than one order of magnitude.

Based on these criteria, we classified the molecules into different categories (see Table \ref{tab_classification}). 
To highlight the change in chemistry during the formation of the disk, we first separated the molecules that have similar initial and final abundances (third column of Table \ref{tab_classification}) and the ones for which the abundance significantly changes between the initial and final times (fourth column of Table \ref{tab_classification}). 
In addition, we separated the molecules with similar initial abundances (rows 1 and 2) and the ones with different initial abundances (rows 3 and 4). Finally, we also separated the species with similar final distributions (rows 1 and 3) and the ones that show significantly different distributions (rows 2 and 4). 
A visual inspection was made to check the classification. NH$_3$ and HNC (indicated in Table \ref{tab_classification} with brackets) have been moved to a different category, as their initial abundances differ by almost an order of magnitude.

From this classification, we can infer:
\begin{itemize}
\item[$\bullet$] All the molecules in the third column of Table \ref{tab_classification} present similar initial and final abundances. Although we cannot exclude the destruction and reformation of these species during the collapse, it is more likely that these species are essentially formed in the molecular cloud or prestellar core. This list includes various types of molecules. Water is one of them, which is not so surprising since both observations and models are in favor of formation at a relatively early stage of the star formation process \citep[e.g.,][]{Jones1984, Coutens2012,Furuya2016,Furuya2017}.
Some ``small'' COMs such as methanol, methyl cyanide, formamide, and their potential precursors (HNCO, H$_2$CO) also belong to this category. Although deuteration is not included in our network, it is interesting to note that these molecules are the organic species with the lowest deuterium fractionation towards the Class 0 protostar IRAS~16293--2422 B, while the most complex species (absent from this category) present a deuteration level at least a factor of two higher \citep{Coutens2016,Jorgensen2016,Jorgensen2018,Persson2018}. The lower deuteration of the ``small'' COMs would consequently be explained by their formation at an earlier stage of the star formation process in the molecular cloud. 

\item[$\bullet$] All the molecules in the fourth column of Table \ref{tab_classification} show different initial and final abundances for at least one of the chosen sets of initial abundances. The physical changes produced during the dense core collapse consequently play an important role in the chemical evolution of these species. In Table \ref{tab_increase}, we separate the molecules that show a clear increase in abundance from the ones showing a clear decrease. The molecules that present an increase in abundances in both cases are ``large'' COMs (CH$_3$CHO, CH$_3$NH$_2$, CH$_3$OCH$_3$, HCOOCH$_3$). This emphasizes the importance of the collapse in the formation of these molecules typically found in ``hot cores'' and ``hot corinos''. Although these molecules are detected towards prestellar cores \citep{Bacmann2012, Vastel2014}, our model shows that they only acquire their highest (gas + solid) abundances during the collapse. 
In contrast, the molecules that show decreasing abundances in both cases are relatively small molecules and radicals. In particular, they include radicals such as HCO and CH$_3$O, which are consumed to form more COMs. The radicals CCH, C$_4$H, CH$_3$, CN, NH, NH$_2$, NO, and OH also show this decrease in abundances. The abundances of several ions also decrease significantly. O$_2$ is also destroyed during the collapse.

\item[$\bullet$] For the molecules in row 1, the impact of the initial conditions on the final abundances cannot be properly tested because the initial abundances of these species are similar. These species seem to have little sensitivity to the history of the physical conditions from the diffuse medium to the disk formation. 

\item[$\bullet$] The molecules in row 2 present different final abundances despite having similar initial abundances. These species are sensitive to the initial abundances of other species that differ in cases A and B; they include three radicals (CH$_3$O, HCO, and OH) and two ions (HCO$^+$ and PCH$_3$$^+$).

\item[$\bullet$] The molecules in row 3 present similar final abundances despite having different initial abundances. These molecules are particularly interesting, as they are not sensitive to the initial conditions. Their abundances are determined during the collapse. Here, we again find some ``large'' COMs: CH$_3$CCH, CH$_3$CHO, CH$_3$OCH$_3$, H$_2$CCO, and HCOOCH$_3$.

\item[$\bullet$] Finally, a large number of species present different initial abundances, but also different final abundances (row 4). The species with similar initial and final abundances ([X]$_{i}$ = [X]$_{f}$) are  likely only sensitive to the evolution of the physical conditions leading to the cold and dense core. Once the collapse starts, they do not evolve further. In contrast, for the few species with different initial and final abundances ([X]$_{i}$ $\neq$ [X]$_{f}$), both the initial prestellar phase and the collapse play a role in their different abundances. This is the case of some large COMs such as CH$_3$COCH$_3$ and CH$_3$NH$_2$, as well as some small molecules such as CO, NO, O$_2$, NH, NH$_2$, and PO.
\end{itemize}

\begin{table*}[!ht]
\begin{center}
\caption{Molecules with increasing and decreasing abundances.}
\label{tab_increase}
\begin{tabular}{c p{0.3\linewidth} p{0.3\linewidth} p{0.3\linewidth}}
\hline \hline
Case & Increasing abundances & Decreasing abundances & Others \\
\hline
A & CH$_3$CHO, CH$_3$COCH$_3$, CH$_3$NH$_2$, CH$_3$OCH$_3$, H$_2$CCO,  HCOOCH$_3$, OCS
& C$_4$H, CCH, CH$_3$, CH$_3$O, CN, CO, CS, H$_3$$^+$, HCO, HCO$^+$, HNO, NH, NH$_2$, N$_2$H$^+$, NO, O$_2$, OH &
C$_2$H$_7$$^+$, CH$_3$CCH, H$_5$C$_2$O$_2$$^+$, HCP, l-C$_3$H, PCH$_3$$^+$, PO \\
B & CH$_3$CHO, CH$_3$NH$_2$, CH$_3$OCH$_3$, HCOOCH$_3$, HCP, PO
& C$_2$H$_7$$^+$, C$_4$H, CCH, CH$_3$, CH$_3$O, CN, H$_3$$^+$, HCO, HCO$^+$,  N$_2$H$^+$, NH, NH$_2$, NO, O$_2$, OH 
& CH$_3$CCH, CH$_3$COCH$_3$, CO, CS, H$_2$CCO, H$_5$C$_2$O$_2$$^+$, HNO, l-C$_3$H, OCS, PCH$_3$$^+$ \\
\hline
\end{tabular}
\end{center}
{\bf Notes:} As the abundances vary from particle to particle, we consider that the abundance increases or decreases if it concerns more than two thirds of the particles (10 000). The molecules in the column "others" are the ones that cannot be classified. For example, they can present  a spread distribution around the value of the initial abundance or they have similar initial and final abundances in the given case. 
\end{table*}%

\subsection{Ionization and resistivity}

\begin{figure*}[!ht]
\begin{center}
\includegraphics[width=0.95\hsize]{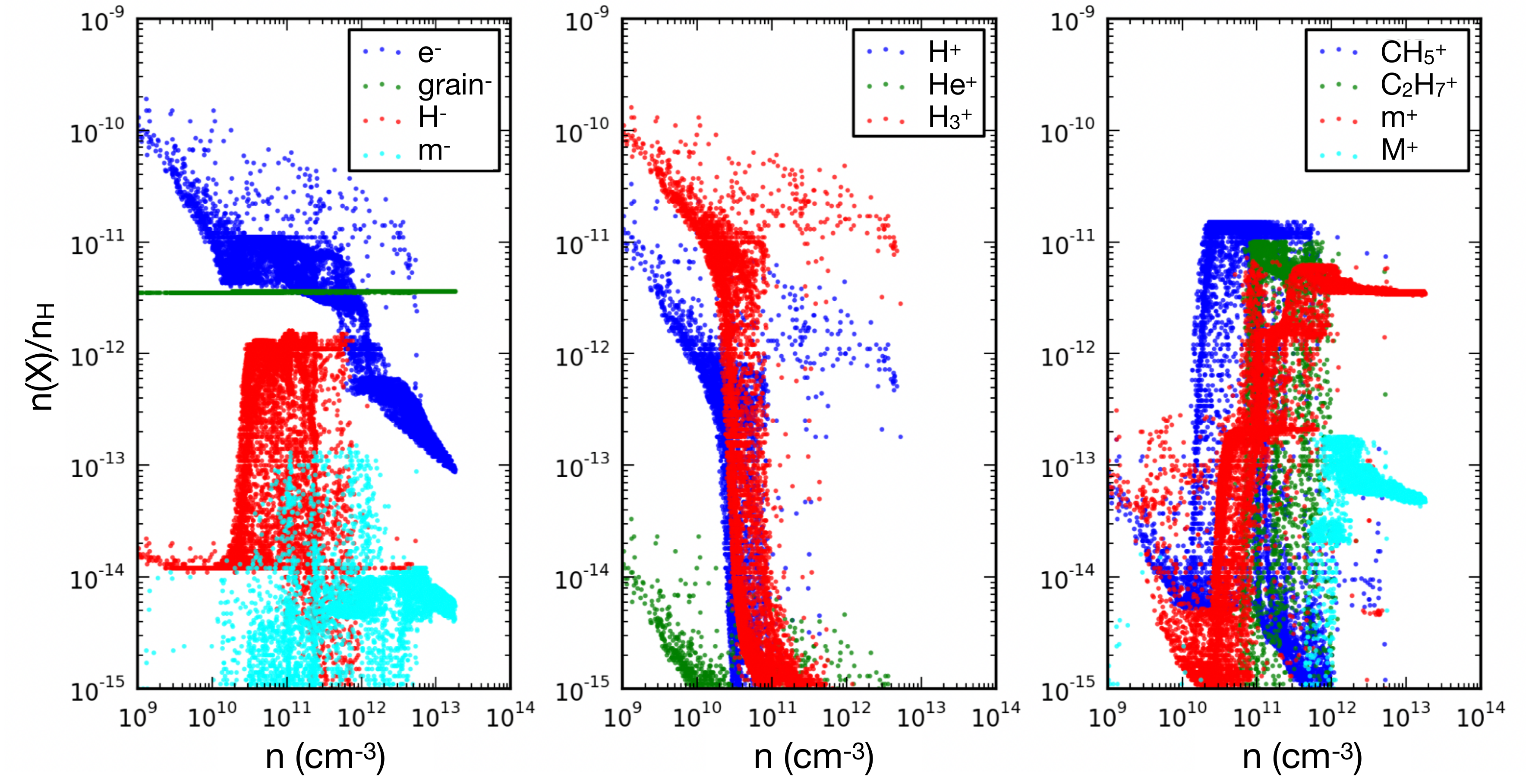}
\caption{Abundances of ions, electrons, and grains as a function of density for all the disk particles at the final time of the simulations (5.83 $\times$ 10$^4$ yr). The set of initial abundances A is used. Following \cite{Umebayashi1990} and \cite{Marchand2016}, M$^+$ represents the metallic ions (i.e., all the ions that include Si, Fe, Mg, and Na). m$^+$ includes all the other positively charged ions (containing H, C, N, O, S, and P atoms) except for H$^+$, H$_3^+$, He$^+$, CH$_5^+$, and C$_2$H$_7^+$, which are shown separately. m$^-$ represents all the anions (containing H, C, N, O, and S atoms) except for H$^-$. }
\label{fig_ab_ions}
\end{center}
\end{figure*}

\begin{figure*}[!ht]
\begin{center}
\includegraphics[width=0.95\hsize]{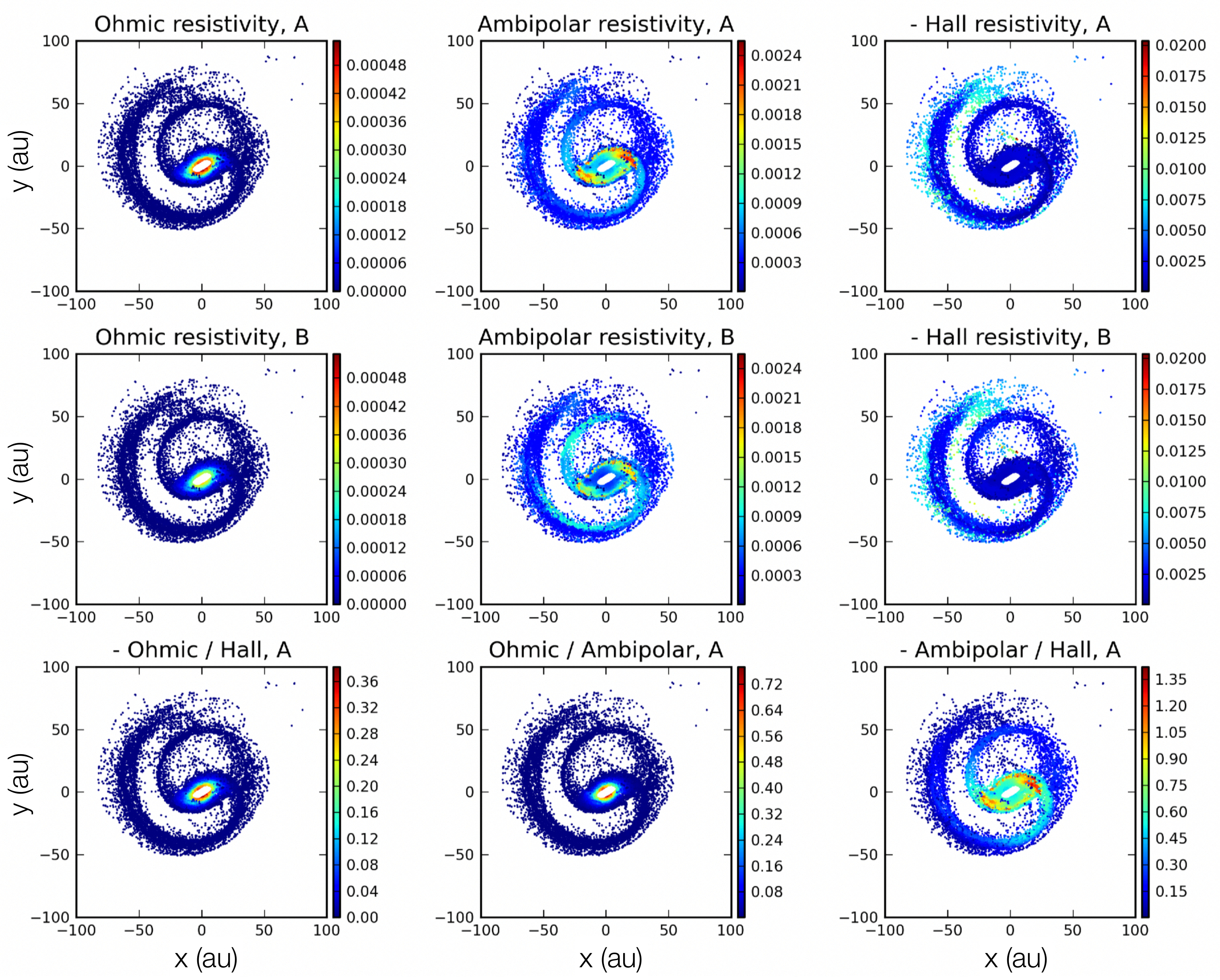}
\caption{Spatial distribution of the Ohmic, ambipolar, and Hall resistivities for all disk particles at the final time of the simulations (5.83 $\times$ 10$^4$ yr). The first two rows are for the sets of initial abundances A and B. The last one shows the ratios between the resistivities for case A. The Hall resistivities are negative. }
\label{fig_resistivity}
\end{center}
\end{figure*}

The chemistry module we used to estimate the ambipolar diffusion resistivity in the models is not consistent with the full gas-grain chemistry we follow on the tracer particles. Currently, performing full MHD calculations with full gas-grain chemistry is not achievable and models can follow the chemical evolution of networks with about 50 species \citep[e.g.,][]{Dzyurkevich2017}. Nevertheless, we can compute the resistivities from the full gas-grain chemistry and compare it with the ones we used from \cite{Marchand2016}.  

At the end of the simulations, the ionization level has significantly decreased compared to its value at the initial time (initial abundances of a few 10$^{-9}$ for the most abundant ions; see Sect. \ref{Sect_ab_init}). The most abundant ions are CH$_5^+$ and H$_3^+$ in both cases with similar average abundances of a few 10$^{-12}$. In case A, C$_2$H$_7^+$ also presents an average abundance of 1\,$\times$\,10$^{-12}$. In case B, H$_5$C$_2$O$_2^+$, PCH$_3^+$, HNS$^+$, and C$_3$H$_9^+$ have abundances of (4--7)\,$\times$\,10$^{-13}$. 
However, these different ions peak in different regions of the disk (see Section \ref{sect_spatial_dist}) and their abundances vary significantly among the different particles. For example, in Figure \ref{fig_ab_ions} we see that H$_3^+$ peaks in the regions of lower densities, while CH$_5^+$ is particularly abundant in the regions with a density between 10$^{10}$ and 10$^{12}$ cm$^{-3}$. Several molecular ions including H$_5$C$_2$O$_2^+$ contribute to the positive charge in the densest regions. Electrons and grains mainly contribute to the negative charge (see Figure \ref{fig_ab_ions}).
 
Ionization has a strong effect on the resistivity of the disk. The nonideal magneto-hydrodynamic resistivities are of three types: ambipolar diffusion, Ohmic diffusion, and Hall effect.
To estimate them, we used the equations 2--8 from \citet{Marchand2016} and calculated them separately for each particle.
Figure \ref{fig_resistivity} shows the spatial distribution of the three resistivities as well as the ratios of the resistivities. The initial abundances do not appear to have any significant effect on the resistivities of the disk. The Hall effect is the dominant mechanism, except for densities of about 10$^{12}$ cm$^{-3}$, where the ambipolar resistivity is sometimes higher than the Hall resistivity. The Ohmic resistivity is always lower.
Our results show that electrons, negatively charged grains, and a large variety of molecular ions (H$_3^+$, CH$_5^+$, H$_5$C$_2$O$_2^+$, C$_2$H$_7^+$, NH$_4^+$, HCNH$^+$, C$_3$H$_9^+$, H$_2$COH$^+$, etc.) need to be included in the chemical network to properly calculate the resistivity of the disk. While at low densities ($\lesssim$ 10$^{10}$ cm$^{-3}$) H$_3^+$ is the most important molecular ion to take into account, more complex molecular ions contribute to the resistivity at the highest densities ($>$ 10$^{12}$ cm$^{-3}$, see Figure \ref{fig_ab_ions}).

 Although we consider a unique size of grains (0.1 $\mu$m) in the chemical modeling, we examined whether or not grain growth, which is known to occur in disks, has an impact on the resistivities.  
Tests were carried out for two particles. The first particle ends in a dense and warm region ($n$ = 1.7\,$\times$\,10$^{12}$ cm$^{-3}$, $T$ = 170 K), the other one in a colder and less dense region ($n$ = 7.0\,$\times$\,10$^{9}$ cm$^{-3}$, $T$ = 14 K). We find that assuming a grain size ten times larger does not change the previous results. The values of the resistivities are  different by less than a factor of 2, except for the ambipolar resistivity of the warmest particle, which differs by a factor of 2.5. In both cases, the Hall resistivity dominates and the species that contribute to the resistivities are also the same ones as for a grain size of 0.1 $\mu$m.

The resistivity we obtain from the full gas-grain chemistry is in qualitative agreement with that obtained in the literature, where MRN grain size distributions are used \citep{Marchand2016,Dzyurkevich2017,Zhao2018}. The ambipolar and Hall resistivities dominate  the Ohmic diffusion coefficient by orders of magnitude, and have similar amplitude, within a factor of ten in the disk. 
We can conclude that, although full gas-grain chemistry is required to follow the chemical history and budget of the forming protoplanetary disk, it is not necessarily required for the MHD resistivity.  Even though we do not see any significant effect on the resistivities for grain sizes of 0.1 and 1 $\mu$m, other studies show that dust size could be a key physical quantity \citep{Zhao2016,Dzyurkevich2017}. 
More work needs to be done in this field, in particular to quantify the effect of dust growth and size distribution on ionization. 

\section{Discussion}
\label{sect_discussion}

\subsection{Comparison to 2D semi-analytical models}

Two-dimensional semi-analytical models were previously used to investigate the chemical history of the material present in the disk at the end of the collapse. In particular, \citet{Visser2009} investigated H$_2$O and CO ices. No chemical reaction was taken into account in their model apart from desorption and adsorption of H$_2$O and CO. However, their study predicted that the infalling material spent enough time at warm temperatures during the collapse to abundantly form COMs on grains, which agrees with our findings. 

Later, \citet{Visser2011} extended the chemical analysis with a gas-phase network, including adsorption onto and desorption from dust grains, and a basic grain-surface network with only hydrogenation reactions. The different physical conditions encountered along the computed trajectories merited division of the disk into several zones with different chemical histories. In our simulations, the situation is somewhat more complex. We see spatial variations in abundances correlated with the temperature, but the trajectories can also vary significantly for particles ending in the same area of the disk (cold or warm regions; see Sect. \ref{Sect_evol}), which can lead to a large range of abundances for some species in the same region. 

Following the same approach, \citet{Drozdovskaya2014} investigated the evolution of methanol with a larger and more comprehensive gas-grain chemical network. These latter authors obtained an abundance of methanol ice of 8\,$\times$\,10$^{-7}$ at the end of the prestellar phase and found that the ice abundance in the disk either increases until a value slightly above 10$^{-6}$ or significantly decreases in some cases depending on the region of the disk and the considered dominant disk-growth mechanism (viscous spreading or continuous infall of matter). In our model, the initial abundance is higher, (1--3)\,$\times$\,10$^{-5}$. 
The reaction--diffusion competition can explain part of the difference. The formation of methanol relies on two activation barriers, which are lowered due to the reaction--diffusion competition. \citet{Drozdovskaya2016} found a higher value of 3.7\,$\times$\,10$^{-6}$ at the end of the prestellar phase, when taking into account reaction--diffusion competition. Different physical conditions and timescales during the precollapse phase could also explain the different initial abundances obtained for methanol \citep{Kulterer2020}.  
The total methanol abundance in our model does not evolve significantly during the disk formation and cannot be enhanced as methanol is already formed abundantly in the cold and dense core. The disk obtained with the 3D MHD simulations is also still young ($\sim$ 5.8\,$\times$\,10$^4$ yr), while the one in \citet{Drozdovskaya2014} reaches 2.5\,$\times$\,10$^5$ yr. More than 6\,$\times$\,10$^5$ tracer particles are still in the envelope at the end of our simulations. No powerful outflow is seen. Consequently, we do not consider UV irradiation from the star contrary to \citet{Drozdovskaya2014}. Once the protostellar object has sufficiently evolved, high stellar UV fluxes should photodissociate both solid and gaseous methanol in the innermost region of the disk as well as on the disk surface, which explains the decreasing abundance of methanol seen in \citet{Drozdovskaya2014}.

With the same model, \citet{Drozdovskaya2016} studied the evolution of various COMs. Their abundances are found to be high in the outer part of the disk and decrease towards the center because of the high UV stellar irradiation. To compare the results of these latter authors with those from the present study, it is consequently better to focus on the abundances in the outer midplane of the disk. The abundance of HCOOCH$_3$ found by these latter authors increases significantly during the collapse similar to our results. They find that the average abundance of methanol is relatively similar at the end of the initial prestellar phase and at the final time in the disk, in agreement with our findings. For the other complex molecules common to the two studies (H$_2$CCO, HCOOH, CH$_3$CHO, CH$_3$OCH$_3$), the situation is different. Their abundances of H$_2$CCO at the end of the initial prestellar phase and in the disk are relatively similar, while we see an increase in H$_2$CCO. HCOOH is more abundant at the end in their model, while no significant change is observed in our study. CH$_3$CHO and CH$_3$OCH$_3$ become more abundant during the formation of the disk in our study, while the average abundance of these molecules decreases more or less depending on the chosen dominant disk growth mechanism in \citet{Drozdovskaya2016}. In spite of these different behaviors, the final abundances of the COMs we obtained are in good agreement with those of these latter authors for an infall dominated disk (except for H$_2$CCO which is significantly less abundant in our study). In their spread-dominated disk, the abundances are generally much lower. Regarding the smaller molecules, the results are also in agreement, as \citet{Drozdovskaya2016} found that most of them (H$_2$O, NH$_3$, CH$_4$, H$_2$S, N$_2$) show relatively similar initial and final average abundances. Differences are seen for CO$_2$.  In our case, CO$_2$ does not evolve significantly, while the abundance of CO$_2$ is found to increase by these latter authors  and to reach higher values than in our study. The stellar FUV irradiation would explain this difference as well. The abundance of H$_2$S is also two orders of magnitude higher in our study than in \citet{Drozdovskaya2016}. Indeed, the initial elemental sulfur abundance is lower in their study. A depleted sulfur abundance is imposed, because their chemical network is not designed to reproduce the observed depletion of sulfur in dark clouds \citep{Vidal2017}.

\subsection{Comparison to (magneto-)hydrodynamic models}

\citet{Aikawa2008} initiated studies of chemical evolution of collapsing cores with radiation hydrodynamic simulations. However, the 1D symmetry adopted in that study prevented any investigation of the molecular evolution during the formation of the disk. \citet{vanWeeren2009} first studied molecular evolution in a forming disk with a 2D hydrodynamical model. This demonstrated that the employed method could be extended to 3D modeling and higher resolution as long as the chemical modeling of the particles is parallelized. 

\citet{Furuya2012} were the first ones to investigate the chemical evolution from a cold core to a first hydrostatic core using a 3D radiation hydrodynamic simulation. Contrary to our study, they did not take into account any magnetic field. The presence of a magnetic field leads to the formation of more realistic disks, which are less inclined to early fragmentation \citep{Commercon2011,Hincelin2013}. As \citet{Furuya2012} mainly focused on the central core ($<$ 10 au), no direct chemical comparison can be made with this work, but their main conclusion was that the total ice and gas abundances of most species were not altered until a temperature of 500\,K, as the warm-up phase was too short to form large organic species. In our model, COMs such as HCOOCH$_3$, CH$_3$CHO, and CH$_3$OCH$_3$ have time to form efficiently.

Later, \citet{Yoneda2016} used a radiation-hydrodynamics model to investigate the evolution of the chemical composition of 1.5 $\times$ 10$^3$ SPH particles from the cold prestellar core to the rotationally supported gravitationally unstable disk. In agreement with our results, they found that molecules such as H$_2$O, CH$_4$, NH$_3$, CO$_2$, H$_2$CO, and CH$_3$OH are already abundant at the onset of gravitational collapse and are later sublimated as the SPH particles migrate inside the water snow line, while COMs such as HCOOCH$_3$ and CH$_3$OCH$_3$ mainly form during the collapse. However, contrary to  \citet{Yoneda2016}, we found that HCOOH, H$_2$CS, and SO$_2$ mainly form before the collapse.

\citet{Hincelin2013} explored the survival of molecules during the collapse of a prestellar core up to the first Larson core using 3D MHD models. Compared to our study, their simulations assumed ideal magnetohydrodynamics and stopped at an earlier time (3.8\,$\times$\,10$^4$ yr). The magnetic fields are also lower ($\mu$=10 and $\mu$=200). The physical structure of the obtained disks differ depending on the intensity and inclination $\Theta$ of the magnetic field. Most of the particles in their simulated disks have low temperatures ($T$ $<$ 100\,K), whereas in our longer simulation the temperature has time to significantly increase in the inner regions (because of the use of the barotropic equation of state, this is due to the density increase in the center). These latter authors concluded that the abundances of the most abundant species do not significantly evolve during the collapse. This is in agreement with our findings relative to the main carriers of the C, N, and O elements. Furthermore, \citet{Hincelin2013} only see strong variations for HNC, CO$_2$, and HNO. In our study, we do not see such changes for these species, but these variations are not particularly important for their model with $\mu$=10, $\Theta$=0 which is the closest to our model ($\mu$=5, $\Theta$=0). It should be noted that, in our study, a significant decrease is observed for CO when the set of initial abundances A is chosen. The abundance of CO can consequently vary when its initial abundance is not particularly high. These latter authors also noted a similar trend for this molecule when testing the impact that an older molecular cloud would have on the chemistry of a particle arriving in the warm region ($\sim$ 30\,K). \citet{Hincelin2016} presented an additional simulation with $\mu$=2, $\Theta$=0, but the strong magnetic field 
in this case prevents the formation of a rotationally supported disk.

\subsection{Comparison to observations}

\begin{figure*}[!ht]
\begin{center}
\includegraphics[width=0.9\hsize]{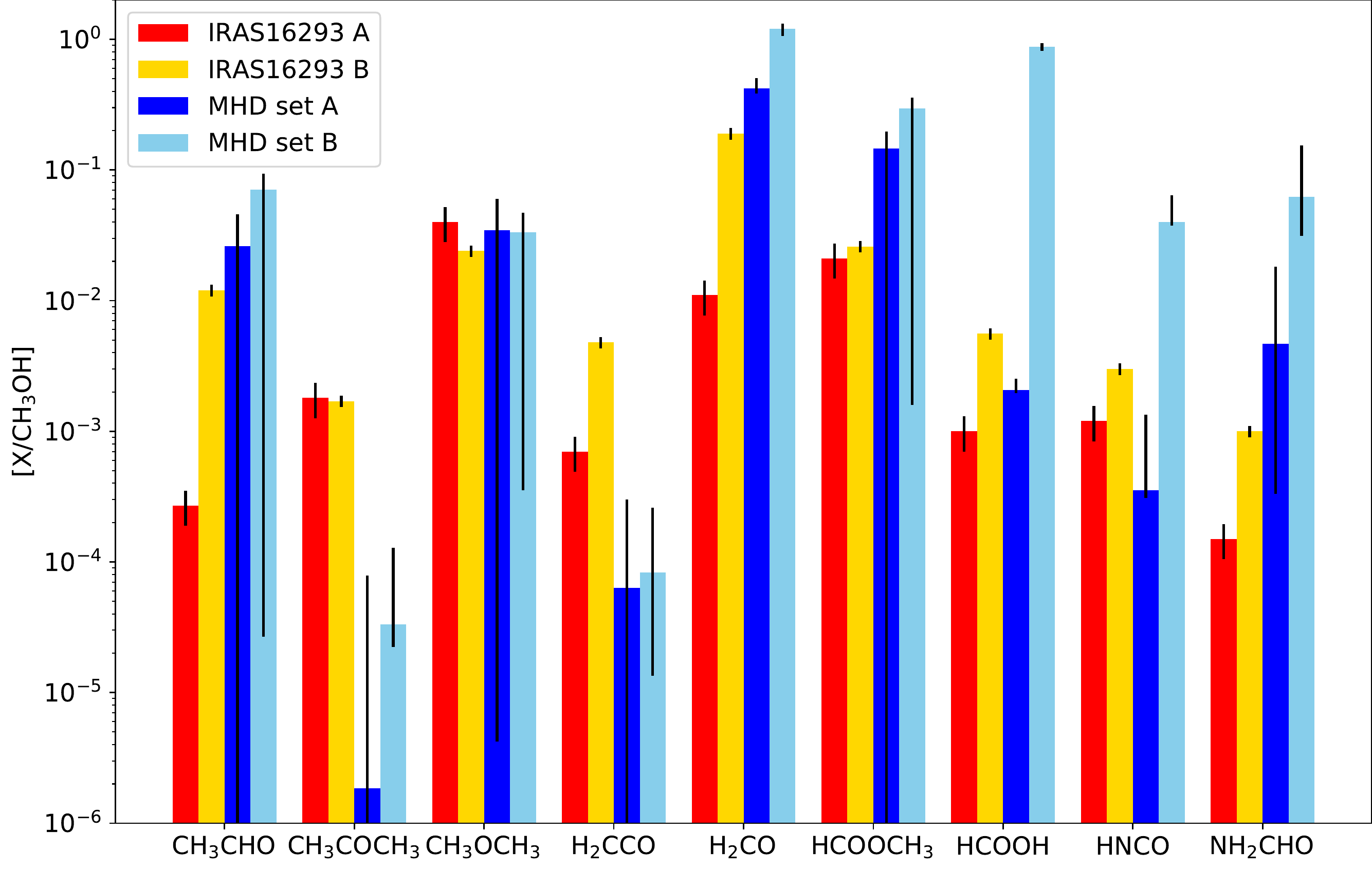}
\caption{Relative abundances of several COMs with respect to CH$_3$OH towards the components A and B of the protostellar binary IRAS16293 (red and yellow, respectively) compared to the mean values found at the final time of our simulations (set of initial abundances A in dark blue and set of initial abundances B in cyan).
The abundance ratios of IRAS16293 come from \citet{Coutens2016,Lykke2017,Persson2018,Jorgensen2018,Manigand2020}. The error bars correspond to an uncertainty of 30\% and 10\%  for the components A and B, respectively. For the simulations, the bars cover the range between the minimum and maximum abundance ratios. }
\label{fig_comp_IRAS16293}
\end{center}
\end{figure*}

The disk obtained at the end of the simulations shows high densities (10$^9$--10$^{13}$ cm$^{-3}$) and covers a large range of temperatures between 11 and 802\,K. The region where the temperature is higher than 100\,K has a diameter of about 50 au. In terms of physical conditions, this region is reminiscent of hot corinos, the warm inner regions of young low-mass protostars, which present a rich complex organic chemistry. 

IRAS~16293-2422 (hereafter IRAS16293) is certainly the most studied of these objects. It has been the target of many observational projects with ALMA. The kinematics were studied at small scales towards the two components A and B of the system. Observations of rotation motions around source A suggested the presence of an almost edge-on disk \citep{Pineda2012,Oya2016}. More recently, \citet{Maureira2020} showed with very high-spatial-resolution observations that component A consists of two compact sources, A1 and A2. A dust disk with a FWHM size of $\sim$12 au is seen towards A2, while an upper limit of $\sim$4 au is derived for the disk around A1. Infall motions are seen towards source B \citep{Pineda2012,Zapata2013}. A face-on disk has been suggested for Source B \citep{Zapata2013,Oya2018}. 
The two components A and B are characterized by a hot corino chemistry \citep[e.g.,][]{Bottinelli2004,Bisschop2008,Caux2011,Jorgensen2016,Manigand2020}. Many COMs were detected for the first time in low-mass protostars towards this source, especially in the framework of the ALMA  Protostellar Interferometric Line Survey (PILS, \citealt{Jorgensen2016}). The emission of COMs is quite compact around the sources, similarly to what we see in our simulations. Indeed, these species are expected to be released in the gas phase through thermal desorption once the temperature is sufficiently high ($T$ $\gtrsim$ 100\,K). 
The relative abundances of the COMs were measured towards both components of the binary \citep[e.g.,][]{Manigand2020}. Given the incompleteness of our network regarding complex organic chemistry, a comparison is only possible for a few COMs. Figure \ref{fig_comp_IRAS16293} shows a comparison of the relative abundances with respect to CH$_3$OH. While some species show good agreement with our model (e.g., CH$_3$OCH$_3$, HCOOH, HNCO), it is clear that other species are underproduced  (CH$_3$OCH$_3$, H$_2$CCO) or overproduced (H$_2$CO), which suggests that some key reactions are missing in this network. A chemical network is currently being developed to include all COMs detected in low-mass star-forming regions. Simulations using this future network would be necessary to model the chemical richness of this source in detail. 

As stated in Section \ref{Sect_evol}, even if our network does not include isotopic fractionation, an interesting correlation is seen between the level of deuteration found for several species in IRAS16293 \citep{Coutens2016,Persson2018,Jorgensen2018,Calcutt2018,Manigand2020} and their period of formation in our simulations. 
On one hand, the species showing a low deuteration of 1--3\% in IRAS16293 B (HNCO, NH$_2$CHO, HCOOH, CH$_3$OH, H$_2$CO, and CH$_3$CN) are the ones presenting similar initial and final abundances in our model. On the other hand, the species with the highest deuteration, 4--8\% (HCOOCH$_3$, CH$_3$OCH$_3$, CH$_3$CHO) are the ones that become significantly more abundant during the collapse.
It should be noted that the abundance of H$_2$CCO, which shows a low deuteration in IRAS16293, increases for the set of initial abundances A, but remains similar for the set B.
 \citet{Jorgensen2018} proposed that this variation in deuteration would reflect the formation time of each species in the ices before or during warm-up and/or infall of material through the  envelope. The results obtained here confirm this interpretation. The species formed during the cold core phase in our simulations also show low deuteration towards IRAS16293 A \citep{Manigand2020}. However, some of the species efficiently formed during the collapse (e.g., HCOOCH$_3$) do not show D/H ratios as high as in source B \citep{Manigand2020}, which would reveal differences stemming from the collapse phases of the components A and B. 

Chemical differentiation was spatially observed on small scales towards IRAS16293 as well as other low-mass protostars \citep[e.g.,][]{Sakai2014a,Oya2016,Okoda2018}. These variations were interpreted as a change of chemistry at the centrifugal barrier separating the infalling and rotating envelope from the disk. In our simulations, molecules also show different spatial distributions (see Section \ref{sect_spatial_dist}). These spatial changes in gas-phase abundances are explained by the increase of temperature and density towards the center. Similarities are seen with these observational studies. CH$_3$OH, HCOOCH$_3$, and SO are concentrated near the protostar in both cases \citep{Sakai2014b,Sakai2014a,Sakai2016,Oya2016,Oya2018}, while CCH and c-C$_3$H$_2$ come from a more distant and colder region: the spiral arms in our simulations and the infalling and rotating envelope in the observational studies by \citet{Sakai2014b,Sakai2014a} and \citet{Okoda2018}. 
Only sulfur-bearing species, CS, H$_2$CS, and OCS, show a different behavior. CS and OCS are observed in the infalling and rotating envelope \citep{Sakai2014b, Oya2016,Oya2018}, while H$_2$CS is seen both in the envelope and inside the centrifugal barrier \citep{Oya2016,Oya2018}. In our simulations, these species are only present in the inner regions. The absence of OCS and CS in the inner regions compared to our simulations suggests missing destruction reactions at high temperature in our network. Their presence in more extended and colder regions in the observations, opposite to our predictions, is certainly due to a difference in the physical conditions. Indeed, in our simulations, we only modeled the disk, which shows very high densities ($>$ 10$^9$ cm$^{-3}$) and low temperatures. In such conditions, these three species are consequently abundantly depleted on the grains. In the envelope, which we did not model in this study, the densities of the particles are lower and these molecules are consequently less depleted onto the grains. This was shown by \citet{Vidal2018}, for example, who modeled the chemistry in the envelope of hot cores and hot corinos and found that the abundances of these three sulfur-bearing species are relatively high in the envelope.

\section{Conclusions}

In this study, we performed 3D MHD simulations of a dense core collapse to investigate the chemical evolution that occurs from the cold core phase to the formation of a young rotationally supported disk. Among the 10$^6$ tracer particles introduced in the simulations, $\sim$1.5 $\times$ 10$^4$ end in the disk. We ran the three-phase gas-grain chemistry code Nautilus to follow the evolution of all the disk particles. Two different sets of initial abundances were considered, taken from SPH simulations of formation of dense cores and carefully chosen to be representative of two prestellar cores with different histories. Comparisons were made between the two final distributions of abundances in the disk as well as with their respective initial abundances in the cold and dense core.
Our main findings can be summarized as follows:
\begin{itemize}
\item[$\bullet$] The molecules were classified into different categories according to their spatial distribution in the disk. The initial abundances can affect the level of abundances of molecules, but have no significant impact on the distribution of most molecules, apart for HCO, HNO, OH, SO, SO$_2$, and CH$_3$. The spatial distribution of the molecules reflects their sensitivity to the temperature distribution. Radicals are usually destroyed in the warm inner region, while molecules such as water and complex organics are present in all regions of the disk, in ices in the cold outer regions, and in the gas phase in the warm inner regions.
\item[$\bullet$] The main carriers of the different chemical elements at the final time of the simulations are usually the same as the initial ones. Significant changes are only seen for phosphorus and sulfur. For one set of initial abundances, atomic P decreases in abundance while the abundances of PO, PN, HCP, and CP increase. For both cases of initial abundances, the main contribution of HS is replaced by that of H$_2$S$_3$. The contribution of radicals such as HCO and CH$_2$OH to the oxygen and carbon budgets is also smaller with time. 
\item[$\bullet$] Despite similar carriers at the initial and final times, tens of molecules show a significant change in abundances during the formation of the disk. In particular, the abundances of ``large'' COMs such as CH$_3$CHO, CH$_3$NH$_2$, CH$_3$OCH$_3$, and HCOOCH$_3$ increase significantly. Despite being present in cold and dense cores, their highest abundances are only reached during the collapse. In contrast, small molecules, especially radicals, see their abundances decrease (CCH, C$_4$H, CH$_3$, CN, NO, NH, NH$_2$, OH, HCO, CH$_3$O, O$_2$). This confirms that the collapse with its warm-up phase promotes molecular complexity. The abundances of some complex species (CH$_3$CHO, CH$_3$OCH$_3$, H$_2$CCO, HCOOCH$_3$) seem to mainly depend on the physical evolution during the collapse, as their final abundances are similar while the initial abundances differ significantly. 
\item[$\bullet$] A high number of species show a similar abundance in the dense core and in the disk. These species are expected to mainly form in the molecular cloud or prestellar core, and include molecules such as H$_2$O, H$_2$CO, HNCO, and ``small'' COMs (CH$_3$OH, CH$_3$CN, NH$_2$CHO) and their possible precursors. It is interesting to note that these COMs are also the ones presenting the lowest deuteration towards the Class 0 protostar IRAS~16293-2422 \citep{Coutens2016, Jorgensen2018, Persson2018, Calcutt2018}.
\item[$\bullet$] In spite of similar initial abundances, some species show different final abundances (CH$_3$O, HCO, OH, HCO$^+$, PCH$_3^+$). These species are consequently sensitive to the initial abundances of other species. 
\item[$\bullet$] The MHD resistivities we obtained from the full gas-grain chemistry are in qualitative agreement with the one we used to estimate the ambipolar diffusion resistivity in the 3D simulations. 
\end{itemize}

In conclusion, the chemical content of prestellar cores has an impact on the final composition of disks, even if it does not affect all species. For some species, the physical evolution during the collapse plays a major role, and for other species both the collapse and the initial abundances of the cold core contribute to the final abundances. 
We classified the species according to their behavior. 
This classification should help observers to interpret the origin of molecules observed in young disks. The results of this study should also be of interest for the chemical modelers of more evolved disks. Indeed, the initial abundances of stand-alone chemical models have to be chosen carefully. Using initial abundances within the range of abundances we obtain here for a young disk should make such models more realistic.

\begin{acknowledgements}
The authors thank the referee M. Drozdovskaya for her constructive comments which helped to improve the paper. 
This project has received funding from the European Research Council (ERC) through the ERC Starting Grant 3DICE (grant agreement 336474). A.C. also acknowledges financial support from the Agence Nationale de la Recherche (grant ANR-19-ERC7-0001-01).  
\end{acknowledgements}

\bibliographystyle{aa} 
\bibliography{Biblio} 
    
\appendix

\section{Supplementary tables}

\onecolumn
\begin{longtable}{l c c c c c c c c c c}
\caption{\label{tab_stat_abundances} Initial and final mean, median, minimum, and maximum abundances (a(b) = a $\times$ 10$^b$) of the species presented in this study for the two sets of initial abundances of all the computed disk particles. The abundances correspond to the total ice + gas abundances. }  \\
\hline \hline
\multicolumn{1}{l}{Species} & \multicolumn{5}{c}{Case A} & \multicolumn{5}{c}{Case B} \\
\hline
 & [X]$_0$ & Mean & Median & Min & Max & [X]$_0$ & Mean & Median & Min & Max \\
\hline
\endfirsthead
\caption{continued.}\\
\hline\hline
\multicolumn{1}{l}{Species} & \multicolumn{5}{c}{Case A} & \multicolumn{5}{c}{Case B} \\
\hline
 & [X]$_0$ & Mean & Median & Min & Max & [X]$_0$ & Mean & Median & Min & Max \\
\hline
\endhead
\hline
\endfoot
c-C$_3$H & 1.3(-07) & 2.6(-08) & 4.2(-08) & 2.8(-24) & 5.4(-08)                 & 3.8(-08) & 1.6(-08) & 2.7(-08) & 3.4(-25) & 3.5(-08) \\ 
c-C$_3$H$_2$ & 2.6(-07) & 7.5(-08) & 1.1(-07) & 1.5(-21) & 1.7(-07)                 & 4.2(-08) & 2.6(-08) & 3.5(-08) & 1.7(-22) & 7.5(-08) \\ 
C$_2$H$_7^+$ & 1.3(-15) & 1.2(-12) & 7.6(-18) & 1.3(-21) & 1.0(-11)                 & 4.7(-17) & 4.2(-14) & 7.1(-19) & 9.2(-23) & 7.0(-13) \\ 
C$_3$H$_8$ & 7.6(-06) & 9.2(-06) & 9.2(-06) & 8.5(-06) & 9.5(-06)                 & 1.1(-06) & 1.1(-06) & 1.1(-06) & 1.0(-06) & 1.1(-06) \\ 
C$_4$H & 1.9(-11) & 2.0(-14) & 2.1(-17) & 3.7(-22) & 5.3(-13)                 & 2.7(-11) & 4.2(-14) & 1.6(-17) & 5.9(-22) & 1.1(-12) \\ 
CCH & 6.3(-10) & 2.3(-14) & 7.0(-16) & 1.7(-20) & 1.0(-13)                 & 8.0(-10) & 2.0(-14) & 8.8(-16) & 1.5(-21) & 8.9(-14) \\ 
CCS & 6.3(-10) & 1.1(-10) & 1.3(-10) & 1.5(-14) & 2.8(-10)                 & 6.6(-10) & 2.3(-10) & 2.9(-10) & 1.2(-15) & 3.8(-10) \\ 
CH$_2$NH & 1.6(-08) & 1.9(-08) & 1.8(-08) & 1.2(-08) & 4.9(-08)                 & 4.5(-07) & 4.5(-07) & 4.5(-07) & 3.4(-07) & 6.8(-07) \\ 
CH$_2$OH & 4.8(-06) & 2.2(-06) & 2.7(-06) & 1.1(-13) & 5.2(-06)                 & 8.4(-06) & 5.1(-06) & 6.3(-06) & 1.1(-12) & 9.2(-06) \\ 
CH$_3$ & 2.7(-06) & 4.8(-08) & 5.5(-12) & 9.9(-14) & 2.5(-06)                 & 4.3(-06) & 7.2(-08) & 1.7(-12) & 3.9(-14) & 4.5(-06) \\ 
CH$_3$CCH & 4.1(-07) & 4.5(-07) & 2.9(-07) & 2.5(-07) & 1.4(-06)                 & 2.8(-08) & 6.1(-08) & 3.3(-08) & 2.7(-08) & 2.4(-07) \\ 
CH$_3$CHO & 8.6(-12) & 3.4(-07) & 3.9(-07) & 1.0(-11) & 5.8(-07)                 & 9.0(-10) & 1.7(-06) & 1.8(-06) & 6.6(-10) & 2.2(-06) \\ 
CH$_3$CN & 1.7(-08) & 1.6(-08) & 1.5(-08) & 1.4(-08) & 6.0(-08)                 & 2.1(-08) & 1.8(-08) & 1.8(-08) & 1.8(-08) & 2.1(-08) \\ 
CH$_3$COCH$_3$ & 2.4(-13) & 2.4(-11) & 8.1(-12) & 2.4(-13) & 1.0(-09)                 & 5.7(-10) & 8.0(-10) & 6.7(-10) & 5.5(-10) & 3.0(-09) \\ 
CH$_3$NH$_2$ & 5.7(-12) & 2.0(-09) & 1.7(-09) & 5.5(-12) & 8.2(-09)                 & 8.0(-10) & 1.4(-08) & 1.1(-08) & 7.6(-10) & 5.8(-08) \\ 
CH$_3$O & 2.9(-06) & 1.7(-07) & 7.3(-08) & 1.4(-16) & 2.6(-06)                 & 6.9(-06) & 3.4(-07) & 1.0(-11) & 1.0(-20) & 8.3(-06) \\ 
CH$_3$OCH$_3$ & 4.5(-11) & 4.5(-07) & 5.1(-07) & 5.6(-11) & 7.6(-07)                 & 8.7(-09) & 8.0(-07) & 9.3(-07) & 8.7(-09) & 1.1(-06) \\ 
CH$_3$OH & 1.1(-05) & 1.3(-05) & 1.3(-05) & 1.2(-05) & 1.4(-05)                 & 2.3(-05) & 2.4(-05) & 2.4(-05) & 2.3(-05) & 2.5(-05) \\ 
CH$_3$S & 1.4(-06) & 6.8(-07) & 8.3(-07) & 6.5(-13) & 1.0(-06)                 & 2.5(-07) & 1.8(-07) & 2.2(-07) & 8.7(-14) & 2.6(-07) \\ 
CH$_3$SH & 1.8(-06) & 1.6(-06) & 1.6(-06) & 1.3(-06) & 1.9(-06)                 & 2.8(-07) & 3.4(-07) & 3.4(-07) & 2.8(-07) & 4.1(-07) \\ 
CH$_4$ & 2.3(-05) & 2.6(-05) & 2.5(-05) & 2.4(-05) & 3.4(-05)                 & 1.9(-05) & 2.2(-05) & 2.0(-05) & 1.9(-05) & 2.9(-05) \\ 
CH$_5^+$ & 7.7(-11) & 4.8(-12) & 1.2(-14) & 1.8(-20) & 1.5(-11)                 & 5.9(-11) & 3.7(-12) & 2.9(-15) & 3.4(-20) & 1.5(-11) \\ 
CN & 1.8(-10) & 9.7(-11) & 3.2(-14) & 1.4(-19) & 2.6(-09)                 & 1.1(-09) & 9.4(-12) & 2.9(-15) & 1.2(-20) & 2.7(-10) \\ 
CO & 5.9(-07) & 9.8(-08) & 7.0(-09) & 1.1(-15) & 1.4(-06)                 & 1.4(-05) & 1.3(-05) & 1.3(-05) & 9.4(-06) & 1.7(-05) \\ 
CO$_2$ & 2.1(-07) & 2.3(-07) & 2.3(-07) & 2.0(-07) & 3.9(-07)                 & 9.9(-06) & 1.0(-05) & 1.0(-05) & 1.0(-05) & 1.1(-05) \\ 
CP & 7.1(-09) & 8.3(-09) & 7.8(-09) & 7.1(-09) & 2.4(-08)                 & 1.0(-10) & 7.0(-09) & 5.7(-09) & 1.2(-10) & 2.4(-08) \\ 
CS & 5.3(-08) & 9.6(-09) & 3.8(-09) & 5.2(-16) & 1.6(-07)                 & 1.0(-07) & 1.9(-08) & 1.7(-08) & 3.4(-14) & 1.1(-07) \\ 
H$_2$CCO & 1.0(-11) & 8.2(-10) & 7.2(-10) & 1.2(-11) & 3.8(-09)                 & 3.9(-10) & 2.0(-09) & 1.9(-09) & 3.3(-10) & 6.1(-09) \\ 
H$_2$CO & 6.6(-06) & 5.5(-06) & 5.5(-06) & 5.1(-06) & 6.4(-06)                 & 2.5(-05) & 2.9(-05) & 3.0(-05) & 2.6(-05) & 3.1(-05) \\ 
H$_2$CS & 4.7(-07) & 5.8(-07) & 2.7(-07) & 2.3(-07) & 2.8(-06)                 & 2.5(-07) & 6.6(-07) & 2.3(-07) & 2.1(-07) & 3.1(-06) \\ 
H$_2$O & 3.0(-04) & 3.0(-04) & 3.0(-04) & 2.9(-04) & 3.1(-04)                 & 1.6(-04) & 1.6(-04) & 1.6(-04) & 1.5(-04) & 1.7(-04) \\ 
H$_2$S & 4.8(-06) & 5.2(-06) & 5.3(-06) & 3.7(-06) & 7.7(-06)                 & 4.5(-06) & 4.5(-06) & 4.8(-06) & 2.6(-06) & 7.5(-06) \\ 
H$_2$S$_3$ & 9.1(-22) & 1.5(-06) & 1.5(-06) & 8.8(-22) & 2.3(-06)                 & 3.5(-08) & 1.7(-06) & 1.6(-06) & 3.4(-08) & 2.4(-06) \\ 
H$_3^+$ & 4.0(-09) & 1.9(-12) & 7.6(-16) & 2.1(-19) & 1.6(-10)                 & 4.4(-09) & 1.9(-12) & 6.1(-16) & 2.7(-19) & 1.6(-10) \\ 
H$_5$C$_2$O$_2^+$ & 1.2(-17) & 4.2(-13) & 3.4(-21) & 1.0(-25) & 1.7(-12)                 & 2.2(-17) & 6.0(-13) & 1.6(-19) & 1.4(-25) & 2.4(-12) \\ 
HC$_3$N & 4.4(-08) & 4.5(-08) & 4.4(-08) & 4.2(-08) & 6.6(-08)                 & 2.1(-08) & 2.1(-08) & 2.1(-08) & 2.0(-08) & 2.9(-08) \\ 
HCl & 3.4(-08) & 3.4(-08) & 3.4(-08) & 3.3(-08) & 3.5(-08)                 & 3.3(-08) & 3.4(-08) & 3.4(-08) & 3.3(-08) & 3.5(-08) \\ 
HCN & 5.0(-05) & 5.0(-05) & 5.0(-05) & 4.9(-05) & 5.1(-05)                 & 2.9(-06) & 3.1(-06) & 3.0(-06) & 2.9(-06) & 3.4(-06) \\ 
HCO & 3.1(-06) & 1.1(-07) & 1.1(-10) & 2.0(-17) & 2.5(-06)                 & 1.8(-05) & 2.1(-06) & 8.2(-09) & 4.1(-15) & 1.8(-05) \\ 
HCO$^+$ & 6.0(-12) & 6.3(-15) & 4.1(-20) & 3.8(-23) & 1.1(-12)                 & 1.7(-11) & 2.5(-13) & 1.6(-16) & 5.9(-20) & 4.3(-12) \\ 
HCOOCH$_3$ & 3.4(-12) & 1.9(-06) & 1.9(-06) & 3.3(-12) & 2.5(-06)                 & 4.0(-08) & 7.1(-06) & 7.2(-06) & 3.9(-08) & 8.4(-06) \\ 
HCOOH & 2.4(-08) & 2.7(-08) & 2.7(-08) & 2.6(-08) & 3.2(-08)                 & 2.1(-05) & 2.1(-05) & 2.1(-05) & 2.0(-05) & 2.2(-05) \\ 
HCP & 4.3(-09) & 5.3(-09) & 4.8(-09) & 4.1(-09) & 2.2(-08)                 & 4.0(-10) & 7.6(-09) & 6.3(-09) & 3.9(-10) & 2.6(-08) \\ 
HCS & 1.5(-07) & 2.4(-08) & 3.0(-08) & 1.3(-18) & 5.6(-08)                 & 1.7(-07) & 6.7(-08) & 8.0(-08) & 8.7(-14) & 1.2(-07) \\ 
HCSi & 9.0(-08) & 8.7(-08) & 8.7(-08) & 8.6(-08) & 8.7(-08)                 & 1.7(-09) & 1.7(-09) & 1.7(-09) & 1.6(-09) & 1.8(-09) \\ 
HF & 1.8(-08) & 1.8(-08) & 1.8(-08) & 1.7(-08) & 1.9(-08)                 & 1.8(-08) & 1.8(-08) & 1.8(-08) & 1.7(-08) & 1.9(-08) \\ 
HNC & 3.3(-08) & 2.0(-08) & 3.1(-08) & 2.6(-12) & 4.7(-08)                 & 1.9(-07) & 1.2(-07) & 1.8(-07) & 8.0(-12) & 2.0(-07) \\ 
HNCO & 4.3(-09) & 4.6(-09) & 4.2(-09) & 4.1(-09) & 1.7(-08)                 & 1.1(-06) & 9.6(-07) & 9.3(-07) & 9.2(-07) & 1.5(-06) \\ 
HNO & 3.2(-10) & 6.9(-11) & 5.4(-16) & 4.1(-20) & 1.8(-09)                 & 5.7(-07) & 3.7(-07) & 4.5(-07) & 1.7(-11) & 1.0(-06) \\ 
HOOH & 2.6(-09) & 2.0(-09) & 2.0(-09) & 1.7(-09) & 2.5(-09)                 & 6.1(-08) & 9.0(-08) & 8.5(-08) & 7.0(-08) & 1.5(-07) \\ 
HSO & 1.6(-09) & 1.2(-09) & 1.2(-09) & 9.4(-10) & 1.7(-08)                 & 3.1(-07) & 2.5(-07) & 2.6(-07) & 1.8(-07) & 2.8(-07) \\ 
l-C$_3$H & 2.5(-11) & 2.3(-10) & 7.2(-13) & 2.2(-24) & 6.8(-09)                 & 1.3(-11) & 2.6(-10) & 3.3(-13) & 2.7(-25) & 8.4(-09) \\ 
l-C$_3$H$_2$ & 1.4(-07) & 4.2(-08) & 6.1(-08) & 8.6(-22) & 7.7(-08)                 & 2.2(-08) & 1.4(-08) & 1.8(-08) & 7.7(-23) & 3.2(-08) \\ 
N$_2$ & 3.6(-06) & 3.6(-06) & 3.6(-06) & 3.5(-06) & 3.7(-06)                 & 3.4(-06) & 4.1(-06) & 4.1(-06) & 3.4(-06) & 4.4(-06) \\ 
N$_2$H$^+$ & 1.9(-13) & 7.4(-17) & 2.4(-17) & 1.7(-21) & 4.5(-16)                 & 3.0(-13) & 9.8(-17) & 3.2(-17) & 3.3(-21) & 5.2(-16) \\ 
N$_2$O & 1.1(-14) & 8.9(-12) & 1.2(-14) & 1.0(-14) & 2.0(-10)                 & 9.0(-11) & 1.7(-09) & 8.7(-11) & 8.6(-11) & 5.0(-08) \\ 
NH & 5.3(-08) & 2.4(-09) & 2.0(-13) & 5.2(-18) & 5.0(-08)                 & 1.5(-06) & 5.2(-08) & 2.8(-15) & 3.7(-17) & 1.5(-06) \\ 
NH$_2$ & 6.0(-08) & 9.4(-10) & 3.2(-12) & 1.3(-17) & 3.3(-08)                 & 1.6(-06) & 5.8(-09) & 3.4(-15) & 2.4(-17) & 4.8(-07) \\ 
NH$_2$CH$_2$OH & 1.3(-07) & 1.2(-07) & 1.2(-07) & 1.1(-07) & 1.2(-07)                 & 6.7(-06) & 6.0(-06) & 6.0(-06) & 5.9(-06) & 6.3(-06) \\ 
NH$_2$CHO & 4.6(-09) & 6.1(-08) & 2.5(-08) & 4.4(-09) & 2.3(-07)                 & 9.3(-07) & 1.5(-06) & 1.1(-06) & 7.7(-07) & 3.6(-06) \\ 
NH$_3$ & 3.9(-06) & 3.9(-06) & 3.9(-06) & 3.8(-06) & 4.0(-06)                 & 3.4(-05) & 3.6(-05) & 3.6(-05) & 3.5(-05) & 3.7(-05) \\ 
NO & 3.2(-10) & 4.6(-11) & 2.6(-13) & 9.8(-21) & 2.1(-09)                 & 6.0(-07) & 1.9(-07) & 1.6(-08) & 4.1(-14) & 1.2(-06) \\ 
NS & 1.2(-07) & 1.1(-07) & 1.2(-07) & 6.5(-08) & 1.4(-07)                 & 1.0(-06) & 1.0(-06) & 1.0(-06) & 9.2(-07) & 1.1(-06) \\ 
O$_2$ & 2.8(-10) & 4.7(-13) & 2.7(-15) & 7.1(-20) & 9.7(-11)                 & 5.6(-08) & 1.5(-09) & 7.0(-10) & 9.7(-17) & 8.3(-08) \\ 
OCS & 1.0(-09) & 3.9(-07) & 4.1(-07) & 6.6(-10) & 5.2(-07)                 & 1.0(-06) & 1.1(-06) & 1.1(-06) & 4.6(-07) & 1.6(-06) \\ 
OH & 1.1(-09) & 1.2(-11) & 1.4(-14) & 2.1(-17) & 3.1(-10)                 & 1.9(-09) & 2.2(-14) & 1.8(-15) & 1.8(-17) & 5.3(-12) \\ 
P & 6.5(-08) & 6.2(-08) & 6.3(-08) & 2.8(-08) & 6.6(-08)                 & 6.6(-08) & 3.1(-08) & 2.9(-08) & 2.9(-13) & 6.6(-08) \\ 
PCH$_3^+$ & 1.1(-17) & 3.7(-14) & 1.5(-17) & 3.9(-25) & 5.3(-12)                 & 1.7(-17) & 5.5(-13) & 1.2(-17) & 2.8(-27) & 6.3(-12) \\ 
PN & 6.9(-10) & 6.8(-10) & 6.8(-10) & 6.7(-10) & 6.9(-10)                 & 1.1(-08) & 1.1(-08) & 1.1(-08) & 1.0(-08) & 1.1(-08) \\ 
PO & 7.7(-12) & 1.1(-10) & 1.7(-11) & 7.7(-12) & 1.1(-09)                 & 1.4(-10) & 2.0(-08) & 1.4(-08) & 1.4(-10) & 6.6(-08) \\ 
SiC$_4$H & 1.3(-07) & 1.2(-07) & 1.2(-07) & 1.1(-07) & 1.3(-07)                 & 5.2(-07) & 4.9(-07) & 4.9(-07) & 4.8(-07) & 5.1(-07) \\ 
SiC$_6$H & 1.5(-07) & 1.4(-07) & 1.4(-07) & 1.4(-07) & 1.5(-07)                 & 1.3(-07) & 1.2(-07) & 1.2(-07) & 1.2(-07) & 1.3(-07) \\ 
SiC$_8$H & 3.1(-07) & 2.9(-07) & 2.9(-07) & 2.9(-07) & 3.0(-07)                 & 5.0(-08) & 4.7(-08) & 4.7(-08) & 4.7(-08) & 4.9(-08) \\ 
SiH$_4$ & 7.5(-07) & 8.0(-07) & 8.0(-07) & 7.9(-07) & 8.3(-07)                 & 3.0(-07) & 3.2(-07) & 3.2(-07) & 3.2(-07) & 3.4(-07) \\ 
SiO & 3.6(-08) & 3.6(-08) & 3.6(-08) & 3.5(-08) & 3.6(-08)                 & 3.8(-07) & 3.8(-07) & 3.8(-07) & 3.7(-07) & 4.1(-07) \\ 
SO & 8.8(-10) & 1.5(-08) & 9.5(-10) & 5.5(-10) & 3.9(-07)                 & 1.5(-07) & 1.2(-07) & 1.3(-07) & 6.7(-08) & 2.6(-07) \\ 
SO$_2$ & 4.0(-11) & 2.9(-10) & 1.0(-10) & 1.2(-11) & 8.8(-09)                 & 1.8(-07) & 1.5(-07) & 1.5(-07) & 7.3(-08) & 1.9(-07) \\ 
\hline
\end{longtable}

\section{Supplementary figures}
\label{sect_app_figures}

\begin{figure*}[!ht]
\begin{center}
\includegraphics[width=0.9\hsize]{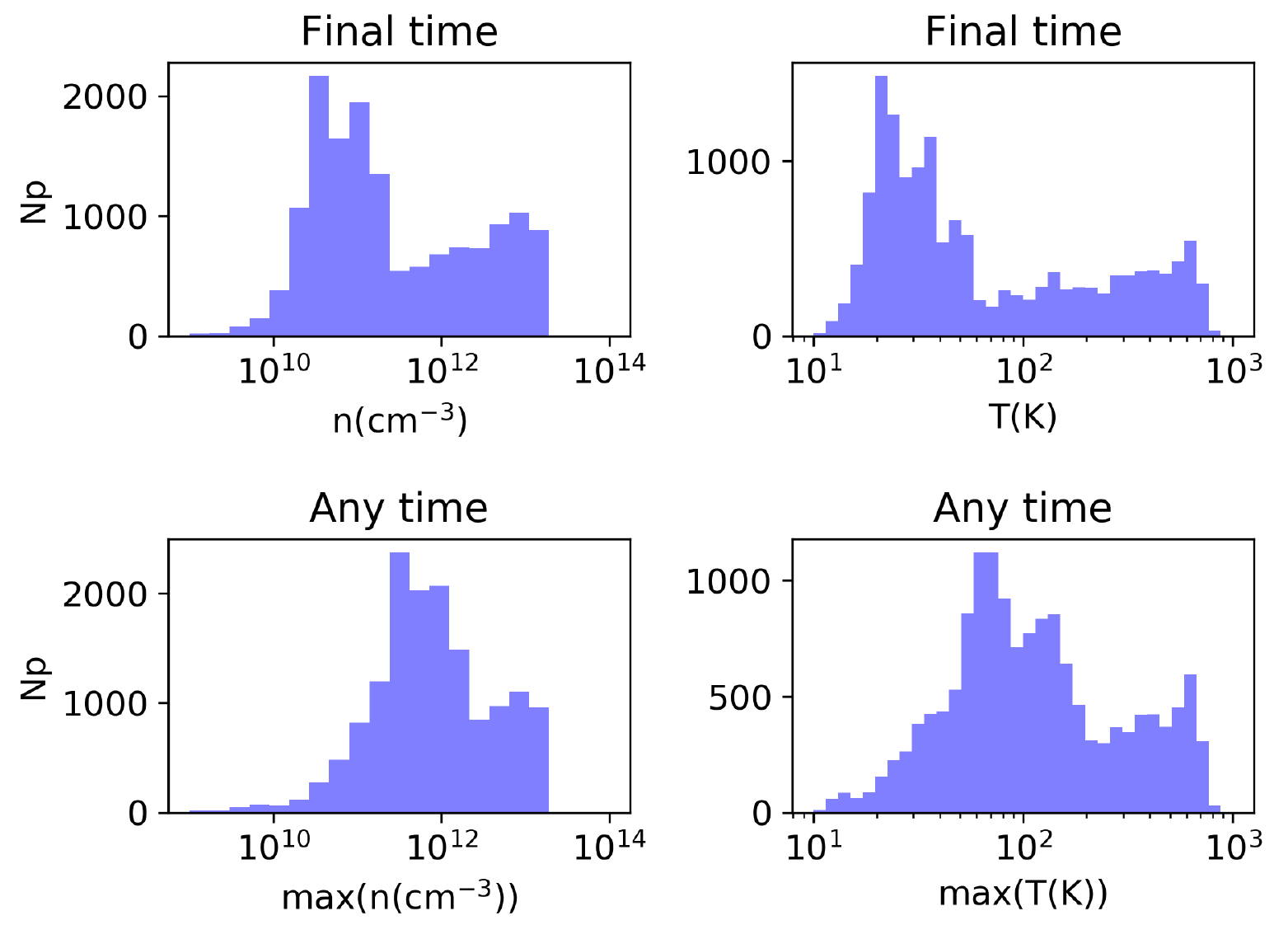}
\caption{Upper panels: Distribution of the number of disk particles as a function of the density (left) and temperature (right) at the final time of the simulations (5.83 $\times$ 10$^4$ yr).
Lower panels: Distribution as a function of the maximum density and temperature reached by the disk particles at any time during their trajectory.}
\label{fig_histo_structure}
\end{center}
\end{figure*}

\begin{figure*}[!ht]
\begin{center}
\includegraphics[width=0.9\hsize]{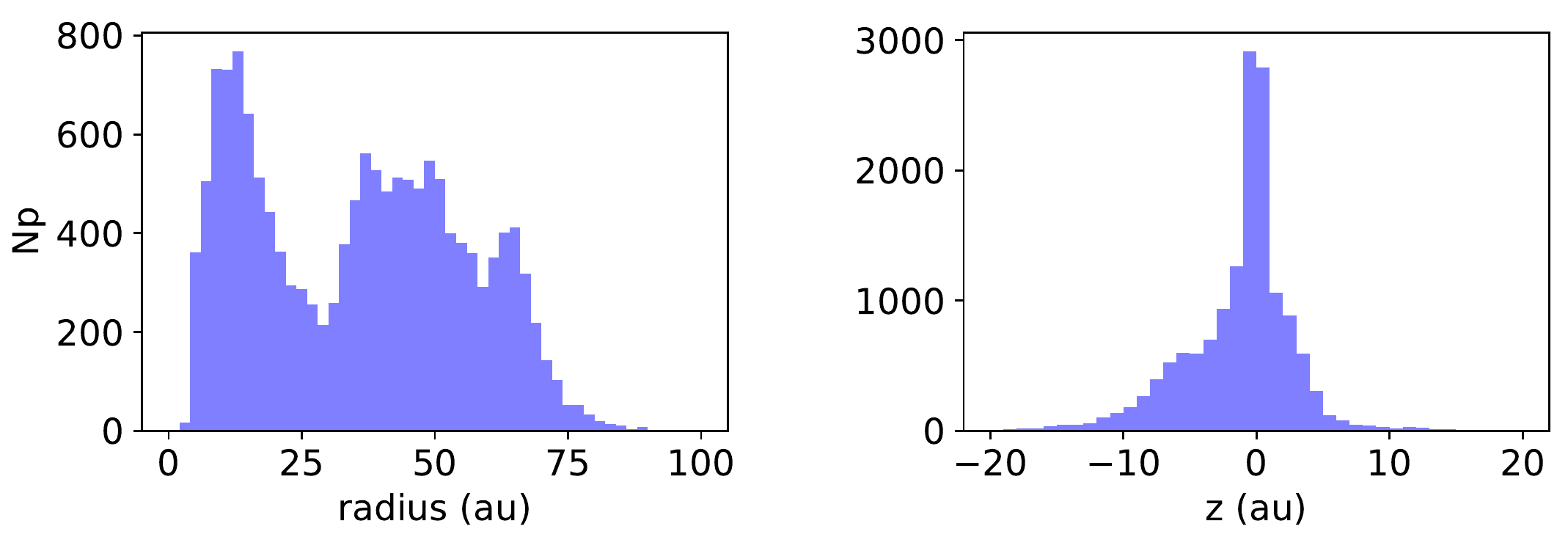}
\caption{Distribution of the number of disk particles as a function of the radius (left) and the height ($z$, right) at the final time of the simulations (5.83 $\times$ 10$^4$ yr).}
\label{fig_histo_radius_vertical}
\end{center}
\end{figure*}

\begin{figure}[!ht]
\begin{center}
 \includegraphics[width=1.\hsize]{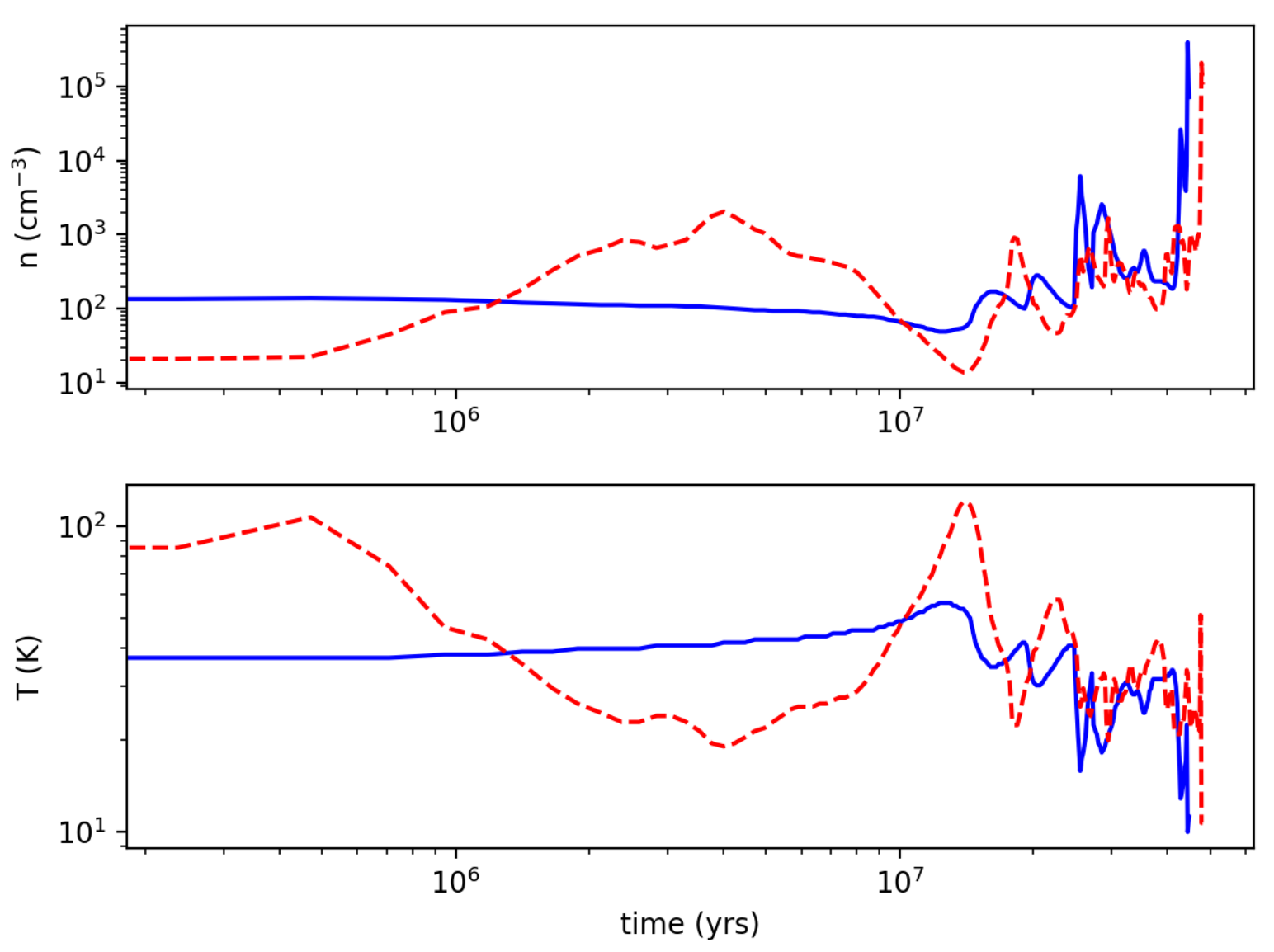}
 \caption{Evolution of the density and temperature of two SPH simulations of the formation of dense cores \citep{Ruaud2018} leading to the two sets of initial abundances A and B used in our simulations (blue solid line for cloud A, red dashed line for cloud B).}
\label{fig_ab_init}
\end{center}
\end{figure}

\begin{figure*}[!ht]
\begin{center}
\includegraphics[width=\hsize]{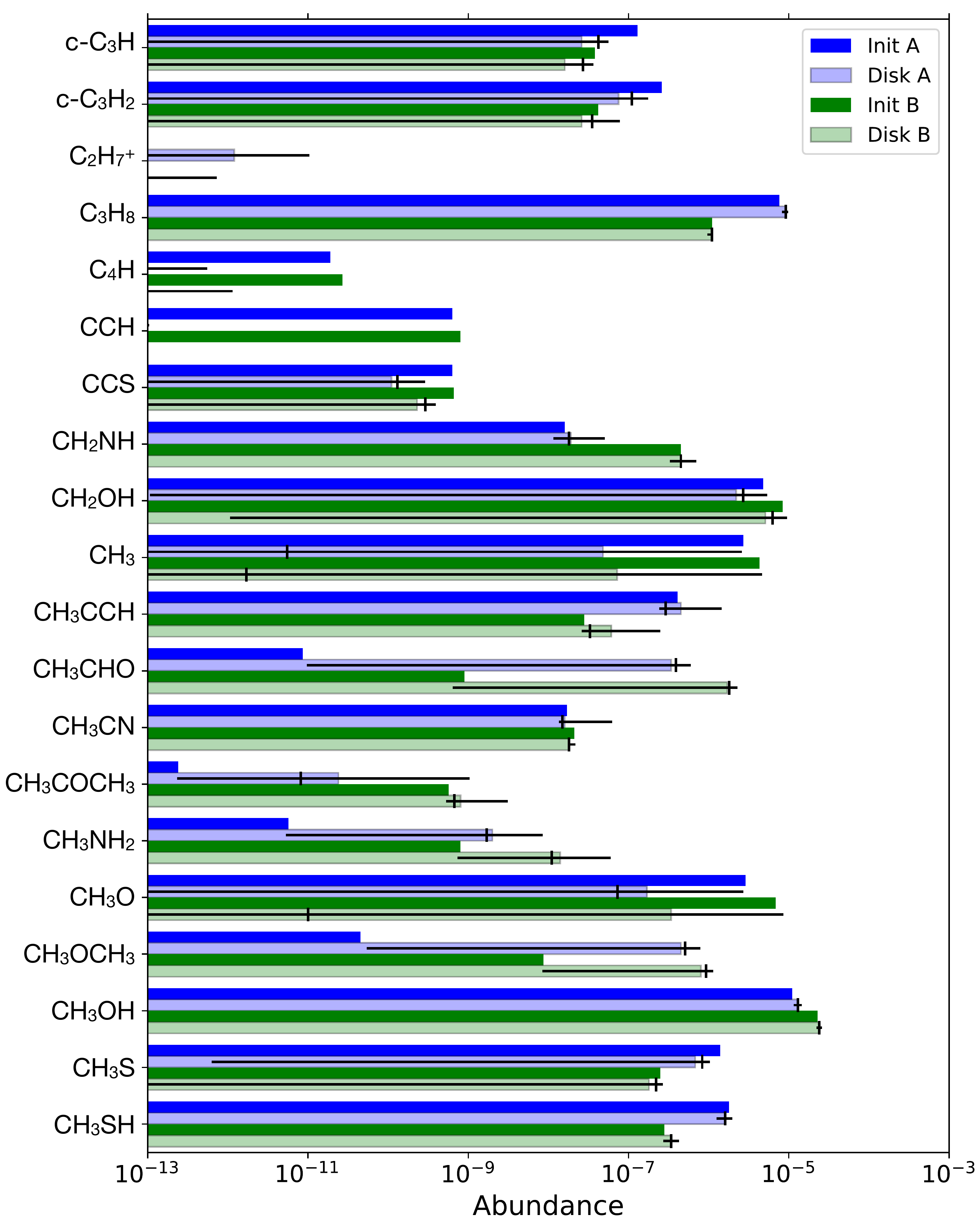}
\caption{Initial and final (gas + grain) abundances of the simulations (case A in blue and case B in green). The bar in light color corresponds to the mean abundance at the final time. The horizontal black line covers the range from the minimum to the maximum abundance among all the disk particles at the final time of the simulations. The vertical black line is the median value.}
\label{fig_stat_1}
\end{center}
\end{figure*}

\begin{figure*}[!ht]
\begin{center}
\includegraphics[width=\hsize]{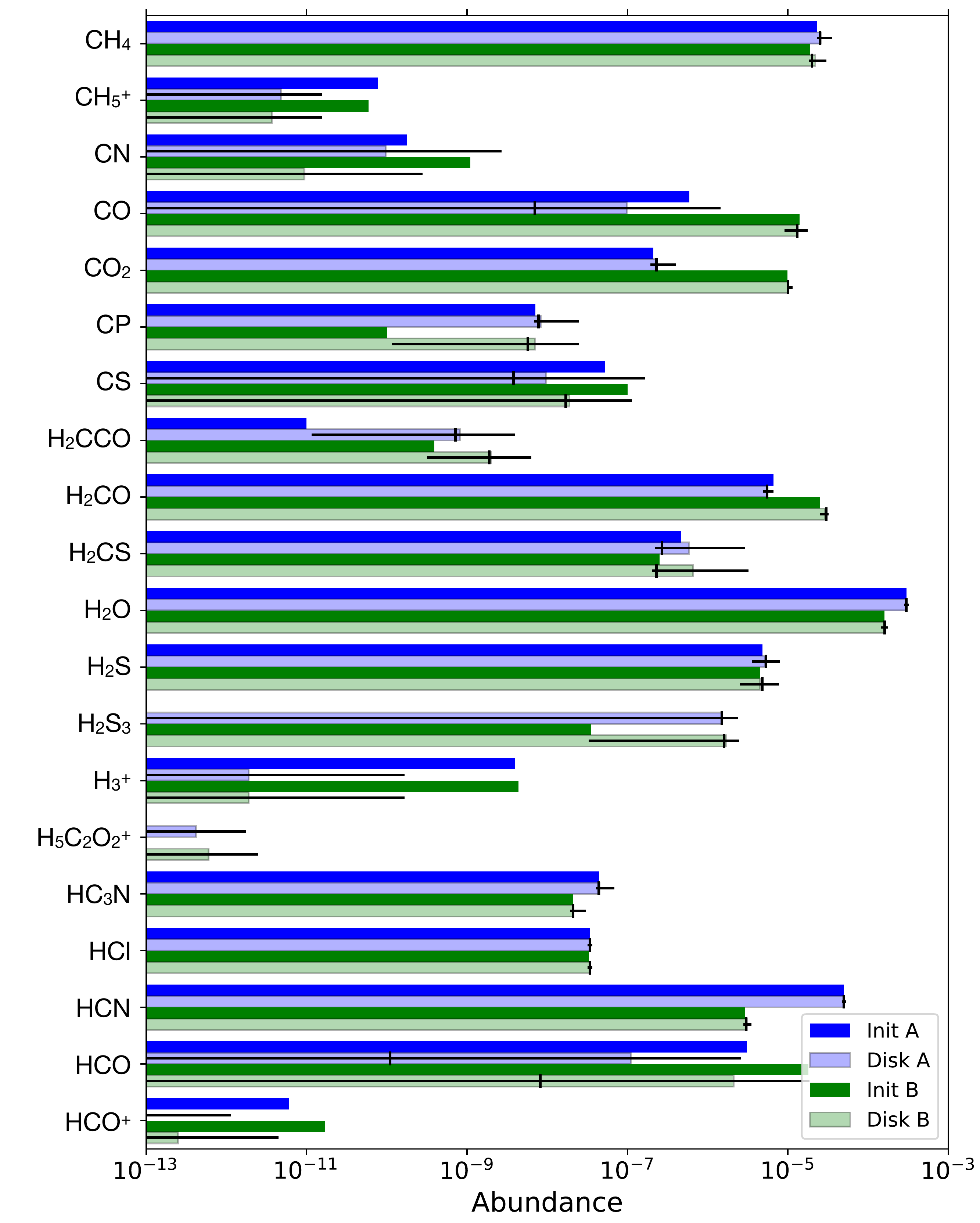}
\caption{Same as Fig. \ref{fig_stat_1}.}
\label{fig_stat_2}
\end{center}
\end{figure*}

\begin{figure*}[!ht]
\begin{center}
\includegraphics[width=\hsize]{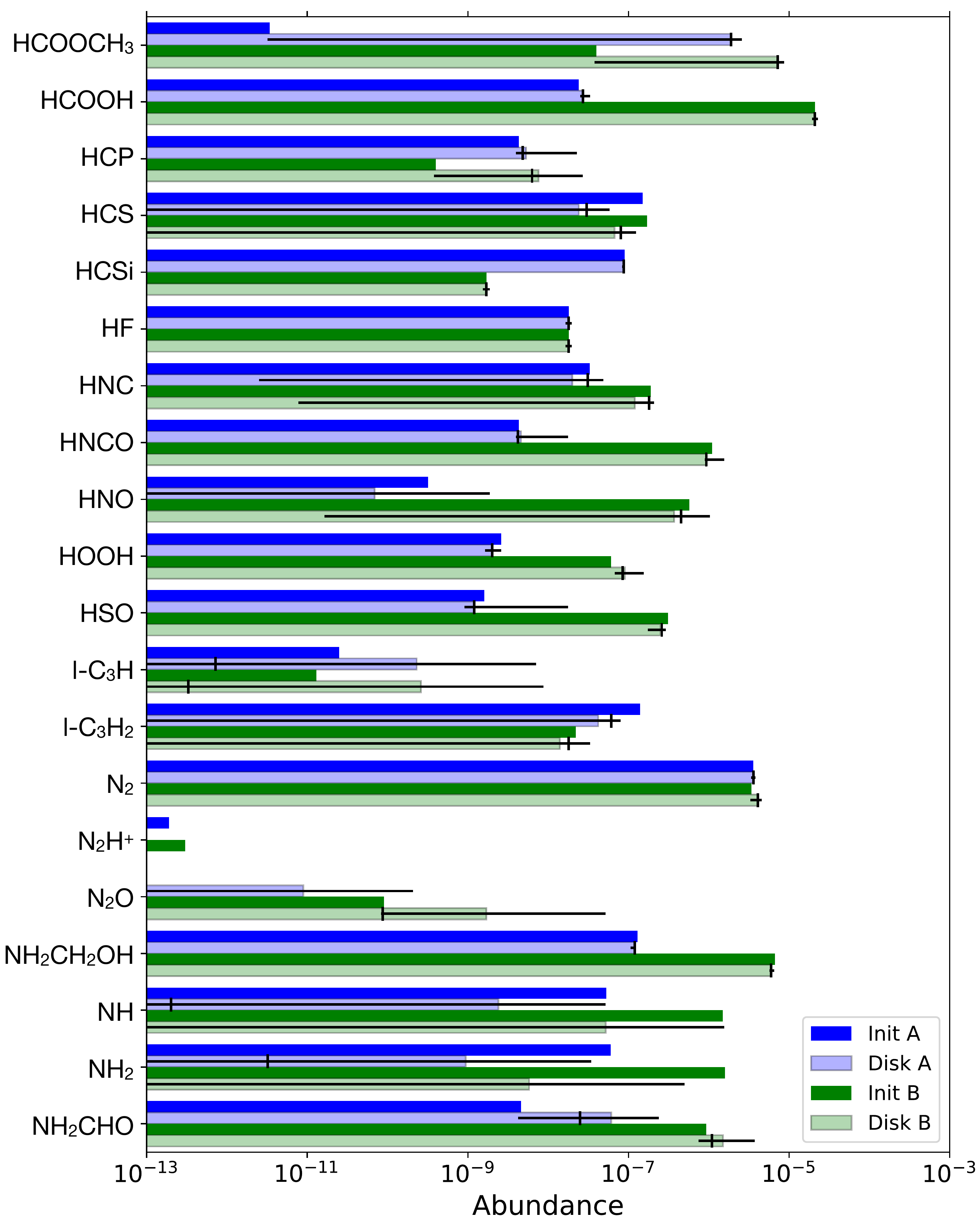}
\caption{Same as Fig. \ref{fig_stat_1}.}
\label{fig_stat_3}
\end{center}
\end{figure*}

\begin{figure*}[!ht]
\begin{center}
\includegraphics[width=\hsize]{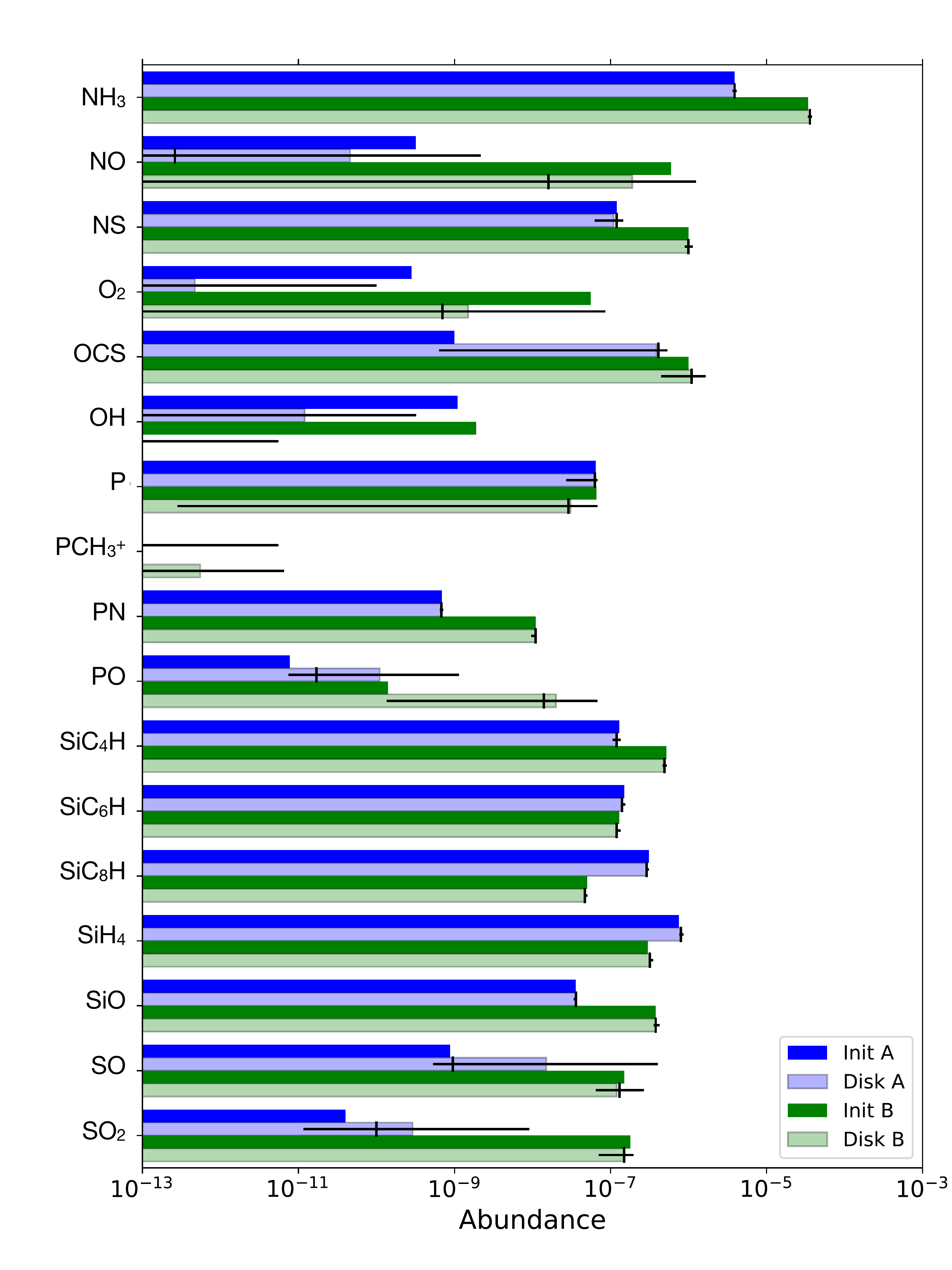}
\vspace{-1cm}
\caption{Same as Fig. \ref{fig_stat_1}.}
\label{fig_stat_4}
\end{center}
\end{figure*}

\begin{figure*}[!ht]
\begin{center}
\includegraphics[width=\hsize]{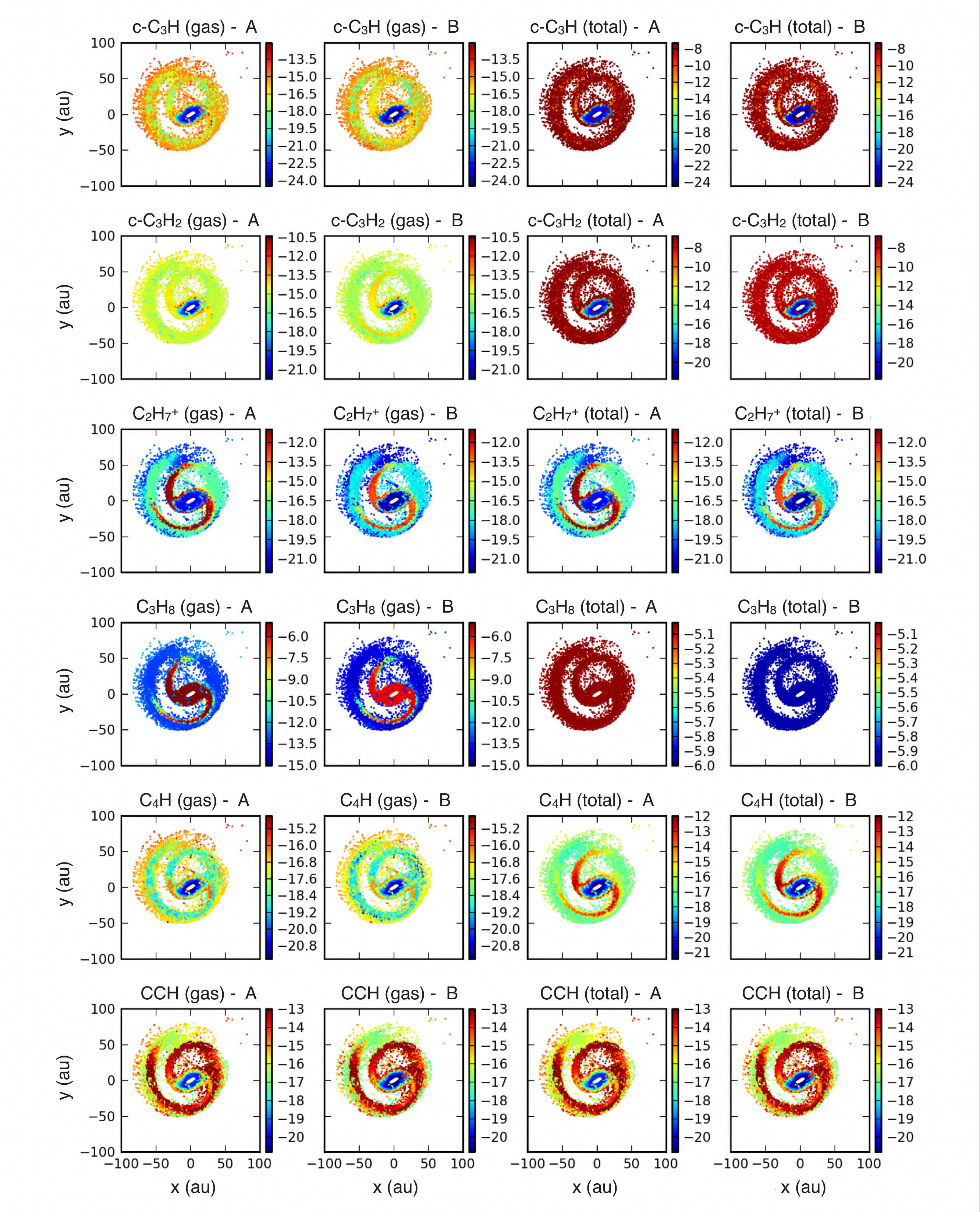}
\caption{Spatial distribution of the molecules in the x-y plane of the disk (stacked z axis) obtained at the final time of the simulations (5.83\,$\times$\,10$^4$~yr) for the two sets of initial abundances A and B. The left panels show the distribution of the gas-phase abundance only, while the right panels show the total (gas + ice) abundance. The color scales represent log$_{10}$(n(X)/n$_{\rm H}$).}
\label{fig_spatial_distrib}
\end{center}
\end{figure*}

\begin{figure*}[!ht]
\begin{center}
\includegraphics[width=\hsize]{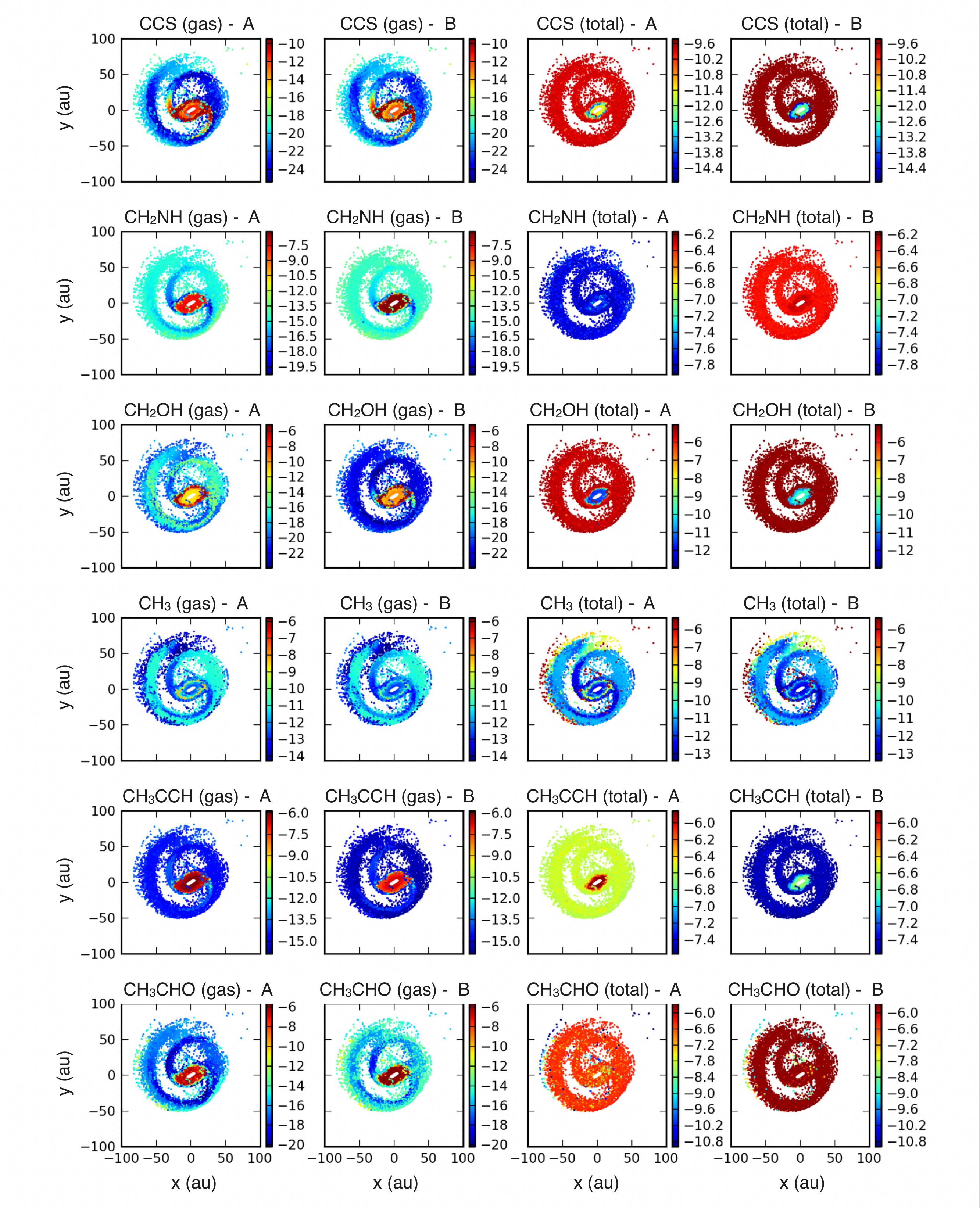}
\caption{Same as Fig. \ref{fig_spatial_distrib}.}
\label{}
\end{center}
\end{figure*}

\begin{figure*}[!ht]
\begin{center}
\includegraphics[width=\hsize]{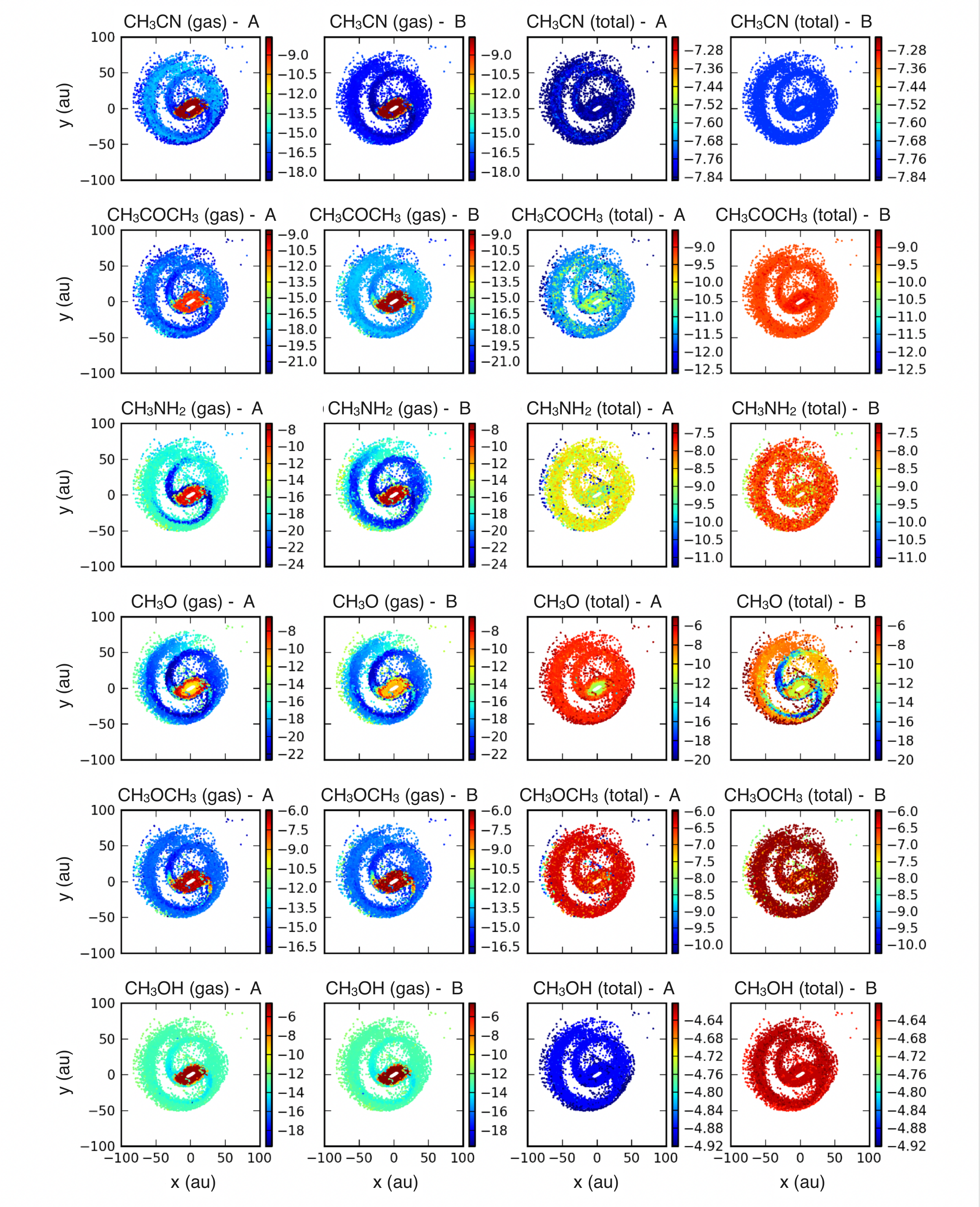}
\caption{Same as Fig. \ref{fig_spatial_distrib}.}
\label{}
\end{center}
\end{figure*}

\begin{figure*}[!ht]
\begin{center}
\includegraphics[width=\hsize]{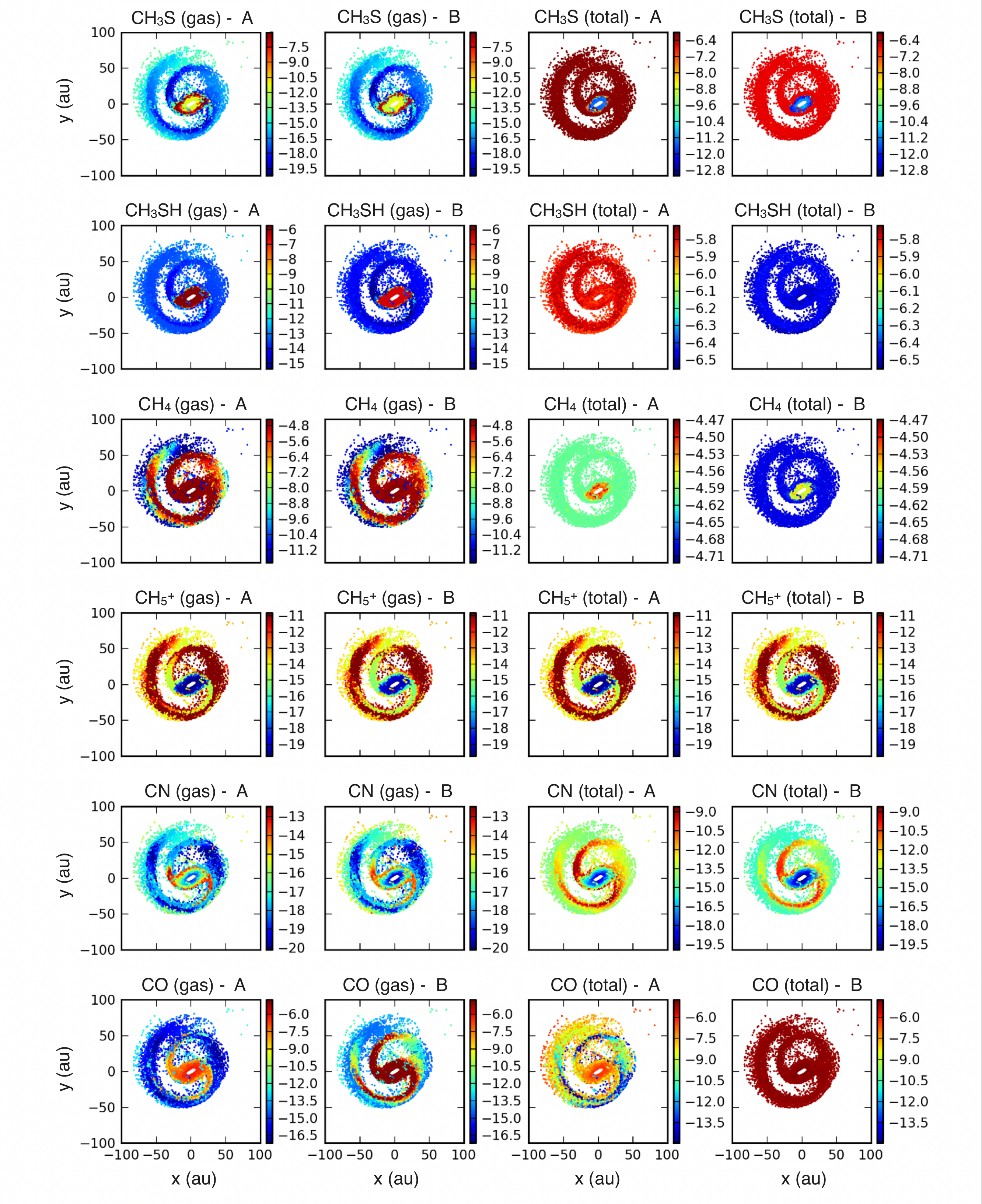}
\caption{Same as Fig. \ref{fig_spatial_distrib}.}
\label{}
\end{center}
\end{figure*}

\begin{figure*}[!ht]
\begin{center}
\includegraphics[width=\hsize]{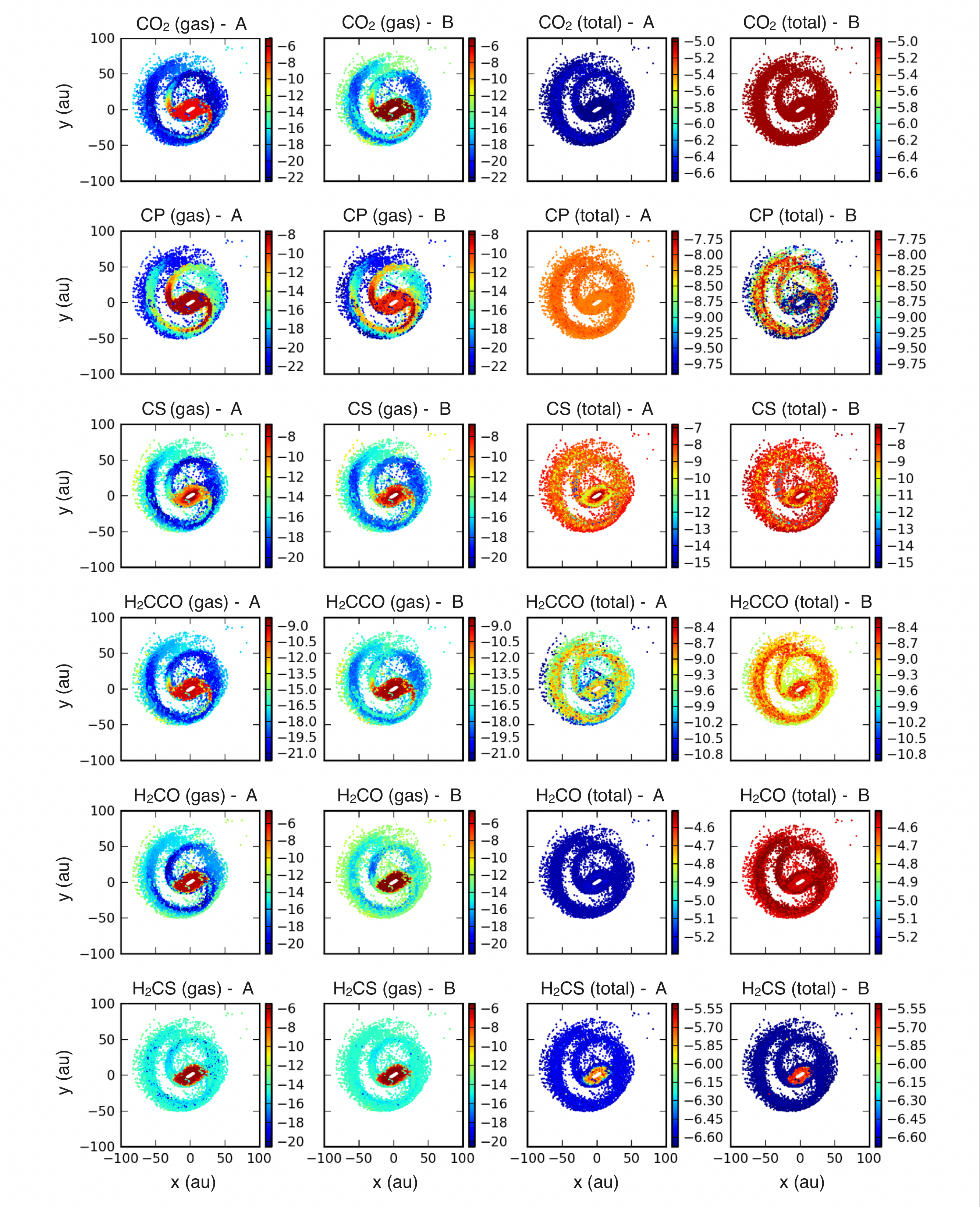}
\caption{Same as Fig. \ref{fig_spatial_distrib}.}
\label{}
\end{center}
\end{figure*}

\begin{figure*}[!ht]
\begin{center}
\includegraphics[width=\hsize]{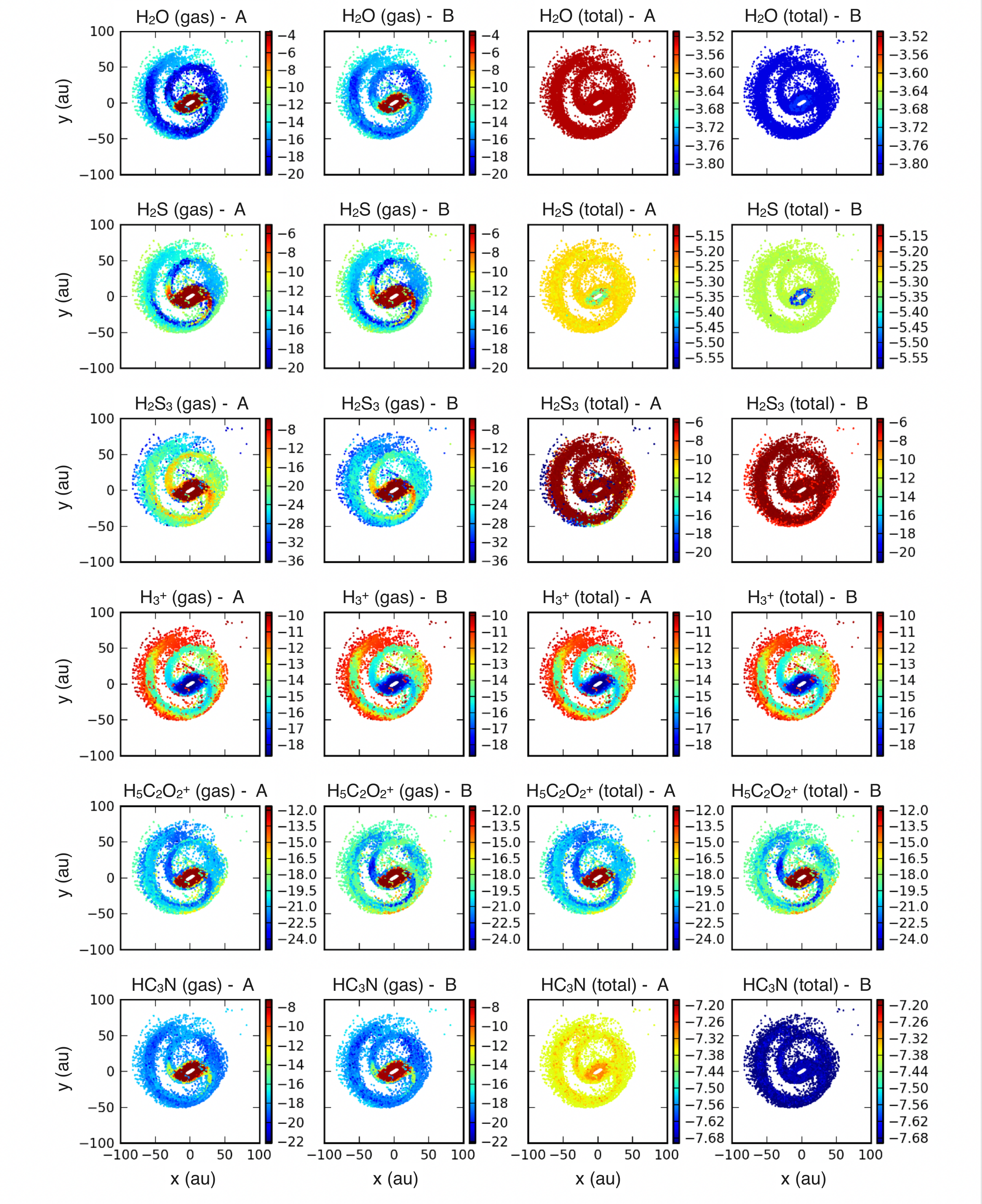}
\caption{Same as Fig. \ref{fig_spatial_distrib}.}
\label{}
\end{center}
\end{figure*}

\begin{figure*}[!ht]
\begin{center}
\includegraphics[width=\hsize]{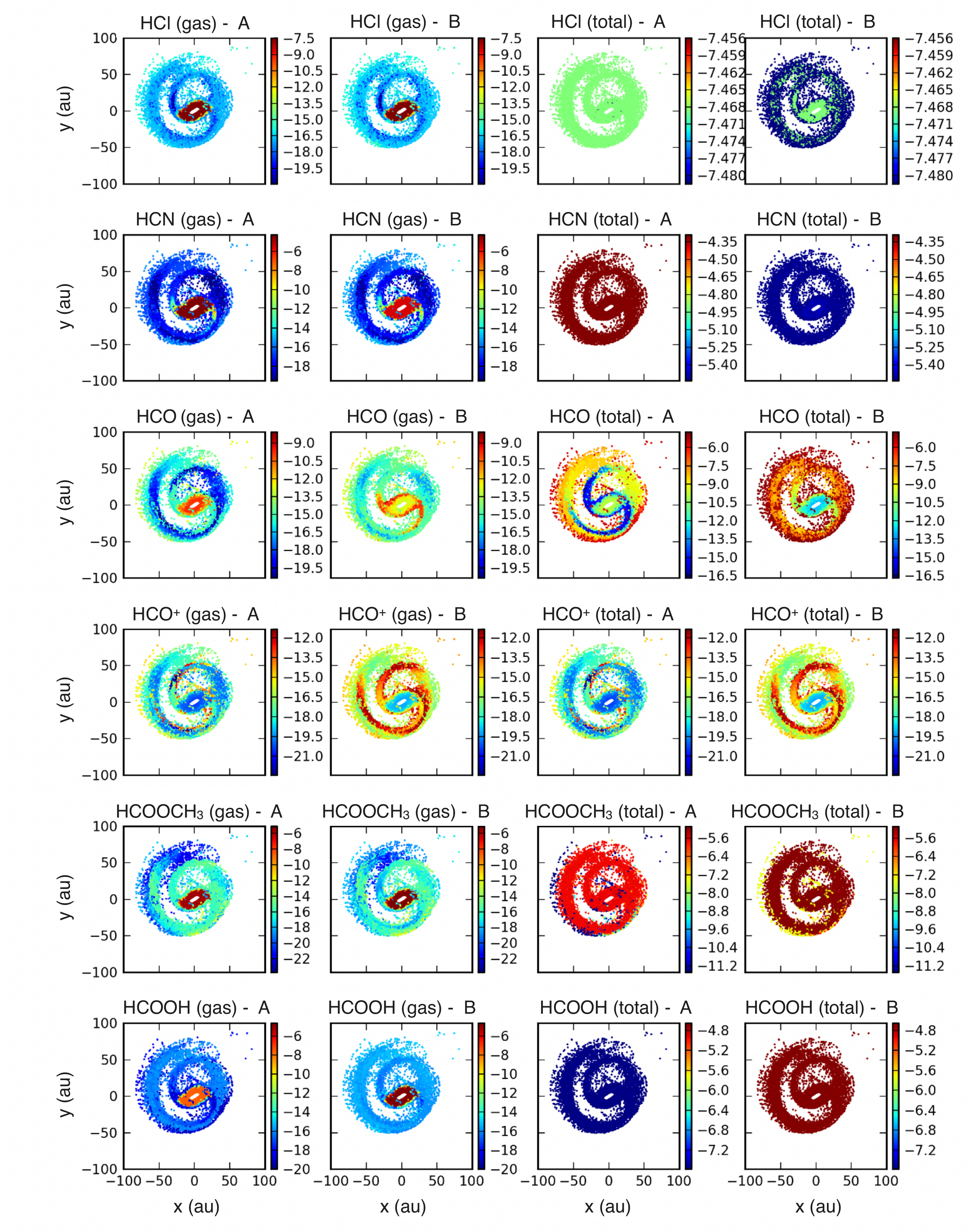}
\caption{Same as Fig. \ref{fig_spatial_distrib}.}
\label{}
\end{center}
\end{figure*}

\begin{figure*}[!ht]
\begin{center}
\includegraphics[width=\hsize]{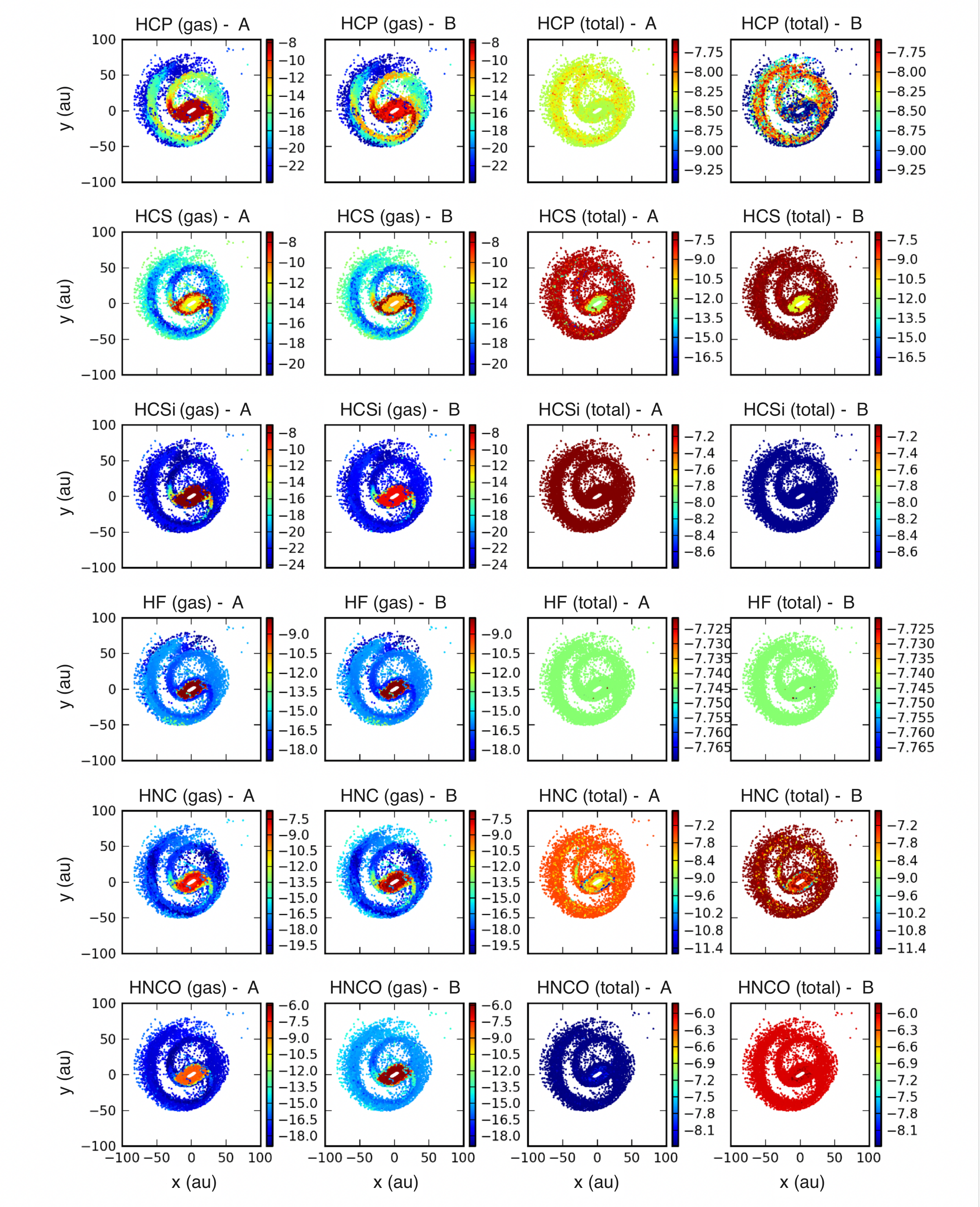}
\caption{Same as Fig. \ref{fig_spatial_distrib}.}
\label{}
\end{center}
\end{figure*}

\begin{figure*}[!ht]
\begin{center}
\includegraphics[width=\hsize]{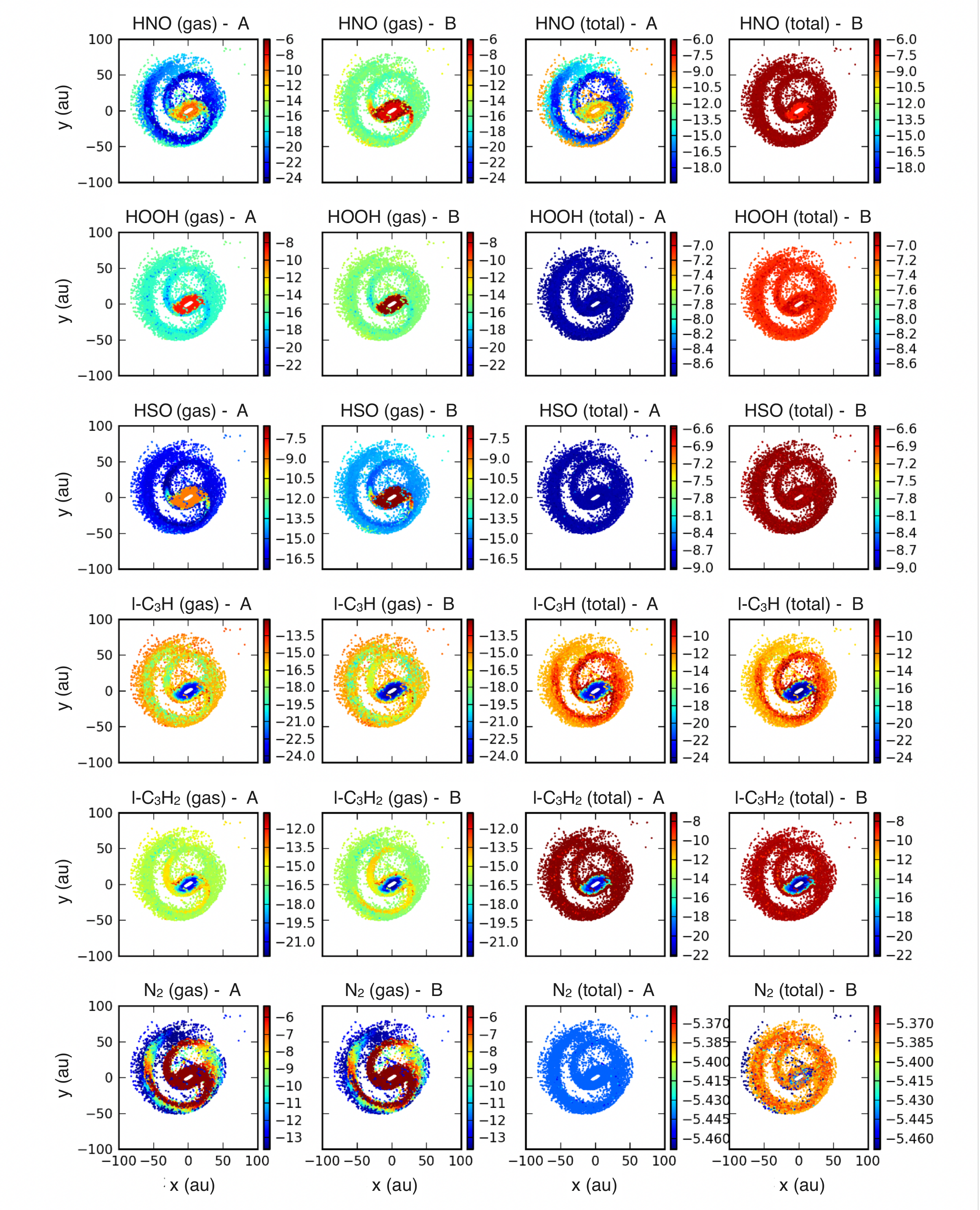}
\caption{Same as Fig. \ref{fig_spatial_distrib}.}
\label{}
\end{center}
\end{figure*}

\begin{figure*}[!ht]
\begin{center}
\includegraphics[width=\hsize]{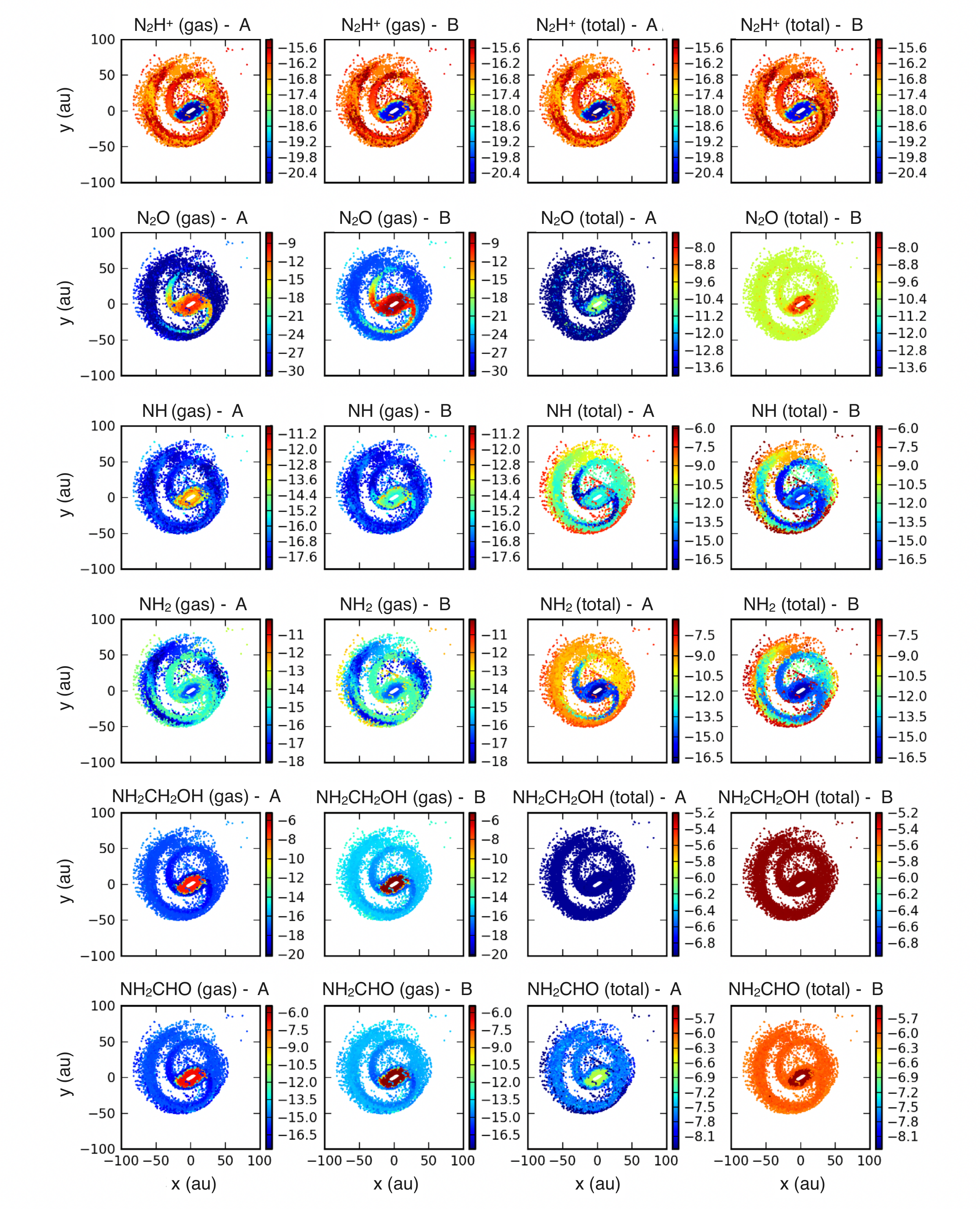}
\caption{Same as Fig. \ref{fig_spatial_distrib}.}
\label{}
\end{center}
\end{figure*}

\begin{figure*}[!ht]
\begin{center}
\includegraphics[width=\hsize]{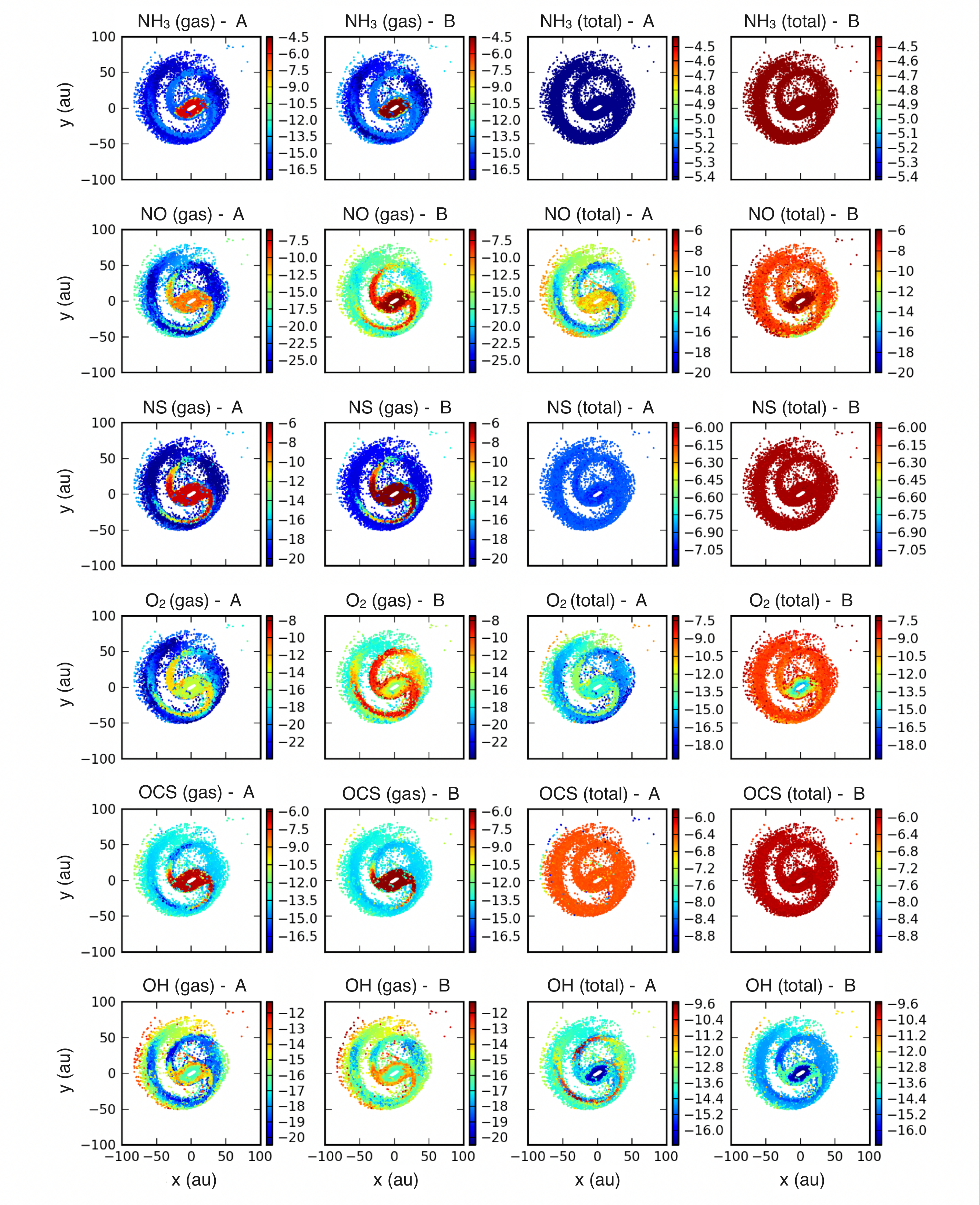}
\caption{Same as Fig. \ref{fig_spatial_distrib}.}
\label{}
\end{center}
\end{figure*}

\begin{figure*}[!ht]
\begin{center}
\includegraphics[width=\hsize]{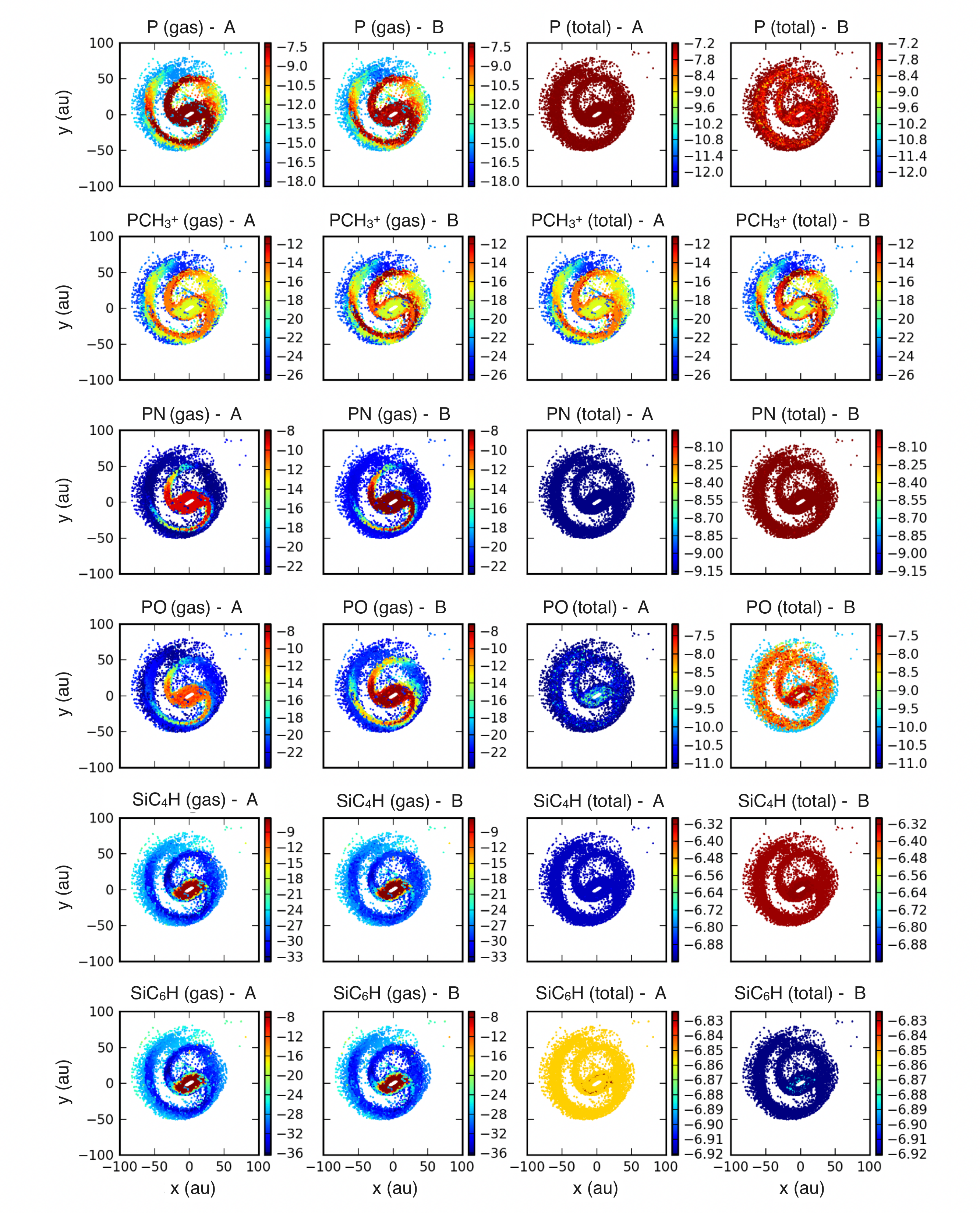}
\caption{Same as Fig. \ref{fig_spatial_distrib}.}
\label{}
\end{center}
\end{figure*}

\begin{figure*}[!ht]
\begin{center}
\includegraphics[width=\hsize]{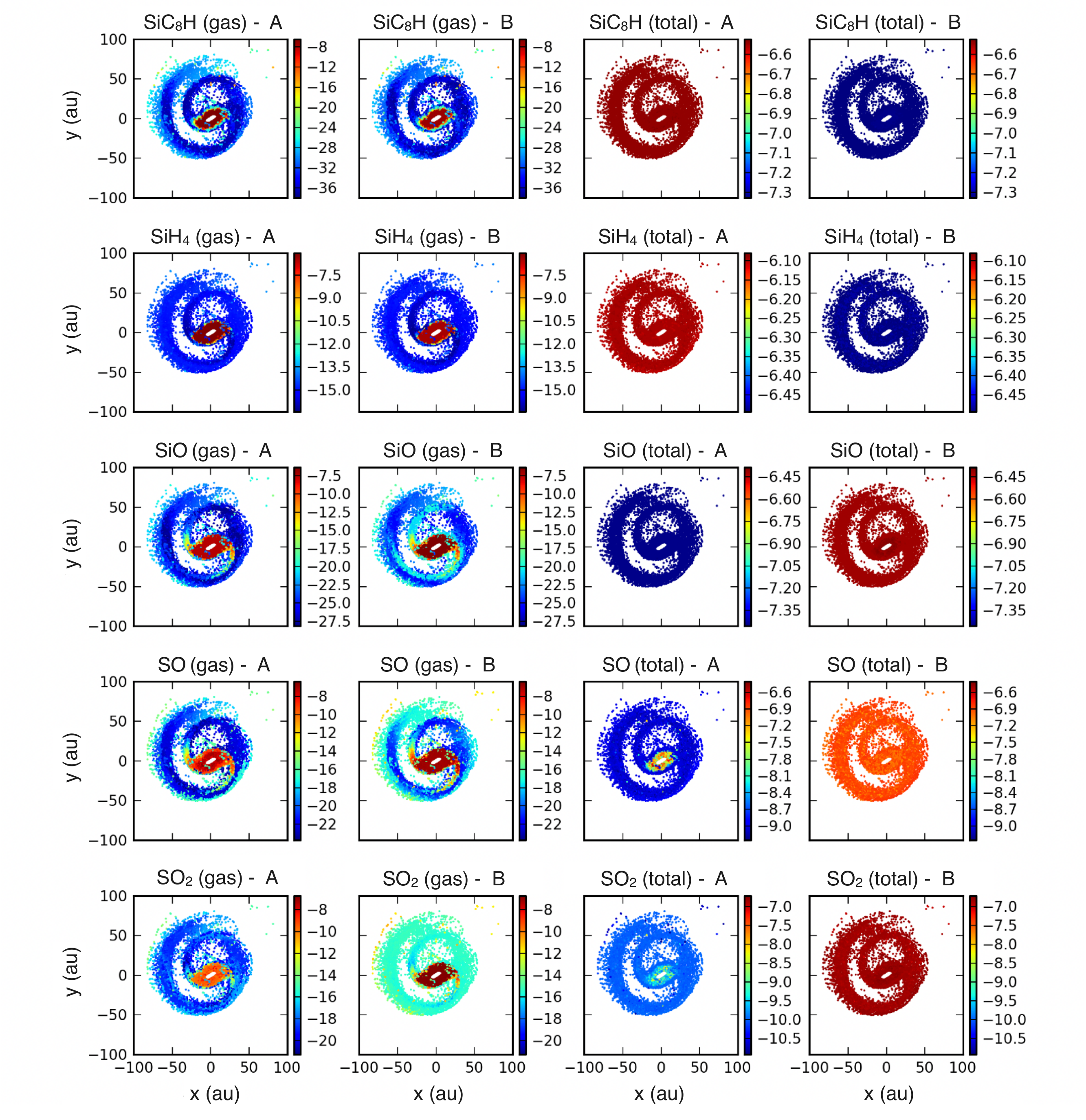}
\caption{Same as Fig. \ref{fig_spatial_distrib}.}
\label{fig_spatial_distribf}
\end{center}
\end{figure*}

\begin{figure*}[!ht]
\begin{center}
\includegraphics[width=\hsize]{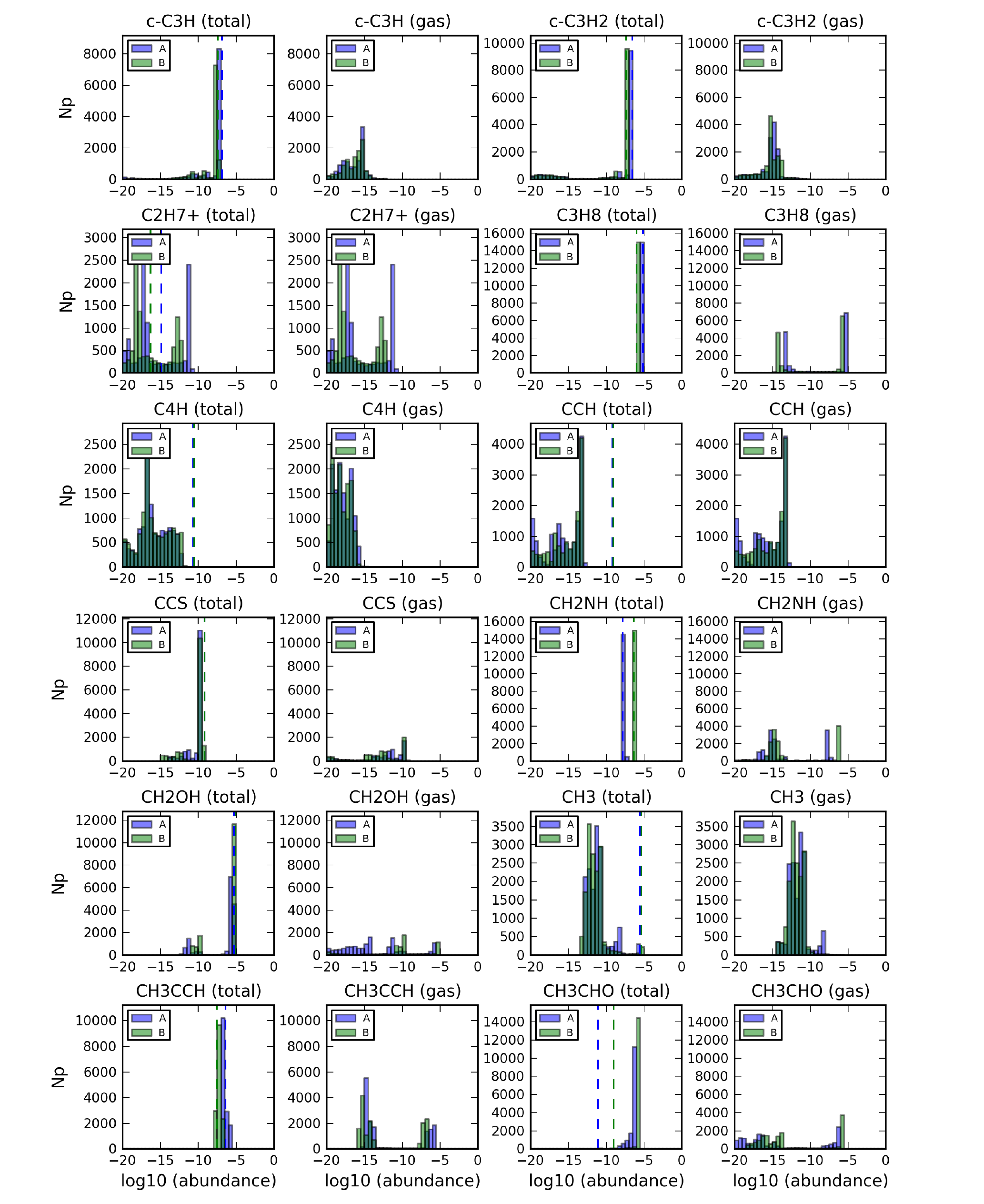}
\caption{Distribution of the number of disk particles as a function of the total and gas-phase abundances at the final time of the simulations (5.83\,$\times$\,10$^4$~yr). The distribution for the set of initial abundances A is in blue, while the one for B is in green. The initial total abundances are indicated with dashed lines. }
\label{fig_histo}
\end{center}
\end{figure*}

\begin{figure*}[!ht]
\begin{center}
\includegraphics[width=\hsize]{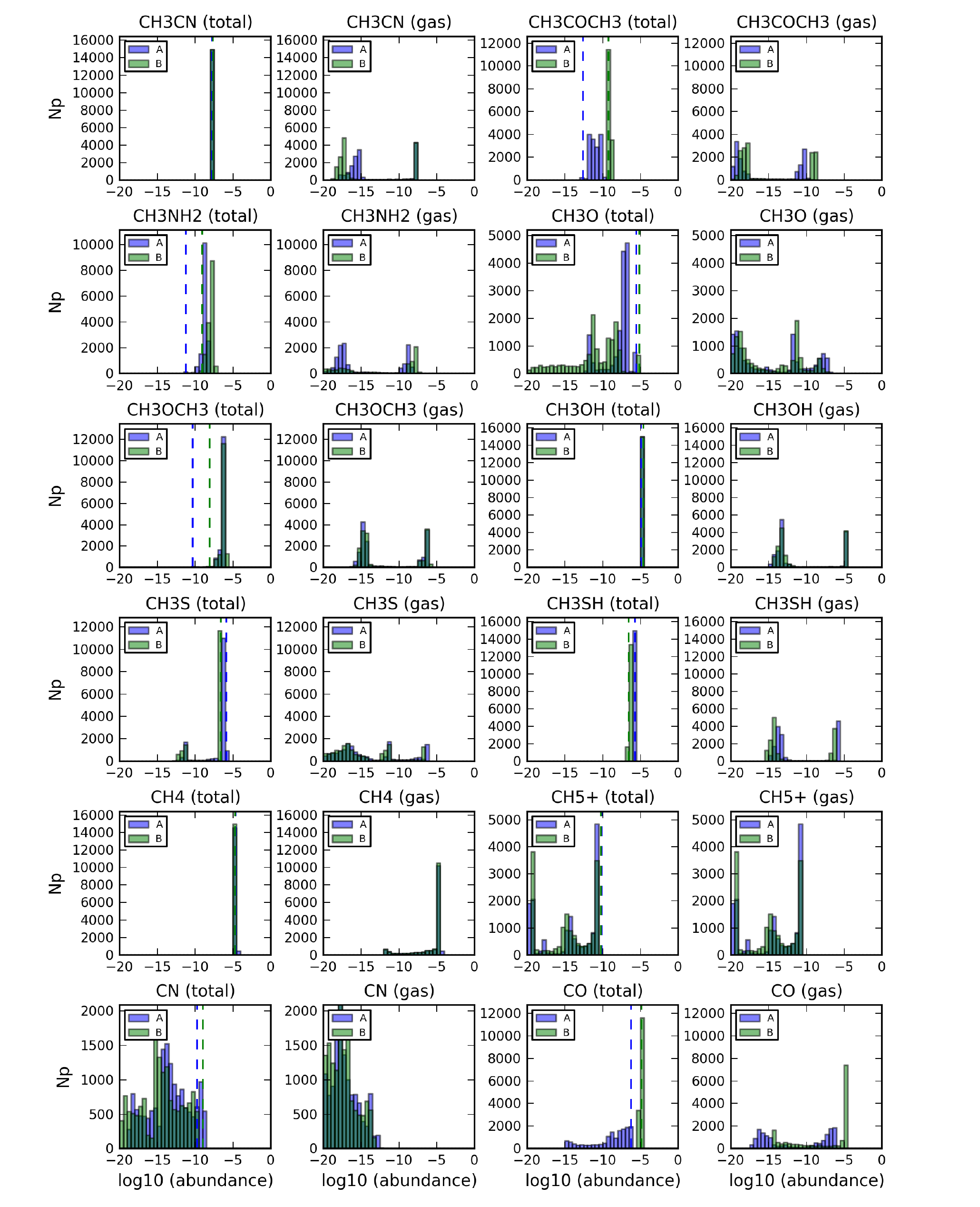}
\caption{Same as Figure \ref{fig_histo}.}
\label{}
\end{center}
\end{figure*}

\begin{figure*}[!ht]
\begin{center}
\includegraphics[width=\hsize]{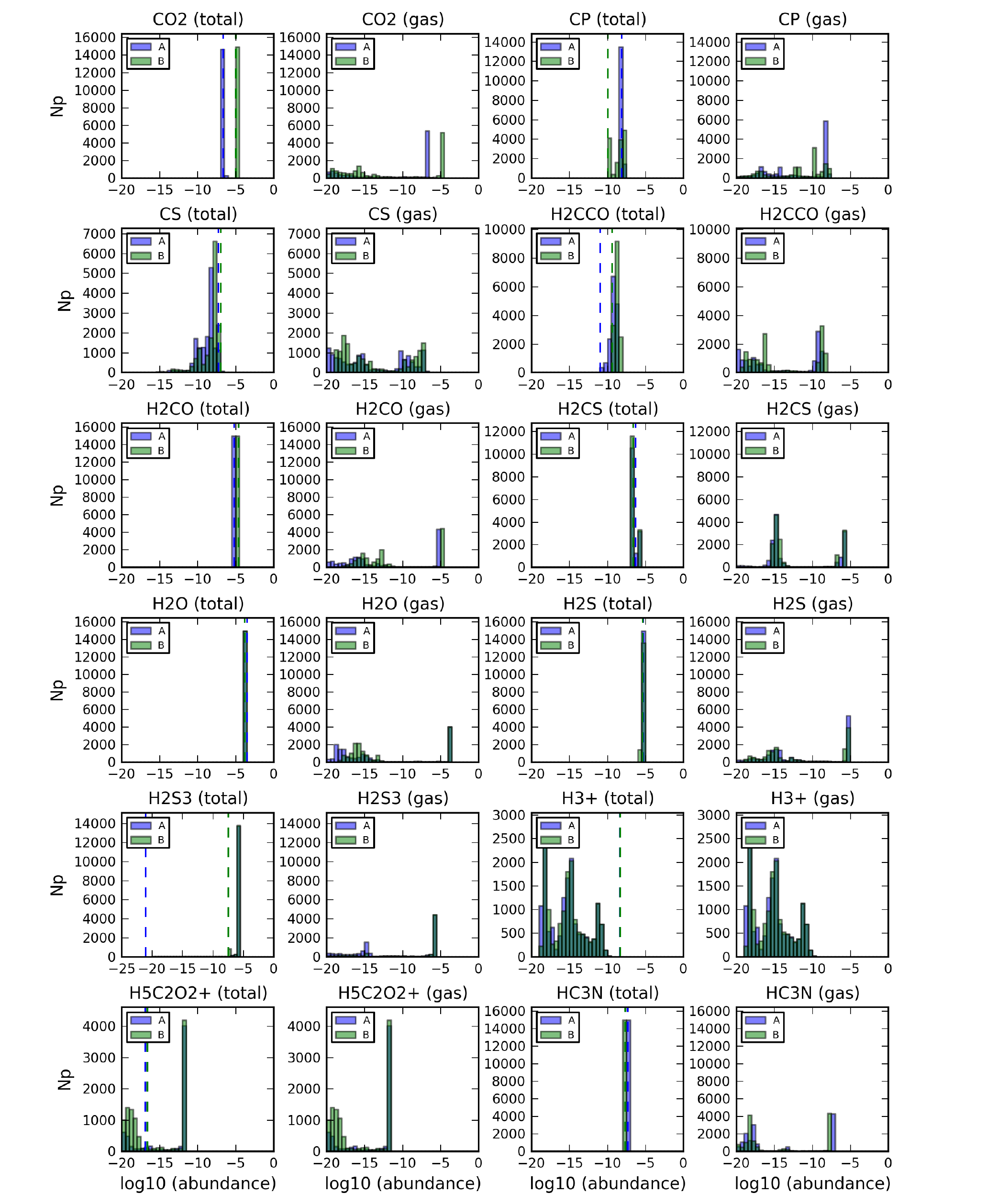}
\caption{Same as Figure \ref{fig_histo}.}
\label{}
\end{center}
\end{figure*}

\begin{figure*}[!ht]
\begin{center}
\includegraphics[width=\hsize]{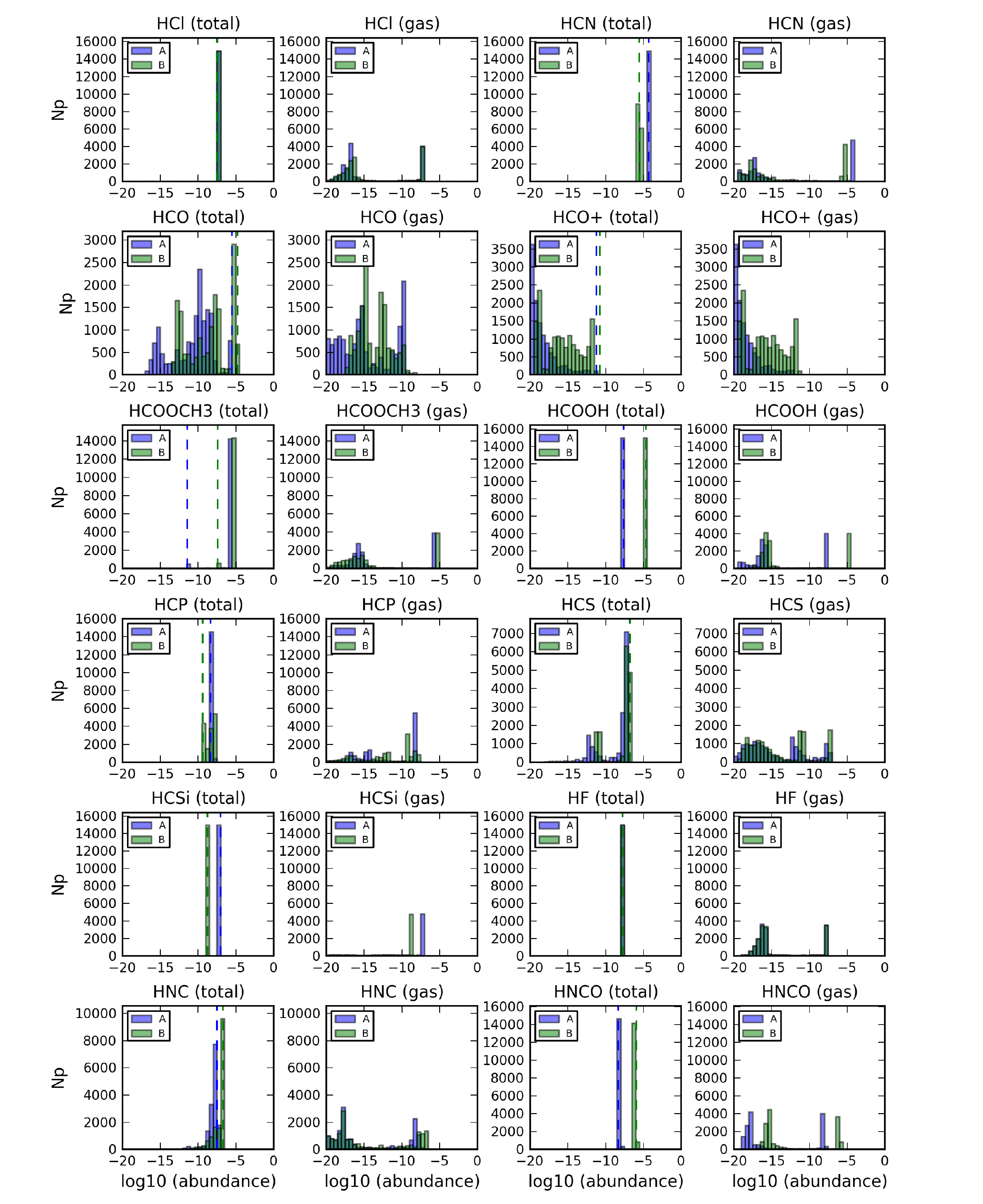}
\caption{Same as Figure \ref{fig_histo}.}
\label{}
\end{center}
\end{figure*}

\begin{figure*}[!ht]
\begin{center}
\includegraphics[width=\hsize]{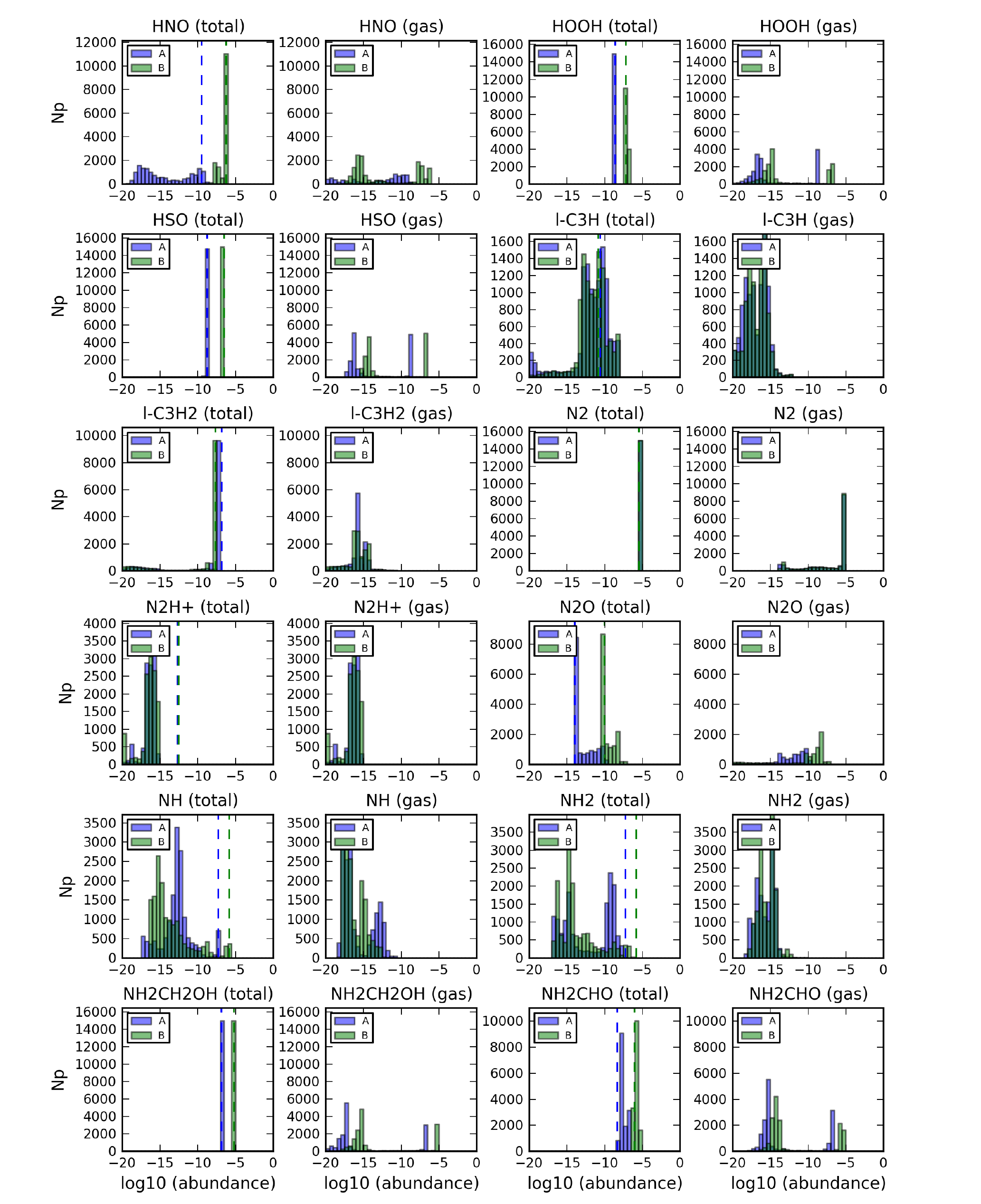}
\caption{Same as Figure \ref{fig_histo}.}
\label{}
\end{center}
\end{figure*}

\begin{figure*}[!ht]
\begin{center}
\includegraphics[width=\hsize]{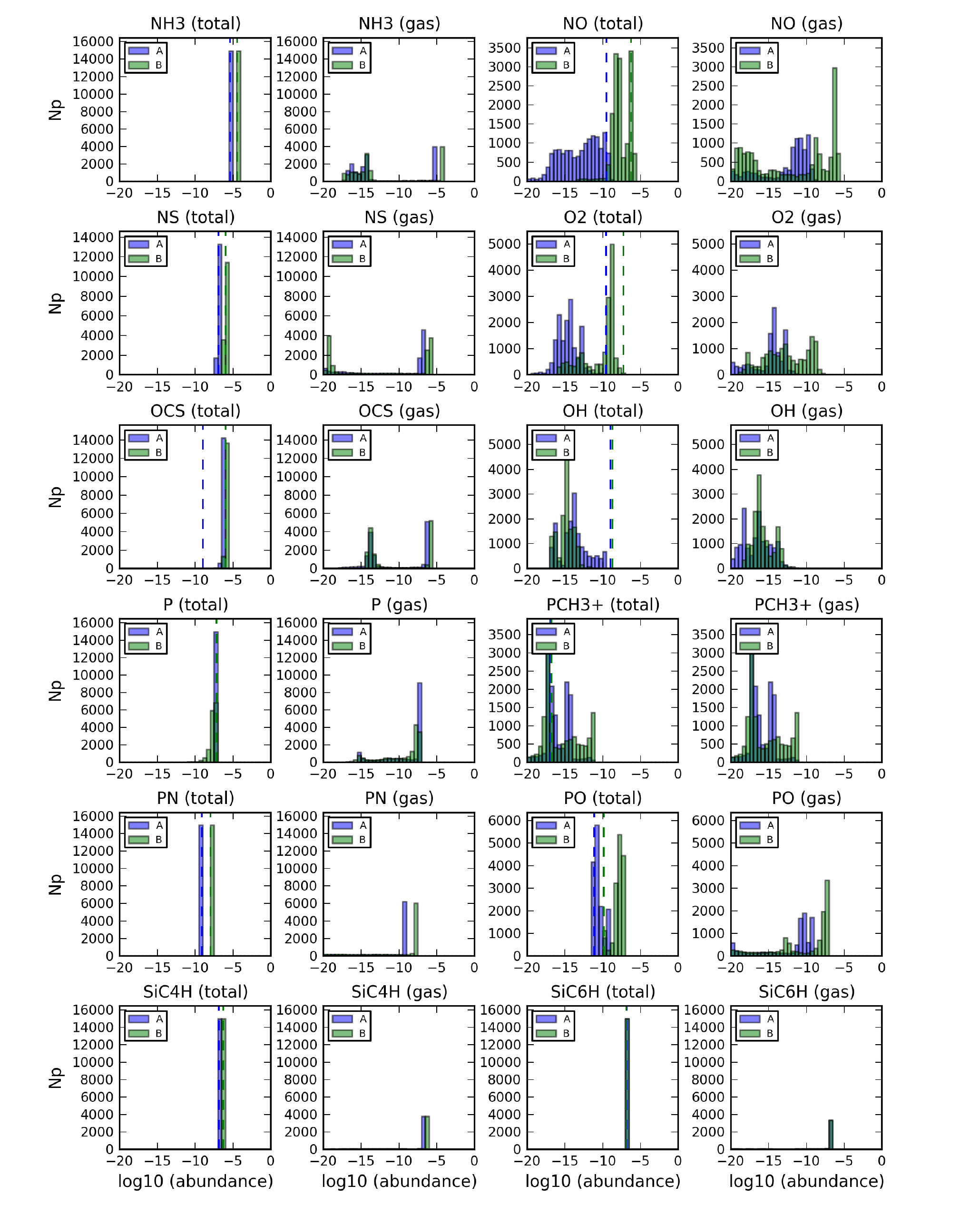}
\caption{Same as Figure \ref{fig_histo}.}
\label{}
\end{center}
\end{figure*}

\begin{figure}[!ht]
\begin{center}
\includegraphics[width=\hsize]{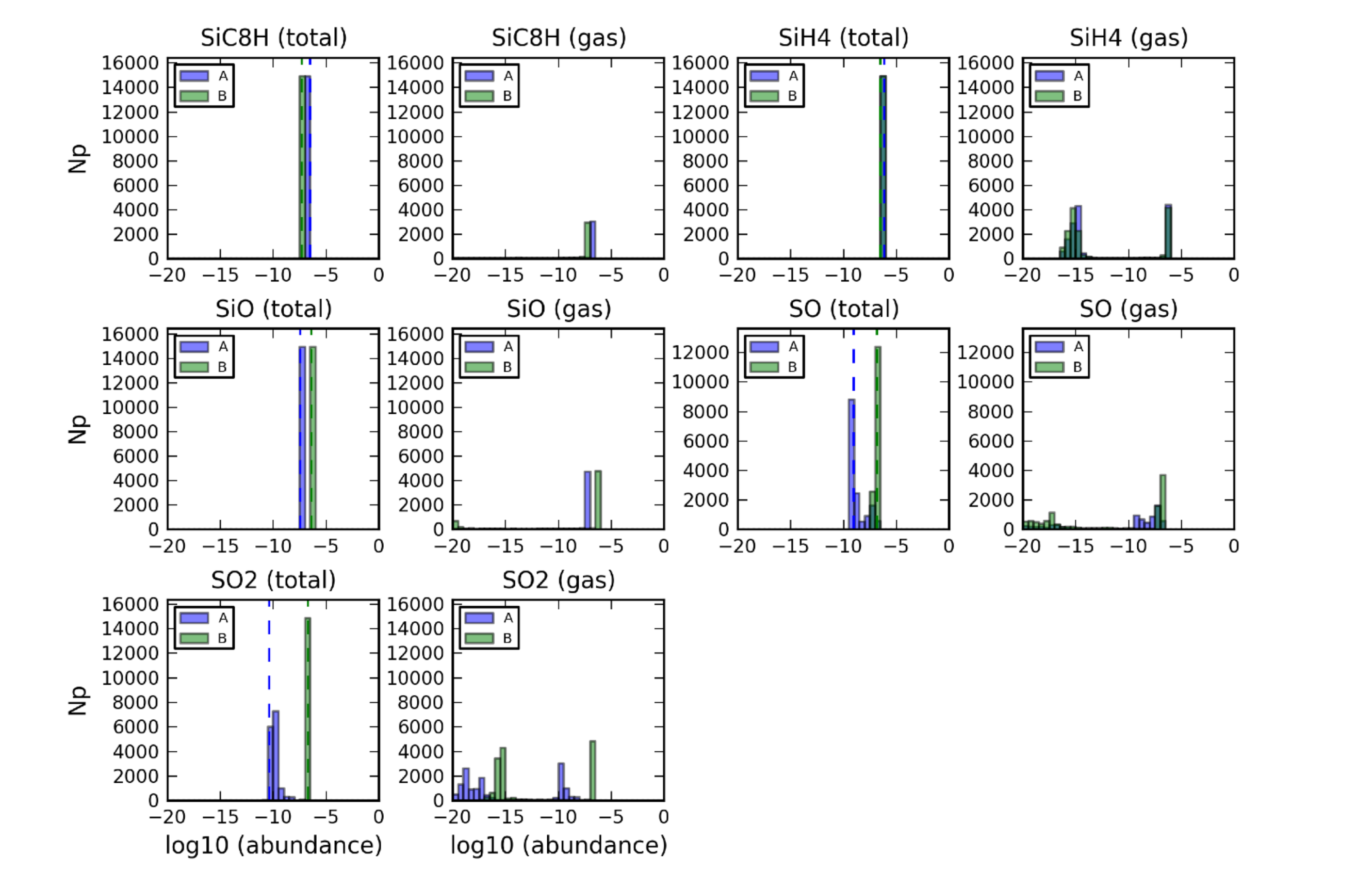}
\caption{Same as Figure \ref{fig_histo}.}
\label{fig_histof}
\end{center}
\end{figure}

\end{document}